\documentstyle[aps,eqsecnum,epsf]{revtex}

\begin{document}

\begin{flushright}
JLAB-THY-97-10 \\
hep-ph/9704207
\end{flushright}

\begin{center}
{\Large \bf  Nonforward   Parton Distributions }
\end{center}
\begin{center}
{A.V. RADYUSHKIN\footnotemark
}  \\
{\em Physics Department, Old Dominion University,}
\\{\em Norfolk, VA 23529, USA}
 \\ {\em and} \\
{\em Jefferson Lab,} \\
 {\em Newport News,VA 23606, USA}
\end{center}
\vspace{2cm}

\footnotetext{Also Laboratory of Theoretical 
Physics, JINR, Dubna, Russian Federation}

\begin{abstract}

Applications of perturbative QCD to
deeply virtual Compton scattering
and hard exclusive electroproduction 
processes require a generalization of 
usual parton distributions for the case when
long-distance information
is accumulated in  nonforward matrix
elements $\langle p' |  {\cal O}(0,z)  |p \rangle$ 
of quark and gluon light-cone  operators.
We describe 
two types of nonperturbative functions 
parametrizing such matrix elements:
double distributions $F(x,y;t)$
and nonforward distribution functions
${\cal F}_{\zeta}(X;t)$,   
discuss their spectral properties,
evolution equations which they satisfy,
basic uses and general aspects of factorization 
for hard exclusive processes.

\end{abstract}

\newpage

\section{Introduction}

 The standard feature of  
applications of perturbative QCD to hard
processes  is the introduction of  phenomenological
functions  accumulating information about
nonperturbative long-distance dynamics.    The well-known 
 examples are the 
 parton distribution functions $f_{p/H}(x)$ \cite{feynman}
used in  perturbative QCD approaches to hard
inclusive processes and 
distribution amplitudes   $\varphi_{\pi}(x), \,
\varphi_{N}(x_1 , x_2 , x_3) $,  which naturally
emerge in the  asymptotic  QCD
analyses of hard exclusive processes 
\cite{cz77,farjack,ar77,tmf,bl,bl80}.
Recently,  it was argued that the same 
gluon distribution function $f_g(x)$ used for description
of hard inclusive processes also determines 
the amplitudes  of  hard exclusive $J/ \psi$ \cite{ryskin}
and  $\rho$-meson 
\cite{bfgms} electroproduction. 
Furthermore, it was proposed \cite{ji} 
to use  another exclusive process of  
deeply virtual Compton scattering $\gamma^* p \to \gamma p' $
 (DVCS) for measuring   
quark  distribution functions inaccessible 
in  inclusive measurements
(earlier discussions of nonforward Compton-like   
amplitudes $\gamma^* p \to \gamma^* p' $ 
with a  virtual photon or $Z^0$ in the
final state  can be found   in refs. \cite{barloe,glr,drm}).
The important feature (noticed  long ago  \cite{barloe,glr})
 is that  kinematics of hard elastic 
electroproduction processes (DVCS can be also treated as one of them) 
requires 
the presence of the longitudinal component
in the  momentum transfer 
$r\equiv p - p'$ from the initial hadron to the final:
$r_{\|} = \zeta p$. For DVCS and $\rho$-electroproduction 
in the region       $Q^2 >> |t|, m_H^2$,
the longitudinal momentum asymmetry (or ``skewedness'') 
parameter $\zeta$
coincides with  the
Bjorken variable $x_{Bj} = Q^2/2(pq)$ associated with 
the virtual photon momentum $q$ \cite{afs}. 
This means that the nonperturbative matrix element
$\langle p' | \ldots | p \rangle$ 
is essentially asymmetric and, strictly 
speaking, the distributions which appear in the 
hard elastic electroproduction amplitudes
differ  from those studied in inclusive 
processes. In the latter case, 
one has  a symmetric situation when the same
momentum $p$ appears in both brackets of the hadronic matrix element
$\langle p | \ldots | p \rangle$. 

For diffractive processes, one deals with a  kinematic situation
when both the  variable $\zeta$ 
specifying the longitudinal momentum asymmetry (skewedness) 
of the nonperturbative
matrix element $\langle p' | \ldots | p \rangle$ 
and the absolute value of the momentum transfer $t \equiv (p'-p)^2$
are small.  Studying  the DVCS process,
one should be able to  consider the whole region 
\mbox{$0 \leq \zeta \leq 1$}  and $t \sim  1\, {\rm GeV} ^2$ \cite{ji2}. 
In this situation, one deals with essentially
{\it nonforward} (or {\it off-forward} in terminology 
of Ref.   \cite{ji})
kinematics for the matrix element $\langle p' | \ldots | p \rangle$. 
The basics of the    PQCD approaches   
incorporating  asymmetric/off-forward  parton distributions 
were formulated in refs.\cite{ji,compton,gluon,ji2}.
A detailed analysis of PQCD factorization 
for hard meson electroproduction processes
was given in Ref.  \cite{cfs}.
Applications of  asymmetric gluon  distributions 
to elastic diffractive $J/\psi$  electroproduction
were discussed in \cite{hbhoy,hz,fks}.
In a recent paper \cite{jims}, the  
off-forward   quark  distributions 
were studied within the MIT bag 
model.  A  numerical study 
of the evolution of the asymmetric gluon distribution
was attempted in Ref.  \cite{ffgs}.  
Thus,  there is an increasing interest in the  
studies of   these new types of hadron distributions,
their general properties and applications.

Our goal in the present paper is to give a detailed
description of the approach outlined in 
our earlier papers \cite{compton,gluon}.
The basic idea of refs.\cite{compton,gluon}  is that 
constructing a consistent PQCD approach  for 
 amplitudes of hard exclusive electroproduction processes
one should treat the initial momentum $p$ and 
the longitudinal part of the momentum transfer $r$ on 
equal footing by  introducing 
double distributions
$F(x,y)$, which specify the fractions of  $p$ 
and  $r_{\|}$ 
carried by the constituents 
of the nucleon.
These distributions have hybrid properties:
they look like distribution functions 
with respect to $x$ and like  distribution amplitudes
with respect to $y$.
Writing  matrix elements of  composite operators 
in terms of   double distributions
is the starting point of constructing 
  the PQCD parton picture.  
Another important step is  taking  into account 
the logarithmic scaling violation. 
The  evolution
kernels $R(x,y;\xi,\eta)$ for 
double distributions have a remarkable property:
they  produce  the DGLAP evolution 
kernels $P(x/\xi)$ \cite{gl,ap,d}
when  integrated  over $y$,  while 
integrating  $R(x,y;\xi,\eta)$ over $x$
one obtains the 
BL-type   evolution kernels $V(y,\eta)$  \cite{bl,bl80} 
for the relevant  distribution amplitudes\footnote{Originally,
the evolution equation for the pion 
distribution amplitude in QCD
was derived and solved in Ref.  \cite{tmf},
where the anomalous dimension matrix 
$Z_{nk}$ was used instead of $V(y,\eta)$ (see also \cite{phlet}).}.
One can use   these properties of the kernels
 to  construct  formal solutions of the 
one-loop evolution equations for  the double  distributions. 
The  longitudinal momentum transfer  
$r_{\|}$ is proportional to $p$: $r_{\|} = \zeta p$
and, for this reason,  it is  convenient to parametrize
matrix elements $\langle p-r | \ldots  |p \rangle$
by  {\it asymmetric distribution functions}
${\cal F}_{\zeta} (X)$  specifying the total light-cone fractions
$Xp$, $(X-\zeta)p$   of  the initial hadron momentum $p$
carried by the ``outgoing'' and ``returning'' partons\footnote{The
 asymmetric  distribution functions 
defined in Ref.  \cite{gluon} 
are  similar  to, but not  coinciding  with
the $t \to 0$ limit of the
off-forward parton distributions 
introduced by X.Ji \cite{ji}, see Sec.  IX.}.  
It should be emphasized that double 
distributions $F(x,y)$ are universal functions in the sense that
they do not depend on the skewedness 
parameter $\zeta$ while the asymmetric  distribution
functions ${\cal F}_{\zeta} (X)$  form a family 
of $X$-dependent functions changing their shape when 
$\zeta$ is changed.
The functions ${\cal F}_{\zeta} (X)$ 
also have hybrid properties. 
In the region $X \geq \zeta$  the
returning parton   carries a positive
fraction $(X-\zeta)p$ of the initial momentum $p$,
and hence ${\cal F}_{\zeta} (X)$  is similar 
to the usual parton distribution $f(X)$.
On the other hand, in the region $0 \leq X \leq \zeta$
the difference $X- \zeta$ is  negative,
${\rm i.e.},$ the second parton should be  treated
as propagating together with the first one.
The partons in this case share 
the longitudinal momentum transfer $r_{\|} =\zeta p$ 
in fractions $Y\equiv X/\zeta$ and $1-Y$. 
This means that in the region $X \leq \zeta$
the function ${\cal F}_{\zeta} (X)$ 
looks like a distribution amplitude.
It is possible to  formulate 
equations governing the evolution
of the asymmetric distribution functions 
${\cal F}_{\zeta}^g (X)$
and establish  relations  between these functions,  
  double  distributions $F(x,y)$ 
and usual distribution functions $f(x)$ \cite{compton,gluon}.

Constructing a QCD  parton picture for 
hard electroproduction processes, 
it is very important to know  spectral 
properties of the relevant parton distributions $F(x,y)$
and ${\cal F}_{\zeta} (X)$. 
Using the approach  \cite{spectral} based on the 
$\alpha$-representation analysis \cite{bosh,analsmat,nakan,oz}, 
it is possible to  prove that  double distributions
$F(x,y)$ have a natural property that both $x$ and $y$ 
satisfy the ``parton'' constraints 
 $0 \leq x \leq 1$, $ 0 \leq y \leq 1$
for  any Feynman  diagram contributing  to $F(x,y)$. 
A less obvious restriction   $0 \leq x+y \leq 1$  guarantees 
that the argument $X=x+y \zeta$ of the  asymmetric distribution
${\cal F}_{\zeta} (X)$ also changes between the limits 
$0$ and $1$.  An important observation
here is that $X=0$ can be obtained only if both $x=0$
and $y=0$.  Because of  vanishing 
phase space for such a configuration,
one may expect that  asymmetric distributions ${\cal F}_{\zeta} (X)$ 
vanish for $X=0$.  This property is very essential,
because the hard subprocess amplitudes usually
contain $1/X$ factors.  When   ${\cal F}_{\zeta} (0) \neq 0$,
one faces a singularity   ${\cal F}_{\zeta} (X)/X$  
at the end-point of the integration region
$0 \leq X \leq 1$. Since such a singularity 
is not integrable, 
factorization of short- and long-distance contributions 
does not work in that case. 

The paper is organized in the following way.
In Sec.  II, we consider parton distributions
in a toy scalar model. 
Despite its simplicity, it shares 
many common features with the realistic QCD case.
In particular,  the spectral properties of 
distribution functions are  not affected 
by the  numerators  of quark and gluon propagators,
derivatives in triple-gluon vertices,  ${\rm etc.}$ 
Hence,   studying a scalar model
we just concentrate on the denominator structure 
of the relevant momentum integrals,
which is the same in  both theories. We start 
with the simplest  example of the usual (forward)
distribution $f(x)$ and then consider more and more complicated 
functions: the double distribution $F(x,y)$, 
asymmetric distribution function  ${\cal F}_{\zeta} (X)$
and nonforward distribution  ${\cal F}_{\zeta} (X;t)$.
Explicit expressions for these functions
at one-loop level are  obtained with the help
of the  $\alpha$-representation.  
Using the latter one can easily establish the
spectral properties of the distribution functions.
The  $\alpha$-representation also provides 
a very effective starting point for a general analysis 
of factorization and large-$Q^2$ behavior of elastic amplitudes. 
In Sec.  III, we outline  the all-order extension 
of the one-loop analysis. 
We give an all-order definition of  the  
 double  distribution function
$F(x,y)$ and demonstrate that it has the spectral 
properties $0 \leq \{ x, y, x+y \} \leq 1$.
We show  how one can use the 
$\alpha$-representation analysis for 
finding integration regions responsible
for the leading large-$Q^2$ contributions.
We also discuss modifications of twist  counting rules
in QCD due to cancellations   between different 
gluonic contributions in  the
Feynman gauge and other complications 
which appear in gauge theories.
In Sec.  IV, we give definitions of nonforward distributions
${\cal F}_{\zeta} (X;t)$
in QCD.  Just like the usual 
distribution functions $f(x)$ and distribution
amplitudes $\varphi(y)$, the new 
distributions depend on the factorization 
scale $\mu$, ${\rm i.e.},$  it is more appropriate to  use the notation  
${\cal F}_{\zeta} (X;t;\mu)$ for the nonforward distributions rather than 
simply ${\cal F}_{\zeta} (X;t)$.   Evolution equations governing the
$\mu$-dependence of the nonforward distributions are discussed 
in  Sec.  V and  Sec. VI.  We  show how one can obtain  
evolution kernels for    nonforward distributions
using already  known kernels $B(u,v)$  of the evolution 
equation for the light-ray operators \cite{brschweig,bb}.
Since this equation has an operator form, 
substituting it into a specific  matrix element one can convert   
$B(u,v)$ into  desired  evolution kernels.
In particular, taking   $\langle p | \ldots |p \rangle $ 
one obtains the DGLAP kernels, choosing $\langle 0 | \ldots |p \rangle $ 
one gets BL-type kernels while  resorting to 
$\langle p' | \ldots |p \rangle $ and 
parametrizing the matrix elements through $F(x,y)$ or 
${\cal F}_{\zeta} (X)$ one ends up with the  
kernels $R(x,y;\xi,\eta)$  and $W_{\zeta}(X,Z)$ 
governing the evolution of double and asymmetric/nonforward distributions,
respectively.  In Sec.  V,  we discuss the derivation of  the evolution 
kernels $W_{\zeta}(X,Z)$ for the nonforward distributions.
 We show, in particular, that in the region $0  \leq \{ X,Z \}  \leq \zeta $,
the kernels $W_{\zeta}(X,Z)$ reduce to the BL-type   
kernels $V(X,Z)$ calculated for rescaled variables $X/\zeta$,
$Z/\zeta$. This result is very natural, since  ${\cal F}_{\zeta} (X)$
can be treated as a distribution amplitude when $X \leq \zeta$.
In the opposite limit $\zeta \leq \{ X,Z \}\leq 1 $,
the evolution is similar to that of the DGLAP equation,
the basic distinction  being the  difference
between the outgoing $X,Z$ and returning 
$X' \equiv X - \zeta, Z'  \equiv Z - \zeta$
momentum fractions. We show that writing 
the kernels  $W_{\zeta}(X,Z)$ in terms of the  fractions $X,X',Z,Z'$
in the region 
$\zeta \leq \{ X,Z \} \leq 1 $ gives the 
functions  $W(X,X';Z,Z')$ which 
have the symmetry property   with respect to the 
interchange of initial and final partons:
$W(X,X';Z,Z') = W(X',X;Z',Z) $. 
For $\zeta =0$ one has $X=X', Z=Z'$ and the kernels
$W_{\zeta=0}(X,Z)$  acquire  the DGLAP form.
In Sec.  VI, we discuss the 
QCD evolution  of the nonforward distributions.
Qualitatively,  the evolution can be described 
in the following way.
Due to the DGLAP-type evolution in the $X > \zeta$ region, 
the momenta of  partons decrease, and distributions
shift  into  the regions of smaller $X$. However, when the parton momentum
degrades  to values smaller than the momentum transfer
$r = \zeta p$, the further evolution is like 
that for a distribution amplitude:
it tends to make the distribution symmetric 
(or antisymmetric) with respect to
the central point $X= \zeta/2$ of the $(0, \zeta)$ segment. 
In section VII, we briefly discuss two basic uses 
of nonforward distributions:
deeply virtual Compton scattering and hard elastic meson electroproduction.
In particular, we show how to combine the definition of the gluon 
distribution through  the  matrix element of the 
 gauge-invariant gluonic operator $ G_{\mu \alpha}^a (0) E_{ab}(0,z;A) 
G_{ \alpha \nu}^a (z)$  with the usual Feynman rules
formulated for the vector potential $A_{\mu}^a$.
In Sec.  VIII, we discuss possible sources of PQCD factorization
breaking for  hard elastic electroproduction processes,
due to  singularities 
at the end-points of the integration region. 
In particular, we emphasize  the importance of establishing the 
${\cal F}_{\zeta} (0)=0$ property for the nonforward distributions.
In Sec.  IX, we compare our notations, definitions and terminology
with those   used by  other authors 
(off-forward parton distributions $H(x,\xi;t)$ 
introduced by X. Ji \cite{ji}  and  nondiagonal 
distributions $f(x_1,x_2)$ defined by 
Collins, Frankfurt and Strikman \cite{cfs}).
Sec.  X contains  concluding remarks.

\section{Forward and nonforward distributions in  scalar toy model}

\subsection{Introductory remarks}

The  parton distributions ${\cal F}_{\zeta}(X;t;\mu)$
parametrizing  nonforward  matrix 
elements $\langle p' | {\cal O}(0,z) |p \rangle$
of composite two-body operators ${\cal O}(0,z)$  on the light cone
$z^2=0$  depend on four parameters.
In addition to the ``usual'' parton variable $X$
specifying  the fraction $Xp$ of the initial 
hadron momentum $p$ carried by the active parton 
(more formally,  $X$ may be treated as 
 the Fourier-conjugate parameter
to $(pz)$),  the functions  ${\cal F}_{\zeta}(X;t;\mu)$ 
also depend on 
the invariant momentum transfer $t=(p'-p)^2$,
the skewedness parameter $\zeta = (rz)/(pz)$ 
(where $r \equiv p-p'$) and  the evolution/factorization scale
$\mu$.   The latter  characterizes the subtraction procedure 
for singularities that appear on the light cone $z^2=0$ 
(in general, $\mu$ may be different from the 
scale $\mu_R$ introduced by the $R$-operation 
for ordinary ${\rm UV}$ divergences, but the usual
convention is to take $\mu = \mu_R$). 
Furthermore, depending on the type
of the composite operator ${\cal O}(0,z) $, 
one would get quark, antiquark, flavor-singlet,
flavor-nonsinglet, gluonic, spin-dependent,
spin-independent,  {\it {\rm etc.}} distributions. 
In this situation, we  propose  to follow  a  step by step approach.
We will start with  simplest examples and then gradually proceed to more 
complicated ones.  For this reason, 
we consider  first a toy scalar model. The lowest 
nontrivial level corresponds to one loop Feynman diagrams.
The relevant integrals are easily calculable, 
and  their study provides  useful information about 
the structure  of the nonforward 
distributions, especially about their spectral properties,
because the  latter are insensitive to
numerators of quark and gluon propagators and other  complications
brought in by the spin structure of the  realistic QCD  case.

\subsection{Forward  distribution functions}

Our starting point is the scalar analogue  of the usual
``forward'' parton distribution functions $f(x)$.
Consider a one-loop box diagram for a scalar
version of the virtual forward 
Compton amplitude (Figs.1$a,b$). Both incoming and outgoing
virtual ``photons'' have momentum $q\equiv q'-\zeta p$, where
$q'$ and $p$ are lightlike momenta $(q')^2=0, p^2 =0$.
The ``photons'' couple with the constant $e$ 
to a massive scalar ``quark'' field $\phi$.
The initial and final hadrons  are imitated 
by massless scalar particles with the momentum $p$.
Their coupling to the quarks is specified by a constant $g$. 
In these  notations, $q^2 \equiv -Q^2 = -2 \zeta (pq')$.
Since $(pq)=(pq')$, the parameter $\zeta$ coincides with the Bjorken
variable $\zeta = x_{Bj} \equiv Q^2/2(pq)$.
Using the $\alpha$-representation for the scalar propagators
\begin{equation}
\frac1{m^2-k^2 - i\epsilon} =
{i} \int_0^{\infty} e^{i\alpha (k^2-m^2+i\epsilon)} 
\, d \alpha
\label{1} \end{equation}
and calculating  the resulting Gaussian integral over the
loop momentum $k$ we obtain for the diagram 1$a$:
\begin{equation}
T_a(p,q) = 
- \frac{e^2 g^2}{16 \pi^2}
\int_0^{\infty}  \exp\left \{  i \left [2(pq') \, 
\alpha_1 {{
\alpha_3 - \zeta(\alpha_2+\alpha_3+\alpha_4)
}\over{\alpha_1+\alpha_2+\alpha_3+\alpha_4}}
-\lambda \, (m^2-i\epsilon)  \right ] \right \}
\frac{d\alpha_1 d\alpha_2 d\alpha_3 d\alpha_4 }{\lambda^2}\, . 
\label{2} \end{equation}
We use the shorthand notation $\lambda \equiv 
\alpha_1+\alpha_2+\alpha_3+\alpha_4$. 
The large-$Q^2$ asymptotics is determined by integration
over the region where the coefficient accompanying $2(pq')$
vanishes. Otherwise, the integrand rapidly 
oscillates and the result of  integration 
is exponentially suppressed. Integration 
over $\alpha_1 \sim 0$ region is evidently 
the simplest possibility. Other 
variants are $\alpha_1+\alpha_2+\alpha_3+\alpha_4 \sim 0$
or $\alpha_3 - \zeta (\alpha_2+\alpha_3+\alpha_4) \sim 0$.
It is easy to check that the leading power behavior 
is generated by the  $\alpha_1 \sim 0$ integration, which gives 
\begin{equation}
T_a(p,q) = 
- \frac{i e^2 g^2}{16 \pi^2} 
\int_0^{\infty} 
\frac1{2(pq')(\alpha_3/  \tilde \lambda  - \zeta+i\epsilon)}
\frac{e^{ - i \tilde \lambda (m^2-i\epsilon)}}{\tilde \lambda^2} \,
 d\alpha_2 d\alpha_3 d\alpha_4 \, + O(1/Q^4) \ ,   
\label{3} \end{equation}
where $\tilde \lambda \equiv  \alpha_2+\alpha_3+\alpha_4$.
Introducing the  distribution function
\begin{equation}
 f(x) = \frac{i  g^2}{16 \pi^2} 
\int_0^{\infty} \, 
\delta \left ( x - \frac{\alpha_3}{\alpha_2+\alpha_3+\alpha_4}
\right   ) \, 
\frac{e^{ - i \tilde \lambda (m^2-i\epsilon)}}{\tilde \lambda^2} \,
 d\alpha_2 d\alpha_3 d\alpha_4 \,  ,
\label{4} \end{equation}
we can write the leading power contribution in the parton form:
\begin{equation}
T_a^{\rm as} (p,q) = -\int_0^1  \frac{e^2}{2(pq')(x-\zeta+i\epsilon)} 
\,  f(x) \, dx = -\int_0^1  \frac{e^2}{(xp+q)^2+i\epsilon} 
\,  f(x) \, dx  \equiv \int_0^1  t_a(xp,q)
\,  f(x) \, dx \, .
\label{5} \end{equation}
At the last step,  we introduced the parton subprocess amplitude
\begin{equation} t_a(xp,q) = - \frac{e^2}{(xp+q)^2+i\epsilon}\, .
\label{6} \end{equation}
Hence, the parameter $x$ can be treated as the 
fraction of the initial momentum $p$ carried
by the quark interacting with the virtual photon.
Note that the limits $0 \leq x \leq 1$ 
necessary for this interpretation of $x$ 
are automatically imposed by the $\alpha$-representation
of $f(x)$.
A similar result holds for the $u$-channel diagram 1$b$:
\begin{equation}
T_b^{\rm as} (p,q) = \int_0^1  \frac{e^2}{2(pq')(x+\zeta-i\epsilon)} 
\,  f(x) \, dx = -\int_0^1  \frac{e^2}{(xp-q)^2+i\epsilon} 
\,  f(x) \, dx \equiv \int_0^1  t_b(xp,q)
\,  f(x) \, dx \, .
\label{7} \end{equation}
 The distribution function $f(x)$  is defined here by the same
$\alpha$-parameter integral (\ref{4}).  The latter can be easily calculated
to give 
 \begin{equation}
f(x) = \frac{g^2}{16 \pi^2 m^2 } \, (1-x) \, \theta(0 \leq x \leq 1) \, .
\label{8} \end{equation}
Note, that $f(x)$  is purely real.
Due to singularity at $x = \zeta$ in Eq.    (\ref{5}),
the total amplitude $T \equiv T_{a}+T_b$ has 
both real and imaginary parts.
Since $x\geq 0$ and $\zeta \geq 0$, its imaginary part 
is given  by the $s$-channel contribution  
$T_a(p,q)$ only:
\begin{equation}
\frac1{\pi e^2} \, {\rm Im} \, T^{\rm as} (p,q) = 
\int_0^1  {\rm Im} \ t_a(xp,q) \, f(x) \,  dx = 
\int_0^1  \frac{1}{2(pq')} \, \delta(x-\zeta)\,  f(x) \, dx = 
\frac{ f(\zeta)}{2(pq')}   =  \frac{1}{2(pq')} \,  
\frac{g^2}{16 \pi^2 m^2 }
(1-\zeta) \,  . 
\label{9} \end{equation}
The real  part of $T$ is given by $T_b$ and by the real part of $T_a$:
\begin{equation}
 {\rm Re } \, T_a^{\rm as} (p,q) = 
\int_0^1  {\rm Re} \ t_a(xp,q) \, f(x) \,  dx = 
- \frac{e^2}{2(pq')}\,  {\rm P} \int_0^1  \, 
\frac{f(x)}{x-\zeta} \,  dx \, , 
\label{10} \end{equation}
where P stands for the principal value prescription.

To translate these results into the OPE language,
we write the  contribution of the diagrams 1$a,b$ 
in the coordinate representation:
\begin{equation}
T(p, q) = \int \langle p |\, \phi(0) \phi(z) \,|p \rangle
\, \left (e^{-i(qz)} + e^{i(qz)} \right ) \, D_m(z^2) \, d^4 z \, .
\label{11} \end{equation}
The  large-$Q^2$ asymptotics  of $T(p, q)$ 
is given by the leading light-cone behavior of both the quark propagator 
 $D_m(z^2) = 1/4i \pi^2 (z^2 -i \epsilon)+ \ldots $ 
and  the matrix element  
$\langle p|\, \phi(0) \phi(z)\, |p \rangle$
\begin{equation}
T(p, q) = \int \, \frac{e^{-i(qz)}+ e^{i(qz)}}
{4 i \pi^2 (z^2-i\epsilon)}  \,
\langle p|\, \phi(0) \phi(z)\, |p \rangle|_{z^2=0}
\,  d^4 z \, +O(1/Q^4).
\label{12} \end{equation}
Defining the parton distribution function $f(x)$ by
\begin{equation}
 \langle p\, | \,  \phi(0) 
 \phi (z) \, | \, p \rangle |_{z^2=0} 
 =    \int_0^1  
   \,\frac1{2} \left ( e^{-ix(pz)} + e^{ix(pz)} \right )
\, f(x) \, dx \,  , 
\label{13}
 \end{equation} 
we rederive the parton formulas (\ref{5}), (\ref{7}).
Basically, the integral (\ref{13})   can be treated as 
 a Fourier representation for the light-cone matrix element 
$\langle p\, | \,  \phi(0) 
 \phi (z) \, | \, p \rangle |_{z^2=0} \equiv \tilde f(pz)$  
which is a function  of the only variable $(pz)$. 
However, to derive the spectral constraint   $-1 \leq x \leq 1$
for the Fourier partner of $(pz)$  
and establish the property $f(x) = f(-x)$,   
one should incorporate the fact that $\tilde f(pz)$  is given by 
 Feynman integrals with specific analytic properties
and that  we have the same scalar field $\phi$
at both points 0 and $z$.
The $\alpha$-representation which we used above is one of 
the most effective (though perturbative) ways to take  these properties 
into account. In Ref.  \cite{spectral} (see also Sec. 
III below),
the $\alpha$-representation was used to 
 prove that  the  constraint  $0 \leq x \leq 1$ 
in Eq.    (\ref{13}) and similar (but more complicated) 
constraints for
multiparton distributions and distribution amplitudes 
 hold for  any Feynman diagram.  
Two other   approaches to studying 
spectral properties of parton distributions 
are described in refs.\cite{efp,jaffe}.  

Anticipating  comparison with the 
nonforward distributions  discussed below,
 it  is worth  emphasizing  here that 
the Bjorken $\zeta$-parameter is not present 
in Eq.    (\ref{13}) defining the
parton distribution function
$f(x)$.  It appears only after  one   calculates the
Compton amplitude $T(p,q)$.

\subsection{Double distributions}

Now, consider a one-loop box diagram for the scalar analogue of 
the deeply virtual 
Compton scattering amplitude (Figs.1$c,d$).  
Using the same basic light-cone momenta $p, q'$ as in the forward case, 
we write  the momentum  of the incoming  
virtual photon as $q\equiv q'-\zeta p$.  The
 outgoing real photon carries the lightlike momentum $q'$.
The  momentum conservation requires that the final hadron 
has the momentum $(1-\zeta)p$, ${\rm i.e.},$ in this kinematics we
have a lightlike momentum transfer $r \equiv \zeta \, p$. 
Since the initial momenta $q,p$ are identical to those of the
forward amplitude,  the parameter $\zeta$ 
coincides with the Bjorken
variable $x_{Bj} \equiv Q^2/2(pq)$.
In the $\alpha$-representation, the contribution of 
  the diagram 1$c $ is 
\begin{equation}
T_c(p,q,q') = 
- \frac{e^2 g^2}{16 \pi^2}
\int_0^{\infty}  \exp\left \{  i \left [2(pq') \, 
\alpha_1 {{
\alpha_3 - \zeta(\alpha_3+\alpha_4)
}\over{\alpha_1+\alpha_2+\alpha_3+\alpha_4}}
-\lambda \, (m^2-i\epsilon)  \right ] \right \}
\frac{d\alpha_1 d\alpha_2 d\alpha_3 d\alpha_4 }{\lambda^2}\, . 
\label{14} \end{equation}
The large-$Q^2$ limit is again governed  by
the small-$\alpha_1$  integration which gives 
\begin{equation}
T_c(p,q,q') = 
- \frac{i e^2 g^2}{16 \pi^2} 
\int_0^{\infty} 
\frac1
{2(pq')(\alpha_3/  \tilde \lambda  - \zeta(1-\alpha_2/
\tilde \lambda )+i\epsilon)}
\frac{e^{ - i \tilde \lambda (m^2-i\epsilon)}}{\tilde \lambda^2} \,
 d\alpha_2 d\alpha_3 d\alpha_4 + O(1/Q^4) \ .   
\label{15} \end{equation}
In the forward case, the  ratio $\alpha_3/  \tilde \lambda$ 
 was  substituted by the variable $x$ which was  interpreted then as 
the fraction of the initial hadron momentum carried by the active quark.
The   result expressed by  Eq.    (\ref{15})   contains  also another ratio
 $\alpha_2/  \tilde \lambda$.
 So, let us introduce the {\it double distribution }
\begin{equation}
F(x,y) = \frac{i  g^2}{16 \pi^2} 
\int_0^{\infty} \, 
\delta \left ( x - \frac{\alpha_3}{\alpha_2+\alpha_3+\alpha_4} \right   ) 
\delta \left ( y - \frac{\alpha_2}{\alpha_2+\alpha_3+\alpha_4}
\right   ) \, 
\frac{e^{ - i \tilde \lambda (m^2-i\epsilon)}}{\tilde \lambda^2} \,
 d\alpha_2 d\alpha_3 d\alpha_4 \, .
\label{16} \end{equation}
It is easy to see that both  variables $x,y$ vary
between 0 and 1. Furthermore, their sum is also  
confined within these limits:
$0 \leq x+y \leq 1$. Hence, $F(x,y) =  \theta(x+y \leq 1) \, F(x,y) $.  
Using Eq.    (\ref{16}), we  write the leading power contribution 
of $T_c(p,q,q')$ in terms
 of the double distribution:
\begin{eqnarray}
&& \lefteqn{T_c^{\rm as} (p,q, q') = -\int_0^1 \int_0^1 
\frac{e^2}{2(pq')(x+y\zeta -\zeta+i\epsilon)} 
\,  F(x,y) \, dx \, dy }
 \nonumber \\
&& = -\int_0^1 \int_0^1 \frac{e^2  }{(xp+yr+q)^2+i\epsilon} 
\, F(x,y)  \, dx \, dy \equiv  \int_0^1 \int_0^1 t_c(xp+yr,q, q')
\,  F(x,y) \,\theta(x+y \leq 1) \, dx \, dy \, .
\label{17} \end{eqnarray}
The parton subprocess amplitude $t_c$ is given by 
\begin{equation} t_c(xp+yr, q, q') = -\frac{e^2}{(xp+yr+q)^2+i\epsilon}\, .
\label{18} \end{equation} 
Hence, the momentum $xp+yr$ of the quark interacting with the virtual photon
originates both from the initial hadron momentum $p$ (term $xp$) 
and the momentum transfer   (term $yr$). 
In a similar way, for  the $u$-channel diagram 1$d$, we get  
\begin{eqnarray}
&& \lefteqn{T_d^{\rm as} (p,q', \zeta) = \int_0^1 \int_0^1 
\frac{e^2}{2(pq')(x+y\zeta -i\epsilon)} 
\,  F(x,y) \,  dx \, dy }
 \nonumber \\
&& = -\int_0^1 \int_0^1 \frac{e^2  }{(xp+yr-q')^2+i\epsilon} 
\, F(x,y)  \, dx \, dy \equiv  \int_0^1 \int_0^1 t_d(xp+yr, q,q')
\,  F(x,y) \,\theta(x+y \leq 1) \, dx \, dy \, , 
\label{19} \end{eqnarray}
with the same double  distribution  $F(x,y)$ given by  Eq.     (\ref{16}).  
In the explicit form, 
 \begin{equation}
F(x,y) = \frac{g^2}{16 \pi^2 m^2 }  \ \theta \, (0 \leq x+y \leq 1) \, .
\label{20} \end{equation}
Again, $F(x,y)$ is purely real.
Comparing the $\alpha$-representations for $f(x)$ and $F(x,y)$,
we obtain the reduction formula for  the double distribution $F(x,y)$:
\begin{equation}
\int_0^{1-x} \, F(x,y)\, dy =  f(x)  \, .
\label{21} \end{equation}

Due to the restrictions $x \geq 0, y \geq 0$,  the imaginary 
part of the total amplitude $T \equiv T_{c}+T_d$ is 
given  by the $s$-channel contribution alone:
\begin{eqnarray} 
 \frac1{\pi e^2 }\, {\rm Im} \, T_c (\zeta, Q^2 ) = 
 \frac{1}{2(pq)} \int_0^1 \int_0^{1}  
\delta( x + y \zeta -  \zeta ) F(x,y) \, 
\theta (x+y \leq 1) \, dx \, dy
\nonumber \\ 
=  
\, \frac1{2 \zeta (pq) } \int_0^{\zeta } F(x, 1-x/\zeta ) \, dx =  
\frac{1}{2  (pq)}  \int_0^1
F( \bar y \zeta ,  y ) \, d  y \equiv \frac{1}{2(pq)} \, \Phi(\zeta). 
\label{22}
 \end{eqnarray} 
The last form is  similar to the expression for 
${\rm Im} \, T$ in the forward case: one should just use 
the function $\Phi(\zeta)$ instead of $f(\zeta)$.
Moreover, the integral defining  $\Phi(\zeta)$ 
looks similar to that appearing in the reduction 
formula (\ref{21}). Still, the two integrals are not identical
and,  in  general, $\Phi(\zeta) \neq f(\zeta)$.  
Using the explicit form of $F(x,y)$ for our toy model, we obtain
\begin{equation}
 \Phi(\zeta) =  \frac{g^2}{16 \pi^2 m^2 }\, \theta(0 \leq \zeta \leq 1).
\label{23} \end{equation}
The factor $(1-\zeta)$ present in $f(\zeta)$ 
(see Eq.    (\ref{8}) ),  does not appear here.
Note,  however, that the difference is small for small $\zeta$.

In the OPE language, the basic change 
compared to the forward case is that we should 
deal now with the asymmetric matrix element 
 $\langle p-r\, | \,  \phi(0) 
 \phi (z) \, | \, p \rangle $.
Our definition of the double distribution $F(x,y)$ corresponds 
to the following parametrization 
\begin{eqnarray} 
 \langle p-r\, | \,  \phi(0) 
 \phi (z) \, | \, p \rangle |_{z^2=0} 
 =    \int_0^1   \int_0^1  \, \frac1{2} 
  \left (  e^{-ix(pz)-iy(r z)} + e^{ix(pz)-i\bar y(r z)}
\right ) F(x,y) \, \theta(x+y \leq 1) \,  dx \, dy \, .
\label{24} \end{eqnarray} 

Taking the  limit $r =0$ in Eq.    (\ref{24})   gives  the matrix
element defining the usual parton distribution function $f(x)$, 
and we reobtain  the reduction formula (\ref{21}).
Again,  this   definition  of $F(x,y)$  can be treated as 
 a Fourier representation for a function of 
two independent variables $(pz)$ and $(rz)$, 
with the  spectral constraints   
  $x \geq 0$, $y \geq 0$, $x+y \leq 1$ dictated by the analytic
structure of the relevant 
 Feynman integrals.
An important  feature  implied by
the representation (\ref{24}) is the absence 
of the $\zeta$-dependence in 
the double distribution $F(x,y)$. 
The asymmetric  matrix element
(\ref{24}), of course, has  $\zeta$-dependence. But  
it appears only through the ratio 
$(rz)/(pz)$ of variables in the exponential factor.
 In  this treatment,  $\zeta$  
  characterizes 
the ``skewedness'' or ``longitudinal momentum asymmetry'' of the matrix elements.
The fact that  for the deeply virtual Compton amplitude $T$ the parameter 
 $\zeta$ coincides with 
the Bjorken variable $x_{Bj}=Q^2/2(pq)$ is  a specific  feature of
a particular  process. 
The matrix element  itself accumulates a  
process-independent information and, hence,
has  quite a general nature.

Thus, despite the fact that the momenta
$p$ and $r$ are proportional to each other $r = \zeta  p$,
there is  a clear distinction between them, 
since   $p$ and $r$  specify 
the momentum flow in two different   channels.
For $r=0$, the  momentum flows only in the $s$-channel 
and the total momentum entering into 
the  composite operator vertex is zero. 
In this case, 
the matrix element coincides with  the 
usual    distribution function.
The  partons  entering the composite vertex 
then carry  the   fractions $x_i$ ($i=1,2$)
of the initial proton momentum.
In general,  $-1 < x_i <1$, but when 
 $x_i $ is  negative, we should  interpret the  parton 
as going out of the composite vertex 
and returning to the final hadron.
In other words,  $x_i$ can be  redefined
to  secure that  
the integral always  runs over  the segment  $0\leq x \leq 1$.
In this parton picture, the spectators take the 
remaining momentum  $(1-x)p$.
On the other hand, if  the 
total momentum flowing through the 
composite vertex is $r$, 
the matrix element has the structure
of the distribution amplitude in which 
the momentum  $r$  
splits into the fractions $yr$  and 
$(1-y)r \equiv \bar y r$ carried by the 
quark fields  attached to  that vertex. 
In a combined situation, when both $p$ and $r$
are nonzero,  the initial quark  has  momentum 
$xp +y r$, while the final one  carries the momentum 
$xp - \bar y r$.
Both the initial active quark 
and the spectator  carry  positive 
fractions of the lightlike momentum $p$:
$x+\zeta y$ for the active quark and 
$\bar x- \zeta y  = (1-x-y) +(1-\zeta )y$ 
for   the spectator. 
However, the total fraction of the initial momentum  $p$
carried by the 
quark   returning the
fraction $xp$ into the hadron matrix element  is given by 
$x - \bar y \zeta $ and it may take  both 
positive and negative values.

\subsection{ Asymmetric distribution  functions}

Since $(rz) = \zeta (pz)$,  the variable $y$ appears 
in eq.(\ref{24}) only in the 
$x+y\zeta \equiv X$ 
combination,
where $X$  can be treated as    the {\it total}  fraction 
of the initial hadron momentum $p$ carried by the active  quark.
Since $\zeta \leq 1$ and  $x+y \leq 1$, 
the variable $X$ satisfies a  natural
constraint $0\leq X \leq 1$.
Integrating the  double   distribution $F(X-y \zeta,y)$ 
over $y$ gives  the {\it asymmetric distribution function} 
\begin{equation}
{\cal F}_{\zeta} (X) = \theta(X \geq \zeta) 
 \int_0^{ \bar X / \bar \zeta } F(X-y \zeta,y) \, dy + 
 \theta(X \leq \zeta)  \int_0^{ X/\zeta} F(X-y \zeta,y) \, dy \,  , 
\label{25} \end{equation}
where $\bar \zeta \equiv 1- \zeta$.
The  basic distinction  between 
the  double  distribution $F(x,y)$ 
and the asymmetric  distribution function 
${\cal F}_{\zeta} (X)$ is that   the former  
is a  universal function in the sense that it 
does not depend on the skewedness  parameter
$\zeta$  while the latter is  explicitly labelled by it.
Hence, we 
deal  now  with a family of 
asymmetric distribution functions  ${\cal F}_{\zeta} (X)$ 
whose shape changes when $\zeta$ is changed.
In our toy model, ${\cal F}_{\zeta} (X)$ 
has the following $\alpha$-representation:
\begin{equation}
 {\cal F}_{\zeta} (X)= \frac{i  g^2}{16 \pi^2} 
\int_0^{\infty} \, 
\delta \left ( X - \frac{\alpha_3+ \zeta \alpha_2}{\alpha_2+\alpha_3+\alpha_4} 
\right   ) \, 
\frac{e^{ - i \tilde \lambda (m^2-i\epsilon)}}{\tilde \lambda^2} \,
 d\alpha_2 d\alpha_3 d\alpha_4 \, .
\label{16cal} \end{equation}
Taking the integrals, we get the explicit form 
\begin{equation}
{\cal F}_{\zeta} (X)
= \frac{g^2}{16 \pi^2 m^2 }\, \left \{ \frac{X}{\zeta} \, 
\theta \, (0 \leq X \leq \zeta ) + \frac{1- X}{1- \zeta}\, 
\theta( \zeta \leq X \leq 1) \right \}.
\label{26} \end{equation}
One can see that when $\zeta  \to 0$, the  limiting curve 
for ${\cal F}_{ \zeta}(X)$ reproduces the 
usual   distribution function:
 \begin{equation}
 {\cal F}_{\zeta=0} \, (X) =  f (X) \  . 
\label{27} \end{equation}
In general, this  formula also follows directly from 
the definition of ${\cal F}_{ \zeta}(X)$  and the reduction 
formula (\ref{21})  for the double distribution $F(x,y)$.

The fraction $(X- \zeta) \equiv X' $ of the
original hadron momentum $p$  carried by the ``returning''
parton differs from $X$ by $\zeta$: $X-X' = \zeta $ \cite{afs}.
Since $X$ changes from $0$ to $1$ and $\zeta \neq 0, 1$, the fraction 
$X'$ can be either positive  or negative, ${\rm i.e.},$  
 the asymmetric 
distribution function has two components
corresponding to the regions $1 \geq X \geq \zeta$ and 
 $0 \leq X \leq \zeta$.
In the region $X > \zeta$ (Fig.  2$a$),
where the initial parton  momentum $Xp$ 
is larger than the momentum transfer $r = \zeta p$,
the function ${\cal F}_{\zeta} (X)$ 
can be treated as  a generalization of the 
usual  distribution function $f(x)$
for the asymmetric case when the final hadron momentum $p'$ 
differs by $\zeta p$ from the initial momentum $p$.  
In this case,  ${\cal F}_{\zeta}  (X)$   describes
 a parton   going out of 
the hadron with a positive fraction  $Xp$  
of the original hadron momentum
and then coming back into the hadron with a changed 
(but still positive) fraction  $(X - \zeta)p$.
The parameter   $\zeta$ 
 specifies  the longitudinal momentum asymmetry  of 
the matrix element.

In the region  $X < \zeta$ (Fig.  2$b$), 
 the ``returning''  parton   has
a negative fraction $(X- \zeta)$ of the light-cone momentum $p$.
Hence, it is more appropriate to  treat it  as a parton 
going out of the hadron and 
propagating  along  with the original parton.
Writing $X$ as $X = Y \zeta$, we see that 
both   partons  carry now 
positive fractions $Y \zeta p \equiv Y r$ and
  ${\bar  Y} r  \equiv (1-Y)\, r $ of 
the momentum transfer $r$. The
asymmetric distribution function 
in the region $X= Y \zeta < \zeta$
looks like a distribution amplitude 
$\Psi_{\zeta}(Y)$ for a  $\phi \phi$-state 
with the  total momentum $r= \zeta p$: 
\begin{equation}
\Psi_{\zeta}(Y) =  \int_0^Y F((Y-y) \zeta , y ) \, dy . 
\label{28} \end{equation}
In our model, 
\begin{equation}
\Psi_{\zeta}(Y) = \frac{g^2}{16 \pi^2 m^2 }\, Y \, \theta(0\leq Y \leq 1).
\label{29} \end{equation}

Both $F(x,y) $  and ${\cal F}_{\zeta} (X)$
in our model are purely real.  This result
is determined, in fact, by general
properties of the definition of the double distribution,
Eq.    (\ref{16}). Indeed, let us introduce the Feynman parameters
$\beta_i$ by $\alpha_i = \tilde \lambda \beta_i$.
After  integrating  over $\tilde \lambda$, the
only possible source of imaginary contributions 
is the denominator factor $1/(m^2 - i \epsilon)$.
However, since $m^2 > 0$, this factor is always positive
and the integral is purely real.  
This would not happen if the initial ``hadron'' has a 
sufficiently large  mass $M>2m$.
In this case, instead of $1/(m^2 - i \epsilon)$
we would get 
$1/(-M^2 \beta_3 (1-\beta_3 ) +m^2 - i \epsilon)$
if the final hadron has the same mass $M$.
Then, the denominator is not positive-definite if $M > 2  m$, 
and the integral has both real and imaginary part.
Clearly, the imaginary part appears because the initial 
hadron can decay into its constituents.
If such a possibility is excluded, the double distributions $F(x,y) $ 
and, hence, the asymmetric distribution functions ${\cal F}_{\zeta} (X)$
 are purely real.
 
The usual parton distributions 
$f(x)$ are often related to 
imaginary parts,  
 or more precisely, $s$- and $u$-channel 
discontinuities 
of parton-hadron amplitudes\footnote{In a  
 recent paper,  L. Frankfurt {\it et al.} 
\cite{ffgs}    discuss also discontinuities
in the context of  the 
nondiagonal distribution functions. }.
Note, that in our approach, the  parton distributions 
are defined by form-factor-type matrix elements  which  depend 
only  on    momentum invariants $p^2$, $p'^2$,  
$r^2$ irrelevant to such discontinuities 
(so far  we even were treating  these invariants
as vanishing). 
The variable $X$ in our definition  only  reflects 
a more complicated structure of the operator  
vertex.
To illustrate this point, 
 we write  ${\cal F}_{\zeta} (X)$
in  the momentum representation (see Fig.  3$a$) 
\begin{equation}
{\cal F}_{\zeta} (X) =  \frac{i}{(2 \pi)^4 }
\int \frac{\delta (X-(k q')/(pq')) \, d^4 k}
{(k^2-m^2 + i \epsilon) ((p-k)^2 -m^2 + i \epsilon)
((r-k)^2 -m^2 + i \epsilon)} \, .
\label{mom}
\end{equation}
The $\delta (X-(k q')/(pq'))$-function 
here corresponds to composite operator 
(denoted by a blob on Fig.  3$a$).
Using the $\alpha$-representation, one can take
the Gaussian $k$-integral  and obtain 
the representation given by  Eq.  (\ref{16cal}),  which 
finally  gives our purely  real result  (\ref{26}).

It is worth emphasizing that 
the parton representations (\ref{5}),  (\ref{17})   and 
(\ref{30}) below
are valid  for the total Compton amplitude:
there is  no need to split the latter   into its 
real and imaginary parts in order
to define the parton distribution.
To make a parallel with the traditional
approach in which the parton distributions
are defined through the discontinuities
of parton-hadron amplitudes, 
let us  calculate the $k$-integral above using 
 the  Sudakov decomposition 
\begin{equation}
k= \xi p + \eta q' + k_{\perp} \  ,  \  2(pq') \equiv s \ , 
\label{sud}
\end{equation}
which gives 
\begin{equation}
{\cal F}_{\zeta} (X) =  \frac{i s}{2(2 \pi)^4 }
\int d^2 k_{\perp} \int_{- \infty}^{ \infty}
\frac{d \eta}{ [X \eta s -  k_{\perp}^2 -m^2 + i \epsilon ]
[(X-1 )\eta s -  k_{\perp}^2-m^2 + i \epsilon ]
[(X - \zeta) \eta s -  k_{\perp}^2 -m^2+ i \epsilon ] }\, .
\label{sudrep}
\end{equation}
Looking at the location of singularities 
for the $\eta$-integral, we immediately
see that  a nonzero result is obtained only 
when $0 \leq X \leq 1$.  Furthermore, in the region 
$\zeta  \leq X \leq 1$,  the integral over $\eta$ 
is given by residue at $\eta = - (k_{\perp}^2 + m^2 - i \epsilon)/(1-X)s$,
which corresponds to substituting the 
ordinary propagator $-1/[(p-k)^2 - m^2 +i \epsilon]$
by the  $\delta((p-k)^2-m^2)$-function for
 the line with momentum $(p-k)$.  
In other words, for $\zeta  \leq X \leq 1$,
our one-loop model for the function   ${\cal F}_{\zeta} (X)$ 
is totally given by the residue corresponding  to the 
$s$-channel cut through the
parton-hadron scattering amplitude (see Fig.  3$b$). 
On the other hand, in the region 
$0 \leq X \leq \zeta$,  the integral over $\eta$ 
is given by residue at $\eta =  (k_{\perp}^2 +m^2  - i \epsilon)/Xs$,
which corresponds to cutting the line with
momentum $k$ (see Fig.  3$c$).
Such a cut cannot  be related to
$s$- or $u$-channel discontinuities \footnote{
I am grateful to L. Frankfurt for attracting
my attention to this point and correspondence.}.
In both cases, one can say that ${\cal F}_{\zeta} (X)$
originates from 
a parton-hadron scattering 
amplitude ${\cal T} = i {\cal F}$ whose imaginary 
part is  given by  one or another type of  discontinuities.
In our treatment,  the only important fact is that
the amplitude  ${\cal T}$ is purely imaginary
so that the distribution 
function  $ {\cal F}_{\zeta} (X)$  is purely real.
As we have seen above, the function 
$ {\cal F}_{\zeta} (X)$ can be written in several  different ways,
$e.g.,$  in the  $\alpha$-representation which can be
integrated without taking 
any residues.

In terms of ${\cal F}_{\zeta}(X)$, 
  the virtual Compton amplitude 
$T_{c+d}(p,q,q')$   can be written as 
\begin{equation}
T_{c+d}^{\rm as} (p,q,q') = - \frac{e^2 \, }{2(pq)} \int_0^1
\left [ \frac1{X-\zeta  + i \epsilon} - \frac1{X - i \epsilon} \right ]
{\cal F}_{\zeta}(X) \, dX.
\label{30} \end{equation}
For a real function ${\cal F}_{\zeta}(X)$, the 
imaginary part of $T_{c+d}^{\rm as} (p,q,q')$ is determined by
that of the short-distance amplitude (terms in square brackets).
Since ${\cal F}_{\zeta}(X)$ linearly vanishes as $X \to 0$, 
the  singularity $1/(X - i \epsilon)$ of the  $u$-channel
diagram 1$d$  gives a vanishing imaginary part. 
As a result, the imaginary part of the 
whole amplitude is generated 
by the $1/(X-\zeta + i \epsilon)$ singularity coming from 
the $s$-channel diagram 1$c$:
\begin{eqnarray} 
 \frac1{\pi e^2 }\, {\rm Im} \, T_c (\zeta, Q^2 ) = 
\frac{1}{2(pq)}  \int_0^1 \, 
\delta( X -  \zeta ) \,  {\cal F}_{\zeta}(X)\, dX
=   \frac{1}{2(pq)}  \,{\cal F}_{\zeta}(\zeta). 
\label{31} \end{eqnarray} 
Hence, the integral $\Phi(\zeta)$ in Eq.    (\ref{22})  is equal to
${\cal F}_{\zeta}(\zeta)$, ${\rm i.e.},$
to the asymmetric distribution function ${\cal F}_{\zeta}(X)$
taken at the point $X = \zeta$. The parameter $\zeta$ 
is present  in ${\cal F}_{\zeta}(\zeta)$ twice: first as 
the parameter specifying the longitudinal momentum asymmetry 
of the matrix element and then as 
the momentum fraction at which  the imaginary part appears.
As one may expect, it appears for $X = x_{Bj} =\zeta$,
just like in the forward case. 
Note, however, that  the momentum $(X- \zeta) p$ of 
the ``returning'' parton vanishes when $X  = \zeta$.
In other words,   the imaginary part 
appears in a highly asymmetric  
configuration in which the
fraction
of the original hadron momentum carried
 by the second parton 
vanishes.
Hence,  
${\cal F}_{\zeta}(\zeta)$ in general 
 differs from the function 
$ f(\zeta)$. The latter   corresponds to a symmetric 
 configuration  in which the final parton 
has   momentum equal to that of  the initial one.
As discussed earlier, in our toy model 
$f(\zeta)/{\cal F}_{\zeta}(\zeta)
=f(\zeta)/ \Phi(\zeta) 
 = 1-\zeta$, ${\rm i.e.},$
${\cal F}_{\zeta}(\zeta)$ is larger than $f(\zeta)$, 
though the difference is small for small values of $\zeta$.

The fact that ${\cal F}_{\zeta}(X)$ vanishes for $X=0$
has a rather general nature. Note, that 
for small $X$  the function 
 ${\cal F}_{\zeta}(X)$ is given by its $X \leq \zeta$ component
\begin{equation}
{\cal F}_{\zeta} (X) |_{X \leq \zeta } = 
  \int_0^{ X/\zeta} F(X-y \zeta,y) \, dy \,  .
\label{32} \end{equation}
The size of the integration region is proportional to $X$
and, as a result,   ${\cal F}_{\zeta}(X)$ vanishes like
const$\times X$ or faster for any double distribution $F(x,y)$ 
which is finite for small $x$ and $y$.

In the coordinate representation, 
the   asymmetric  distribution function 
can be  defined 
through the matrix element
\begin{eqnarray} 
&& \langle p'\, | \,  \phi(0) 
 \phi (z) \, | \, p \rangle |_{z^2=0} 
 =    \int_0^1    \frac12 \, 
 \left ( e^{-iX(pz)}  +  e^{i(X-\zeta)(pz)} \right ) \, 
 {\cal F}_{\zeta}(X)  
 \,  dX , 
\label{a32} \end{eqnarray} 
with $\zeta = 1-(p'z)/(pz)$. 
To re-obtain the  relation between ${\cal F}_{\zeta}(X)$
and  the double distribution 
function $F(x,y)$, one should combine this definition
with Eq.    (\ref{24}).
The $\zeta \to 0$ reduction formula 
(\ref{27})   trivially follows from Eq.    (\ref{a32}). 

Using translation invariance,
we can write representation for a more general light-cone operator:
\begin{eqnarray} 
&& \langle p'\, | \,  \phi(uz) 
 \phi (vz) \, | \, p \rangle |_{z^2=0} 
 =    \int_0^1    \frac1{2}
\left (  e^{-iXv(pz)+i(X-\zeta)u(pz)} +
 e^{-iXu(pz)+i(X-\zeta)v(pz)} \right ) \, 
 {\cal F}_{\zeta}(X)  
 \,  dX . 
\label{b32} \end{eqnarray} 
This  formula explicitly shows that if the parton
corresponding to  $\phi (vz)$ has momentum $Xp$, then 
the momentum of the parton  related to  $\phi (uz)$
is $(X - \zeta)p$ and $vice$ $versa$.

\subsection{Nonforward distributions}

Writing the momentum of the virtual photon 
as $q= q'-\zeta p$ is equivalent to using the
Sudakov decomposition in the light-cone ``plus''
($p$) and ``minus'' $(q')$ components in 
a situation when
there is no transverse momentum.
An essential advantage of  expressing  the amplitudes in
 the  $\alpha$-representation 
is that it explicitly shows the dependence of 
 the diagram on 
the relevant momentum invariants. 
This means that  we can 
derive the  parton picture both for zero and 
nonzero invariant momentum transfers 
$t=(p'-p)^2$ without bothering 
about an optimal 
choice of the basic vectors 
 for  the external momenta. 
Maintaining  for simplicity $p^2 = p'^2 =0$, we get for the diagram 1$c $  
\begin{equation}
T_c(p,q,q') = 
- \frac{e^2 g^2}{16 \pi^2}
\int_0^{\infty}  \exp\left \{  i \left [
 {{\alpha_1(2(pq)\alpha_3 
-Q^2 (\alpha_3 +\alpha_4))+t \alpha_2 \alpha_4
}\over{\alpha_1+\alpha_2+\alpha_3+\alpha_4}}
-\lambda \, (m^2-i\epsilon)  \right ] \right \}
\frac{d\alpha_1 d\alpha_2 d\alpha_3 d\alpha_4 }{\lambda^2}\, . 
\label{33} \end{equation}
The small-$\alpha_1$ integration then gives 
\begin{equation}
T_c^{\rm as} (p,q,q') = 
- \frac{i e^2 g^2}{16 \pi^2} 
\int_0^{\infty} 
\frac{e^{ i t \alpha_2 \alpha_4/ \tilde 
\lambda - i \tilde \lambda (m^2-i\epsilon)  }}
{2(pq)(\alpha_3/  \tilde \lambda  - \zeta(1-\alpha_2/
\tilde \lambda )+i\epsilon)}
\frac{d\alpha_2 d\alpha_3 d\alpha_4}{\tilde \lambda^2} \,
  + O(1/Q^4) \ ,   
\label{34} \end{equation}
where $\zeta =Q^2/2(pq) \equiv x_{Bj}$.
Hence,  introducing  the $t$-dependent    double distribution
\begin{equation}
F(x,y;t) = \frac{i  g^2}{16 \pi^2} 
\int_0^{\infty} \, 
\delta \left ( x - {\alpha_3}/{\tilde \lambda } \right   ) 
\delta \left ( y - {\alpha_2}/{\tilde \lambda }
\right   ) \, 
e^{  i t \alpha_2 \alpha_4/ \tilde \lambda - i 
\tilde \lambda (m^2-i\epsilon)}
 \,
\frac {d\alpha_2 d\alpha_3 d\alpha_4 }{\tilde \lambda^2} 
\label{35} \end{equation}
we obtain the same parton formula, but
 with a modified parton distribution $ F(x,y;t)$
\begin{eqnarray}
T_c^{\rm as} (p,q, q') = -\int_0^1 \int_0^1 
\frac{e^2}{2(pq)(x+y\zeta -\zeta+i\epsilon)} 
\,  F(x,y;t) \, dx \, dy  \, .
\label{36} \end{eqnarray}
Moreover, the  dependence on $t$ appears $only$ through 
the $t$-dependence of $ F(x,y;t)$. 
Similarly, we can 
write down  the $\alpha$-representation for the 
$u$-channel diagram 1$d$:  
\begin{equation}
T_d(p,q,q') = 
- \frac{e^2 g^2}{16 \pi^2}
\int_0^{\infty}  \exp\left \{  i \left [
 {{\alpha_1(-2(pq')\alpha_3 
-Q^2 (\alpha_2 +\alpha_4))+t \alpha_2 \alpha_4
}\over{\alpha_1+\alpha_2+\alpha_3+\alpha_4}}
-\lambda \, (m^2-i\epsilon)  \right ] \right \}
\frac{d\alpha_1 d\alpha_2 d\alpha_3 d\alpha_4 }{\lambda^2}\, .
\label{37} \end{equation}
Using $2(pq')= 2(pq) +t$ and  integrating over small $\alpha_1$ gives
the  parton formula 
\begin{eqnarray}
T_d^{\rm as} (p,q,q') = \int_0^1 \int_0^1 
\frac{e^2}{2(pq)(x+y\zeta +xt/2(pq) -i\epsilon)} 
\ F(x,y;t) \,  dx \, dy 
\label{38} \end{eqnarray}
with the same $t$-dependent function  $F(x,y;t)$. In our model, 
\begin{equation}
F(x,y;t) = \frac{g^2}{16 \pi^2  }  \ \frac{\theta \, 
(0 \leq x+y \leq 1)}{m^2 -ty(1-x-y)} \, .
\label{39} \end{equation}
The parton subprocess amplitude in this case has the 
$O(t/2(pq)) = O(\zeta t/Q^2)$
correction term which can be neglected in the large-$Q^2$,
fixed-$t$  limit.
Then the parton amplitude again depends only on the combination
$x+y \zeta$, and it makes sense to 
 introduce the {\it nonforward distribution}
\begin{equation}
{\cal F}_{\zeta} (X;t) = 
 \int_0^{ {\rm min} \{ X/\zeta, \bar X / \bar \zeta  \} }
 F(X-y \zeta,y;t) \, dy \,  , 
\label{40} \end{equation}
which can be treated as the finite-$t$ generalization 
of the asymmetric distribution function ${\cal F}_{\zeta} (X)$
(or more precisely, ${\cal F}_{\zeta} (X)$ is the $t=0$ idealization
of ${\cal F}_{\zeta} (X;t)$). 
In our simple model, it  can be calculated analytically:
\begin{equation}
{\cal F}_{\zeta} (X;t) = \frac{g^2}{16 \pi^2  }
\left \{ \frac{4 \tau \, \theta \, (X \geq  \zeta) } {-t \bar X 
\sqrt{1+\tau^2}} 
\ln \left ( \tau + \sqrt{1+\tau^2} \right )
\,  + \theta \, (X \leq  \zeta) \, \int_0^{X/\zeta} 
\frac{dy}{m^2 -t y \, (\bar X - y \bar \zeta)} \, 
\right \}
\label{41} \end{equation}
where $\tau = \sqrt{(-t/4m^2) (1-X)/ (1- \zeta)}$. 
The function ${\cal F}_{\zeta} (X;t)$
 falls off with increasing $|t|$ like  a form factor.

The  $t$-dependent distributions 
$F(x,y;t) $  and ${\cal F}_{\zeta} (X;t)$
in our model are purely real. 
 Indeed,  introducing again  the Feynman parameters
$\beta_i$ by $\alpha_i = \tilde \lambda \beta_i$ and 
  integrating  over $\tilde \lambda$ gives 
 the denominator factor $1/(-t \beta_2 \beta_4 +m^2 - i \epsilon)$.
However, since $t \leq 0$, this factor is always positive
and the integral is purely real.  
Imaginary part for $F(x,y;t) $  and ${\cal F}_{\zeta} (X;t)$
would appear only if the initial hadron mass 
 satisfies  $M^2> 4 m^2 $.

For real distributions, the imaginary part of the total 
Compton amplitude can be calculated by taking   the imaginary part
of the short-distance amplitude which picks out 
the function ${\cal F}_{\zeta} (\zeta;t)$
\begin{equation}
{\cal F}_{\zeta} (\zeta;t) = \frac{g^2}{16 \pi^2 m^2 T \sqrt{1+T^2}}
\ln \left ( T + \sqrt{1+T^2} \right ) \,  ,
\label{42} \end{equation}
where $T = \sqrt{(-t  /4m^2)(1- \zeta)}$.

In the OPE approach, the  nonforward  distribution 
is given by  the matrix element
\begin{eqnarray} 
&& \langle p'\, | \,  \phi(0) 
 \phi (z) \, | \, p \rangle |_{z^2=0} 
 =    \int_0^1    
\frac12  \left ( e^{-iX(pz)}  +  e^{i(X-\zeta)(pz)} \right ) \, 
 {\cal F}_{\zeta}(X;t)  
 \,  dX .
\label{a42} \end{eqnarray} 

Taking  the local limit $z=0$, we obtain 
the following   sum rule for $ {\cal F}_{\zeta}(X;t)  $
\begin{equation}
 \int_0^1    {\cal F}_{\zeta}(X;t) \,  dX = \langle p'\, | \,  \phi(0) 
 \phi (0) \, | \, p \rangle = F(t) \, ,
\label{43} \end{equation}
where $F(t)$ is the toy model analogue of a hadronic form factor.

\subsection{Time-like photon in the final state}

To  give an example in which 
the skewedness parameter
$\zeta$ does not coincide with the Bjorken 
parameter  $x_{Bj}$, 
let us discuss  a situation considered in refs.\cite{barloe,glr,drm}, 
when the initial photon 
is spacelike $q_1= q' -\zeta_1 p$ while the final photon  is 
timelike  $q_2 = q' + \zeta_2 p$.
Here $q'$ is  a basic  lightlike  vector defining the Sudakov
decomposition 
rather than the momentum of the final photon.
Now, the Bjorken ratio  given by 
$x_{Bj}= - q_1^2/2(pq_1)$  coincides with 
$\zeta_1$. 
 However, by momentum conservation,
the longitudinal part of the 
final hadron momentum differs from that of the initial one
by $(\zeta_1 + \zeta_2)p$, ${\rm i.e.},$ the $\zeta$-parameter
is given in this case   by $\zeta = \zeta_1 + \zeta_2 $,
with $\zeta_2 = q_2^2 /2(pq_1)$.
As a result, the scalar analogue of the timelike photon electroproduction
 amplitude  can be written as 
\begin{equation}
T(p,q_1,q_2) = - \frac{e^2 \, }{2(pq')} \int_0^1
\left [ \frac1{X-\zeta_1  + i \epsilon} - \frac1{X - \zeta_2 -i \epsilon} \right ]
{\cal F}_{\zeta_1+\zeta_2}(X) \, dX
\label{300} \end{equation}
(for simplicity,  we  took  here $t=0$). 
Its imaginary part is proportional to the sum 
${\cal F}_{\zeta_1+\zeta_2}(\zeta_1) + {\cal F}_{\zeta_1+\zeta_2}(\zeta_2)$.
Hence, having information about the imaginary part
of such an amplitude for different values of $\zeta_1$ and $\zeta_2$,
one can directly ``measure'' the asymmetric distribution function 
${\cal F}_{\zeta}(X)$ in the region $X \leq \zeta$,
where ${\cal F}_{\zeta}(X)$ is similar  to a distribution amplitude.

\section{All-order analysis}

\subsection{Handbag diagram to all orders}

Using the $\alpha$-representation, one can write 
down the contribution of any diagram in terms of 
functions of the $\alpha$-parameters 
 specified by the structure  of the diagram.
Since the object of our interest is 
the matrix element of a two-body operator,
we can   extract  it  from  the simplest
handbag diagrams, ${\rm i.e.}, $ those in which the $q$ vertex
is connected to the $q'$ vertex by a single 
propagator. The contribution of any diagram of this type
can be written as  (see, $e.g.,$ \cite{bosh,oz})
\begin{eqnarray}
&&\lefteqn{T^{(i)} (p,q,q') = i^{l} \, \frac{{\rm P} ({\rm c.c.})}{(4\pi i)^{zd/2}}
\int_0^{\infty} \prod_{\sigma=1} d\alpha_{\sigma} D^{-d/2}(\alpha)}
\nonumber \\ &&
\exp \left \{ i q^2 \frac{\alpha_1 A_L( \alpha )}{D(\alpha) }
+i s \frac{\alpha_1 A_s(\alpha)}{D(\alpha) }
+ i  u \frac{\alpha_1 A_{u}(\alpha)}{D(\alpha)} 
+i t \frac{ A_{t}(\alpha)}{D(\alpha) }
- i  \sum_{\sigma} \alpha_{\sigma} (m_{\sigma}^2- i\epsilon) \right \} \,  ,
\label{44} \end{eqnarray}
where $s=(p+q)^2 , u= (p-q')^2$ and $ t=(p-p')^2$ 
are the Mandelstam variables, 
$d$ is the space-time dimension,  ${P({\rm c.c.})}$ is the relevant
 product of the coupling constants, 
 $z$ is the number of loops of the diagram and $l$
is the number  of its
internal lines.
Finally,  
$D,A_s,A_u,A_t, A_L$ are  functions  of the $\alpha$-parameters 
uniquely determined for each  diagram.

 To  describe  them, we need  definitions 
of a tree and a 2-tree of a graph.
A tree (2-tree)  of a graph  G is a subgraph of G 
which consists of  one (two) connected components  each 
of which  has no loops.
Any tree $G_1^{k}$  (2-tree $G_2^{l}$)  of G is 
determined by the set of lines $\sigma$ 
which  should be removed from the initial graph 
$G$ to produce  $G_1^{k}$ ($G_2^{l}$). The product of the 
$\alpha_{\sigma}$-parameters associated with  these lines 
will be referred to as  $\alpha$-tree ($\alpha$-2-tree). 
The function $D(\alpha)$ is called the determinant of the graph.
It is given by the sum of all 
$\alpha$-trees of the graph $G$.
 By $B(i_1, \ldots , i_m | j_1, \ldots , j_n)$ 
we denote the 
sum of all  $\alpha$-2-trees possessing the property that the vertices 
$ i_1, \ldots , i_m$ belong to one component, 
$j_1, \ldots , j_n$ to the other, while the 
vertices  not indicated explicitly may belong to
either  component. 
In these notations, 
\begin{equation}
\alpha_1 A_{L}( \alpha ) = B(q|p,q',p'); \   \ 
 \alpha_1 A_{s}(\alpha)= B(q,p |q',p');  \  \ 
\alpha_1 A_{u}(\alpha)= B(q,p' |p, q'); \  \ 
A_{t}(\alpha)= B(q,q' |p, p')\,  .
\label{45} \end{equation} 
The mnemonics is straightforward: the square of the total momentum 
entering into one 
of the components  (due to momentum conservation,
it does not matter which one) just gives the relevant momentum 
invariant (see Fig.  4). To get all the 2-trees corresponding to 
this invariant, one should make all possible cuts 
resulting in such a separation of external momenta.
 Note, that $\alpha_1$ must be present in all terms of $B(q|p,q',p')$,
$B(q,p |q',p')$ and $B(q,p' |p, q')$
because the vertices $q$, $q'$  in these cases 
 belong to  different components.
On the other hand, for $B(q,q' |p, p')$ these vertices are 
in the same component. As a result, 
there are   terms in $A_{t}(\alpha)$ 
which do not contain   $\alpha_1$ as a factor,
${\rm i.e.},$ $A_{t}(\alpha) =  
\alpha_1 A_{t}^{(1)}(\alpha) + A_{t}^{(0)}(\alpha)$ 
with   $A_{t}^0(\alpha) \neq 0$ and $A_{t}^{(1)}(\alpha) \neq 0$
for $\alpha_1 =0 $.
Similarly, the function $D(\alpha)$ can be written as 
$D(\alpha)= \alpha_1 D_1(\alpha) + D_0(\alpha)$,
where $D_1(\alpha)$ is the determinant for the graph $G_1$
obtained   from $G$ by deleting the line $\sigma_1$,
while  $D_0(\alpha)$ is that  for the graph $G_0$ 
resulting from $G$ by contracting the line $\sigma_1$
into a point (and gluing the vertices $q$, $q'$ into
a single  point). One can see that the
 function $D_0(\alpha)$   can  also be  written in 
terms of the same $\alpha$-2-trees:
\begin{eqnarray}
 D_0(\alpha) = \{ B(q|p,q',p')+B(q,p |q',p')+B(q,p' |p, q')+
B(q,p,p'|q') \} /\alpha_1 \nonumber  \\
 = A_{L}( \alpha )+A_{s}(\alpha)
+A_{u}( \alpha )+A_{R}(\alpha) \, ,
\label{46} \end{eqnarray}
where $A_{R}(\alpha)$ is the function corresponding 
to the cut separating out the momentum invariant $q'^2$.
To get the leading large-$Q^2$ asymptotics,
we integrate over the region $\alpha_1 \sim 0$. This gives 
\begin{eqnarray}
T^{\rm as} (p,q,q') = && \sum \limits_{\rm diagr.}
i^{l+1} \frac{P({\rm c.c.})}{(4\pi i)^{zd/2}}
\int_0^{\infty} \prod_{\sigma=2} d\alpha_{\sigma} D_0^{-d/2}(\alpha)
\left [ q^2 \frac{ A_L( \alpha )}{D_0(\alpha) }
+ s \frac{ A_s(\alpha)}{D_0(\alpha) }
+   u \frac{ A_{u}(\alpha)}{D_0(\alpha)}
+  t \frac{ A_{t}^{(1)}(\alpha)}{D_0(\alpha) }
 -m^2+i\epsilon \right ]^{-1}
 \nonumber \\ &&
 \times \exp \left \{ 
i t \frac{ A_{t}^{(0)}(\alpha)}{D_0(\alpha) }
- i  \sum_{\sigma=2} \alpha_{\sigma} 
(m_{\sigma}^2- i\epsilon) \right \} 
\, .
\label{47} \end{eqnarray}
Using   $s=-Q^2+2(pq), u= -2(pq)+t$ and neglecting
$t$ and $m^2$ compared  to $O(Q^2)$ terms in the
denominator factor,
we transform  it  into 
$$
 -Q^2 \frac{ A_L( \alpha) +A_s( \alpha)}{D_0(\alpha) }
+ 2(pq) \frac{ A_s(\alpha)-A_u(\alpha)}{D_0(\alpha)}
 + i\epsilon \, .
$$
This expression  has  the  structure similar to that of  the one-loop 
contributions (\ref{34}), (\ref{37}).
In particular, it can be   converted   into the form of 
the $s$-channel term (\ref{34}) 
if we  denote 
$[A_s(\alpha)-A_u(\alpha)]/D_0(\alpha)$ by $x$
and $[A_L( \alpha) +A_s( \alpha)]/ D_0(\alpha) $ by $1-y$.
Analogously, to make it look like the $u$-channel
term (\ref{37}), we should take  
$[A_s(\alpha)-A_u(\alpha)]/D_0(\alpha) =-x $
and $[A_L( \alpha) +A_s( \alpha)]/{D_0(\alpha) }=y$.
If we want to have {\em positive} $x$, we should perform 
the first identification in the region where 
$ A_s(\alpha)>A_u(\alpha)$ 
and use the second one in the region where 
$ A_s(\alpha)<A_u(\alpha)$.
In other words, we define the $t$-dependent 
double distribution
by 
\begin{eqnarray}
F(x,y;t) = &&  \sum \limits_{\rm diagr.} i^{l-1} 
\frac{P({\rm c.c.})}{(4\pi i)^{zd/2}}
\int_0^{\infty} \prod_{\sigma=2} d\alpha_{\sigma} D_0^{-d/2}(\alpha)
\exp \left \{ 
i t \frac{ A_{t}^{(0)}(\alpha)}{D_0(\alpha) }
- i  \sum_{\sigma=2} \alpha_{\sigma} 
(m_{\sigma}^2- i\epsilon) \right \} \,
\nonumber \\  &&
\left [ \delta \left (1-y - \frac{ A_L( \alpha )
+A_s(\alpha)}{D_0(\alpha) } \right)  
 \delta \left ( x- \frac{ A_s(\alpha)-
 A_u(\alpha)}{D_0(\alpha) } \right )
\, \theta (A_s(\alpha)>A_u(\alpha) )
\right.  \nonumber \\  &&    \left. 
+ \, \delta \left (y - \frac{ A_L( \alpha )+A_s(\alpha)}{D_0(\alpha) } \right)  
 \delta \left ( x- \frac{ A_u(\alpha)-
 A_{s}(\alpha)}{D_0(\alpha) }\right)
\, \theta (A_s(\alpha)<A_u(\alpha) ) \, \right ] \,  .
\label{48} \end{eqnarray}
An intuitive 
interpretation  is that when  $A_s(\alpha)>A_u(\alpha)$,
the quark  
{\it takes}   the momentum 
 $xp$ from 
the initial hadron. Its total momentum is $xp+yr$.
Alternatively,  when $A_s(\alpha) < A_u(\alpha)$, the quark 
{\it returns} the  momentum $xp$  to the final state, 
and its total returning momentum is
$xp-(1-y)r$.
Due to Eq.    (\ref{46}), we automatically have $0 \leq  x \leq 1$,
$0 \leq  y \leq 1$. Furthermore, since $x+y = 
[A_L( \alpha )+A_u(\alpha)]/D_0(\alpha)\leq 1$ in the  first 
  region and $x+y = 
[A_R( \alpha )+A_s(\alpha)]/D_0(\alpha)\leq 1$ in the second one,
we always have the restriction $x+y \leq 1$.

Again, introducing the Feynman parameters $\beta_i = \alpha_i/\lambda$
and the common  scale $\lambda$ given by the sum of
all  $\alpha_i$-parameters, we can integrate over
$\lambda$ to see that the resulting denominator 
factor $1/(-t A_{t}^{(0)}(\alpha)/D_0(\alpha) + m^2)$ 
is  positive for $t \leq 0$, and  the double distribution 
is purely real.

The same definition of the parameters $x,y$ 
based on the $\alpha$-representation can be used 
in the   realistic case of spin-$\frac12$ quarks.
However, one should take into account that
  the quark lines in that case 
are oriented. Depending   on  
their direction, we should interpret the  
parton with momentum $xp+yr$ either  
as a quark or as an  antiquark.

The nonforward distributions ${\cal F}_{\zeta}(X;t)$ can be 
obtained from the double distributions using Eq.    (\ref{40}).
 The restrictions $x,y \geq 0$, $x+y \leq 1$ guarantee  
that the total fraction $X$
satisfies the basic parton constraint $0 \leq X \leq 1$.
  Furthermore, if the  double distribution
$F(x,y;t)$ is finite for all relevant $x,y$,
the nonforward distribution ${\cal F}_{\zeta}(X;t)$ vanishes
(at least linearly) 
as $X \to 0$.

\subsection{Alpha-representation and  factorization}

Using the $\alpha$-representation, we can write each  perturbative  
diagram contributing to the virtual Compton scattering amplitude 
$T(p,q,q')$ in any  field theory model, including QCD
(see Fig.  5)
\begin{eqnarray}
&&\lefteqn{T^{(i)}(p,q,q') = i^l \frac{P({\rm c.c.})}{(4\pi i)^{zd/2}}
\int_0^{\infty} \prod_{\sigma} d\alpha_{\sigma} D^{-d/2}(\alpha)\, 
G(\alpha,p,q,q' ;m_{\sigma})}\nonumber \\ &&
\exp \left \{ - i Q^2 \, \frac{B_L( \alpha )+B_s(\alpha )}{D(\alpha) }
+ 2i(pq)\,  \frac{B_{s}(\alpha)-B_u(\alpha)}{D(\alpha) } \right \} \nonumber \\ &&
\exp \left \{i\,  t \, \frac{B_t(\alpha)+B_u(\alpha)}{D(\alpha)} 
+i M^2 \frac{B_{1}(\alpha)+B_{2}(\alpha)}{D(\alpha)} 
-i \sum_{\sigma} \alpha_{\sigma} (m_{\sigma}^2- i\epsilon) \right \} \,  .
\label{49} \end{eqnarray}
The only  difference is the presence of the preexponential factor 
$G(\alpha,p,q,q' ;m_{\sigma})$ due to the numerator structure
of the QCD propagators and vertices. It has a polynomial dependence on 
the momentum invariants.
The functions $B(\alpha)$ are defined by the relevant 2-trees,
$e.g.,$ $B_L( \alpha )=B(q|p,q',p')$, ${\rm etc.}$

In the region where $Q^2$ and $2(pq) = \zeta Q^2$ are large,
all the contributions having a  powerlike behavior on $Q^2$ 
can only come from the integration region  inside which all the ratios
$A_L/D, A_s/D, A_u/D$  vanish: if any of them 
 is  larger than some constant $\rho$, 
the integrand rapidly oscillates 
and the resulting contribution from such an integration region
is  exponentially suppressed.

Since  $A_L, A_s, A_u$ and $D$ are given by sums of products
of nonnegative $\alpha$-parameters,  there are 
two basic possibilities to arrange
$A_i/D=0$. In the  first case, called the ``short-distance regime'',
$A_i$  vanishes faster than
$D$ when some of the $\alpha$-parameters tend to zero
(small $\alpha$ correspond to large virtualities $k^2$, ${\rm i.e.},$
to ``short'' distances).
The second possibility, called the
``infrared regime'', occurs if  $D$ goes to infinity faster than
$A_i$ when some of the $\alpha$-parameters tend  to infinity
(large $\alpha$ correspond to small  momenta $k$, ${\rm i.e.},$
to the infrared limit).
One can also imagine a combined regime, when $A_i/D=0$
because some $\alpha$-parameters vanish and some are infinite.

There exists  a simple rule using which one can easily find the
lines $\sigma$ whose $\alpha$-parameters
 must be set to zero and those whose $\alpha$-parameters
  must be taken infinite
 to assure that  $A_i/D=0$.  First, one should realize that
$A_i/D=0$ means that the corresponding diagram
of a scalar theory (in which $G=1$)  has no dependence
on the relevant momentum invariant ($Q^2, s$ or $u$ in our case).
As the second step, one should incorporate the well-known analogy between
the Feynman diagrams and electric circuits \cite{BjDrell}:
the $\alpha_{\sigma}$-parameters may  be interpreted  as the resistances
of the corresponding lines $\sigma$.
In other words,  $\alpha_{\sigma}=0$ corresponds to 
short-circuiting the line $\sigma$ while
$\alpha_{\sigma}= \infty$ corresponds to its removal from
the diagram.
Hence, the problem is to find the sets of lines 
$\{\sigma\}_{{\rm SD}}$, $\{\sigma\}_{{\rm IR}}$
whose contraction into point (for $\{\sigma\}_{{\rm SD}}$)
or removal from the diagram (for $\{\sigma\}_{{\rm IR}}$)  produces the diagram
which in a scalar theory  does not depend on $p_i^2$.
Thus,  the rule determining possible
 types of the powerlike 
contributions is the following:
after  the part of the diagram corresponding to
a  short-distance subprocess is contracted into a  point
and the part corresponding to soft exchange is removed from
the diagram,
the resulting  diagram (``reduced diagram'', cf.
\cite{libbyst,cfs})  should have no dependence
on large momentum invariants.

Some examples are shown in Fig.  6. 
The simplest possibility is to contract into point
some subraph $H$ containing the photon vertices $q,q'$ (Fig.  6$a$).
The reduced diagram depends only on small invariants $t$, $M^2$
and masses $m$. The long-distance part
corresponds to a nonforward distribution.
This is the standard OPE configuration.
However, since $q'^2$ is not a large momentum invariant:
$q'^2 =0$, there is a  less trivial possibility shown in Fig.  6$b$.
In this case, there are two long-distance parts:
one is given by a 
nonforward distribution again and the other 
can be interpreted as the distribution amplitude
(hadronic component) 
of  a real photon.  Exchange of soft quanta between
the two long-distance parts of Fig.  6$b$ corresponds 
to a combined ${\rm SD}$-${\rm IR}$ regime (Fig.  6$c$):
the $\alpha$-parameters of lines inside $H$
vanish   while those belonging to the soft subgraph
$S$  tend to infinity.

One can easily invent other, more complicated configurations.
Fortunately, not all of them are equally important:
different configurations have different
$Q^2$-behavior. The power counting is based on the observation
that in the essential region of integration
$\alpha_{\sigma} \sim 1/Q^2$ for lines in the short-distance 
subgraph $H$
and $\alpha_{\sigma} \sim Q^2/p^4$ for lines 
in the soft subgraph $S$ ($p^2$ is some generic small 
scale, say, $M^2$ or $m^2$). In the momentum representation,
this corresponds to $k \sim Q$ for the $H$-lines and 
$k \sim p^2/Q$ for the $S$-lines.
As a result, in a theory with dimensionless coupling constants,
we can use the dimensional analysis 
to derive that the contribution due to $H$ behaves like  $ Q^{4-d_H}$, 
where
$d_H$ is the sum of dimensions (in mass units) 
of the fields associated with the external 
lines of $H$. We should also take into account extra 
numerator  factors
brought  by these external lines. For instance,  
 each external quark line  adds  a Dirac spinor
 $u(p)$, two of them  give $u(p)\bar u(p) \sim \hat p$,
and $\hat p$ can combine with $\hat q$ from $H$ to give $(pq) \sim Q^2$.
This means that  each external quark line can bring an extra $Q^{1/2}$ factor.
Note, that 1/2 is the spin of the quark.
Similarly, an external gluon line can add a $p^{\mu}$ factor.
Combined with $q_{\mu}$  from $H$ it gives $(pq) \sim Q^2$,
${\rm i.e.},$   the gluon line can bring an  extra $Q=Q^1$ factor 
for the whole amplitude.
Again, ``1'' is the spin of the gluon. 
Hence, each external quark or gluon line 
can give the factor $Q^{s_i-d_i} = Q^{- t_i}$ where 
$t_i = d_i-s_i$ is its twist. 
Note also that  calculating 
the virtual Compton ampitude  we do not convolute
the vector indices $\mu, \nu$ of 
the initial and final photon lines with momentum-dependent vectors.
Hence, each  external photon line gives only the factor $Q^{-1}$ 
due to its dimension.
Thus,  the counting rule  for the
contribution of  the hard subgraph $H$ is 
\begin{equation}
t_H(Q) \lesssim  Q^{4-N-\Sigma_i t_i} \,  , 
\label{50} \end{equation}
where $N$ is the number of external photon lines of the hard subgraph 
and summation is over quark and gluon external lines of $H$.
For the simplest hard subgraph with  two external quark lines
this gives $t_H(Q) \lesssim  Q^0$, a scaling behavior as expected.
For the configuration 6$b$, the estimate is  $t_H(Q) \lesssim  Q^{-1}$.
 Hence, the contribution of Fig.  6$b$ is power-suppressed compared
to that of Fig.  6$a$.  Note that since  the gluons have zero twist, 
the hard  subgraph can have  an arbitrary number of extra  gluon
lines without changing its power behavior. 
A similar power counting estimate \cite{echaja} 
based on $k \sim p^2/Q$
 can be obtained 
for the soft subgraph $S$:
\begin{equation}
t_S(Q) \lesssim  Q^{-\Sigma_j t_j} \, , 
\label{51} \end{equation}
where the summation is over the external lines of $S$.
Hence, exchanging a  soft quark (Fig.  6$d$) 
produces the $1/Q^2$ suppression
($S$ has then two external quark lines each having $t=1$), 
while the exchange of any number of soft gluons
is not necessarily accompanied by a suppression factor,
at least on  diagram by diagram level (for more details, 
see discussion in the next subsection).
For the combined ${\rm SD}$-${\rm IR}$ configuration, the power counting estimate is
\begin{equation}
t_{HS}(Q) \lesssim  Q^{4-N-\Sigma_H t_i}Q^{-\Sigma_S t_j} \, . 
\label{52} \end{equation}

It is convenient to describe the power-low behavior of $T(Q^2)$ 
in terms of the Mellin transformation
\begin{equation}
T(Q^2) = \frac1{2 \pi i} \int \limits_{-i \infty}^{i\infty} 
\left ( \frac{Q^2}{M^2} \right )^J
\Phi(J) \,  dJ \, .
\label{53} \end{equation}
Then the statement that $T(Q^2)\sim (1/Q^2)^n$  is equivalent 
to saying that the Mellin transform $\Phi(J)$ has a pole at $J = -n$.
Take as an example the Mellin transform of the scalar diagram shown in Fig.  7$a$
(it is essentially identical to the diagram 1$c$): 
\begin{equation}
\Phi_c(J) = 
- \frac{e^2 g^2}{16 \pi^2} \, \Gamma (-J)
\int_0^{\infty}   \left [i 
\alpha_1 {{  (\alpha_3+\alpha_4) - \alpha_3/\zeta
}\over{\alpha_1+\alpha_2+\alpha_3+\alpha_4}} \right ]^J
\exp\left \{ i\, t \, \alpha_2 \alpha_4 /\lambda -
i\lambda \, (m^2-i\epsilon)   \right \}
\frac{d\alpha_1 d\alpha_2 d\alpha_3 d\alpha_4 }
{(\alpha_1+\alpha_2+\alpha_3+\alpha_4)^{d/2}}\, . 
\label{54} \end{equation}
Small-$\alpha_1$ integration corresponds to
the simplest ${\rm SD}$-regime 6$a$ and generates  the pole $1/(J+1)$
corresponding to the $1/Q^2$ asymptotic behavior.
The relevant reduced graph is shown in Fig.  7$b$. 

Another possibility to kill the dependence on large variables
is to take $\alpha_3 = \alpha_4 =0$
which corresponds to the  reduced graph shown  in Fig.  7$c$.
To describe a simultaneous vanishing  of two 
 $\alpha$-parameters,
we use  the common scale $\rho = \alpha_3+\alpha_4$
and the Feynman  parameters $\gamma_i = \alpha_i/\rho$.
The resulting $\rho$-integral  $\rho^J \rho d \rho$ gives the pole
$1/(J+2)$ corresponding to  a nonleading behavior $1/Q^4$.

Furthermore,  contracting the whole diagram into point (${\rm i.e.}$ taking $\alpha_i=0$
for all $\alpha$-parameters) we 
also obtain a  reduced graph  which does not depend on large variables.
In this case, we introduce the  
 common scale $\lambda = \alpha_1+\alpha_2+\alpha_3+\alpha_4$
and the relative parameters $\beta_i = \alpha_i/\lambda$.
In $d=4$ dimensions, the  integrand  behaves like
 $ \lambda^J \,  \lambda^3 d \lambda/\lambda^2$
which produces the pole $1/(J+2)$ generating a nonleading behavior 
$1/Q^4$. However, if we  take  a scalar model in $d=6$  space-time dimensions, 
then the integrand  behaves like  $ \lambda^J \,  \lambda^3 d \lambda/\lambda^3$
and small-$\lambda$ integration generates the leading pole $1/(J+1)$.
Note that in this case after the $\lambda$-integration we still have the
factor $\beta_1^J$ capable of producing another $1/(J+1)$ pole due to 
small-$\beta_1$ integration.
Hence, the total singularity of this diagram  in 6 dimensions is $1/(J+1)^2$,
which gives $T(Q^2) \sim (\ln Q^2)/Q^2$.
This corresponds, of course, to the scaling violation ${\rm i.e.},$ to evolution
of the  nonforward distribution. One can even extract the relevant 
evolution kernel from the remaining integral over 
$\beta_2, \beta_3, \beta_4=1-\beta_2- \beta_3$ (the result, in fact,
can be read off Eq.    (\ref{26}) ).
Another observation is that if we simply integrate over small-$\alpha_1$
region, the remaining integral  $d\alpha_2 d\alpha_3 d\alpha_4 /\tilde \lambda^3$
logarithmically diverges in the region of 
small $\tilde \lambda \equiv \alpha_2+\alpha_3+\alpha_4$. 
This  is the standard {\rm UV} divergence of a  matrix 
element of a light-cone operator in a theory with dimensionless
coupling constants.

Taking $\alpha_2 \to \infty$, we  incorporate the ${\rm IR}$ regime
corresponding to the reduced graph 7$d$.
If the quark corresponding to the $\sigma_2$ line is massless, 
the  $\alpha_2$ integral in this limit is 
$\alpha_2^{-J} d\alpha_2/\alpha_2^2$. It produces 
the $1/(J+1)$ pole corresponding to the leading $1/Q^2$ behavior.
In the previous section,  we did not see this contribution because 
the  quark masses were assumed to be nonzero for all the lines.
For nonzero mass,  the factor $\exp[-i\alpha_2 m^2]$ 
suppresses the large-$\alpha_2$ integration and no
poles in the $J$-plane are produced.
In other words, the ${\rm IR}$ regime should be taken into account 
only for massless (or nearly massless) fields. 
Note,  that in QCD the ${\rm IR}$ regime for the
virtual Compton amplitude also  
gives $1/Q^2$ behavior for massless quarks (see Eq.    (\ref{51}) ).
However, in QCD this is a nonleading contribution 
 compared to the scaling behavior 
produced by  the  purely ${\rm SD}$ regime 7$b$.

\subsection{QCD and gauge invariance}

{\it After the SD-dominance is established}, 
the next step is to write the contribution of the ${\rm SD}$ configuration in the 
coordinate representation (Fig.  8$a$) 
\begin{eqnarray} 
T(p,q,q') = \int e^{-i(qz)} \, d^4z \int 
\, \langle p' \, | \, \phi(z_2) C(z,z_1,z_2) 
\phi(z_1) \, | \, p \rangle \, 
d^4z_1 \, d^4 z_2
\label{55} \end{eqnarray} 
(where $\phi$ is a  generic notation for the quark fields
$\psi, \bar \psi$ and the gluon field $A$) and  
expand the bilocal matrix element 
$ \langle p' \, | \, \phi(z_2) \ldots  \phi(z_1) \, | \, p \rangle$
in powers of $(z_2 -z_1)^2$.
Since we already know from the
$\alpha$-representation analysis that 
the virtualities inside the
${\rm SD}$-subgraph are $O(Q^2)$,
 extra powers of 
 $(z_2 -z_1)^2$ for simply dimensional reasons
result in extra powers of $1/Q^2$, and 
the leading large-$Q^2$ 
behavior will be given by the lowest term of this expansion
corresponding  to the lowest-twist composite operator.
Parametrizing the nonforward matrix elements of 
the light-cone operators by formulas analogous  to
Eq.    (\ref{b32})  gives the parton formulas similar to Eq.    (\ref{30}).
Of course, this is just a general idea how to obtain the QCD 
parton picture for the ${\rm SD}$-dominated amplitudes.
 Its practical implementation depends on specific 
properties of  a particular process under consideration.

The most important complication in QCD is 
due to the  gauge nature of the gluonic field.
In the Feynman gauge, the gluon vector potential
$A_{\mu}$ has zero twist,  and we should perform 
an infinite summation over  the external
gluonic lines both for   the ${\rm SD}$-subgraphs 
$H$ and infrared  subgraphs $S$. 
Consider the sum of gluon
insertions into  the quark propagator.
It  is well-known (see, $e.g.,$ 
\cite{rivnc,g^3,echaja,echaja2}) that 
after  summation
\begin{equation} 
S^c(\xi - \eta) + \int S^c(\xi - z) \gamma^{\mu} g A_{\mu}(z) S^c(z - \eta) \, d^4z
+ \ldots = E(\xi, \eta;A) \, S^c(\xi - \eta) \bigl [ 1 + O(G) \bigr ]
\label{1E}  \end{equation} 
all the $A$-fields  can 
be accumulated in the  path-ordered exponential 
\begin{equation}
E(\xi, \eta;A)\equiv P \exp \left 
( ig \int_{\eta}^{\xi} A_{\mu}(z) dz^{\mu} \right )
\label{56}
\end{equation} 
while   $O(G)$ term depends on the gluonic fields 
only through the  tensor
$G_{\mu \nu}$ and its covariant  derivatives.
Since  $G_{\mu \nu}$ is asymmetric with respect to 
the interchange of the indices $\mu$, $ \nu$, 
it  should  be treated as  a twist-1 field.
For the simplest ${\rm SD}$ configuration
possessing  a single long-distance part,
combining  the $E$-factors of  all internal lines
of the ${\rm SD}$-subgraph,
one gets gauge-invariant operators, $e.g.,$
$\bar q (z_1)  \gamma_{\nu}
E(z_1,z_2;A) q(z_2)$.

If the lowest-order  ${\rm SD}$-configuration 
contains two long-distance parts
(like in Fig.  6$b$), the gluonic corrections
include insertions into the external lines of the 
${\rm SD}$ subgraph 8$b$.  The resulting path-ordered 
exponentials  $E_n( \infty, \xi,;A)$ then 
go to infinity along the relevant  light-cone 
directions, $e.g.,$ $q'$ or $p$ in  case
of hard electroproduction processes.
However,  for color-singlet channels there are
at least two such exponentials and 
their long-distance tails
 cancel each other so that   only the 
factors $E(\xi, \eta;A)$ related to 
${\rm SD}$-subgraph vertices $\xi, \eta$ remain. 
The basic effect of the exponential factor 
$E(\xi, \eta;A)$ 
is that  expanding operators 
${\cal O}(\xi , \eta)$
into the   Taylor series, $e.g.,$  
\begin{equation} 
\bar q (\xi)  \gamma_{\nu}
E(\xi, \eta;A) q(\eta) = \sum \limits_{n=0}^{\infty} 
\frac1{n!} \Delta^{\nu_1} \Delta^{\nu_2} 
\ldots \Delta^{\nu_n}  
\bar q (\xi)  \gamma_{\nu} D_{\nu_1}D_{\nu_2}
\ldots D_{\nu_n} q(\xi) \quad ;  \quad   \Delta = \eta - \xi
\label{Add3}  \end{equation} 
one gets local operators 
$\bar q   \gamma_{\nu} D_{\nu_1}D_{\nu_2}
\ldots D_{\nu_n} q $ 
containing  covariant derivatives $D^{\nu}= 
\partial^{\nu} -igA^{\nu}$ 
rather than ordinary ones.

The  cancellation of $E_n(\xi, \infty ;A)$ 
factors is very important 
for the success of the standard factorization program.
Only after such a cancellation, the long-distance 
information is accumulated in universal
matrix elements of  gauge invariant light-cone operators.
  To illustrate the  difference between
 color-singlet and 
  color-nonsinglet channels,
consider matrix element 
$J(p,q') = \langle 0| E_{q'} (\infty,0;A)\psi(0)|p \rangle$ 
of the quark field $\psi(0)$ coming out
of a state with momentum $p$
and taken together with the
accompanying gluonic field $A$ which is then absorbed 
by a  $q'$ channel quark  
collecting the gluonic $A$-fields into 
 the $E_{q'} (\infty,0;A)$ factor (see Fig.  8$c$). 
Note that   the latter can be written as 
\begin{eqnarray} 
E_{q'}(\infty ,0 ) = 
P\exp\left(\int_{0}^{\infty} {\cal A}(t) \, dt \right) = 
1+ \int_{0}^{\infty} {\cal A}(t) \, dt  + 
 \int_0^{\infty} {\cal A}(t) \, dt  
\int_0^{t} {\cal A}(t_1)\, dt_1  + \ldots \nonumber \\ =
1+ \int_{0}^{\infty} {\cal A}(t) \, E_{q'}(t,0) \, dt 
\end{eqnarray}
where ${\cal A}(t) = ig q'_{\mu}A^{\mu}(tq')$.
  Substituting this result into the matrix element
\begin{equation} 
J(p,q')=\langle 0| \psi(0)|p\rangle + 
\int_{0}^{\infty} 
\langle 0| {\cal A}(t) E_{q'}(t,0) \psi(0) |p\rangle \, dt \,  , 
\end{equation} 
shifting the arguments of all fields in the second term by $tq'$
and  performing  the Taylor expansion 
\begin{equation} 
E_{q'}(0,-t) \, \psi(-tq') = \sum_0^{\infty} \frac{(-t)^n}{n!}(q'D)^n \psi(0) \, 
\end{equation} 
we can take the integral over $t$ to get
\begin{equation} 
J(p,q')= \langle 0| \psi |p\rangle - 
\sum_{n= 0}^{\infty} \frac{i^{n+1}}{(pq')^{n+1}} 
\langle 0| {\cal A} \, (q'D)^n \psi |p\rangle  \,  , \label{318}
\end{equation} 
where all the fields are taken at the origin.
In fact,  since
\begin{equation} 
\langle 0| A_{\nu} D_{\nu_1} \ldots D_{\nu_n} \psi |p\rangle 
= p_{\nu}p_{\nu_1} \ldots p_{\nu_n} a_n(\mu^2) \,  , \label{319}
\end{equation} 
the right-hand side of Eq.    (\ref{318}) does not depend on $(pq')$ (cf. \cite{gk}). 
Note, that the new representation for $J(p,q')$, unlike the original one, 
is not explicitly gauge invariant.
However,  the  $\psi$-term can be represented
as
\begin{equation} 
\langle 0| \psi |p\rangle = \langle 0|i \frac{(q'\partial)}{(pq')}
\psi |p \rangle 
\end{equation} 
  and we  can combine it with the first term from the sum to get   a 
term containing a covariant derivative $D= \partial - ig A$:
\begin{equation} 
J(p,q')= \langle 0| i \frac{(q'D)}{(pq')}\psi  |p\rangle -
\sum_{n=1}^{\infty} \frac{i^{n+1}}{(pq')^{n+1}} 
\langle 0| {\cal A} \, (q'D)^n \psi |p\rangle  \, . 
\end{equation} 
Repeating this trick, ${\rm i.e.},$  representing the term outside the sum as
\begin{equation} 
\langle 0| i \frac{(q'D)}{(pq')}\psi |q\rangle  =
 \langle 0| i^2 \frac{(q'\partial)}{(pq')}
\frac{(q'D)}{(pq')}\psi |p\rangle 
\end{equation} 
and combining this term with the $n=1$ term from the sum 
one obtains 
\begin{equation} 
J(p,q') = \langle 0| i^2 \frac{(q'D)^2}{(pq')^2}\psi |p\rangle -  
\sum_{n=2}^{\infty} 
\frac{i^{n+1}}{(pq')^{n+1}} \langle 0| {\cal A} \,(q'D)^n \psi |p\rangle \, .
\end{equation} 
It is clear now that we  can write $J(p,q')$  in a manifestly
 gauge-invariant
form (cf.\cite{mue}): 
 \begin{equation} 
J(p,q') = \lim_{n \to \infty} 
\frac{i^n}{(pq')^{n}} \langle 0| (q'D)^n \psi |p \rangle \equiv 
 \langle 0| \left( \frac{i(q' D)}{(pq')}\right)^{\infty} \psi |p\rangle  \, .
\end{equation} 
In perturbation theory,
 matrix elements $\langle 0| (q'D)^n \psi |p \rangle$  
for finite $n$ have ultraviolet divergences 
which can be regulated in a standard way, $e.g.,$ by
the dimensional regularization. After renormalization,
we get one-loop terms like
 $g^2 \gamma_n \log  \mu^2$. 
However, the anomalous dimension $\gamma_n $ 
contains the usual $(\sum^n_j \, 1/j)$ term \cite{gl} which 
behaves like $\log n $ 
for large $n$. Hence, taking the formal limit $n \to \infty$ one encounters
a logarithmic singularity, which requires an additional
regularization on top of dimensional regularization (cf. \cite{colt}).
The parameter characterizing the extra regularization
can be taken proportional to $\mu$, ${\rm i.e.},$ matrix element 
$\langle 0| E_{q'} (\infty,0;A)\psi(0)|p \rangle$ 
is the simplest example of a long-distance  object 
with a double-logarithmic dependence on the {\rm UV} cutoff \cite{erad378}.
Such objects (``collinear'' or ``jet'' factors \cite{coll,gk,css}) 
play an important role in PQCD studies of  Sudakov effects.
However, within the standard factorization approach,
presence of noncancelling double logarithms
of $Q^2$ (reflected by  double logarithms $\log^2 \mu^2$
in long-distance matrix elements) 
is treated as a failure of the factorization 
program, since   the amplitudes in that case cannot be written 
through a convolution with   parton distributions
defined through matrix elements of light-cone operators
[which have a single-logarithmic dependence on $\mu$].

 Another signature of Sudakov effects is the presence
of the ${\rm IR}$ contributions (see Fig.  6$c$). Again, since  all 
the hadrons participating in a hard   exclusive 
scattering process are color singlets,
summing over all soft gluon insertions 
one would get a path-ordered  exponential over 
a closed contour, and by  Stokes theorem
\begin{equation}
\langle 0 | P \, \exp \left \{ ig  \oint A_{\mu}(z) \, dz^{\mu} \,\right \} 
|0 \rangle 
= 1 + \langle 0 | O(G) |0 \rangle \, , 
\end{equation} 
where $O(G)$ depends on the gluon field only through 
the field strength tensor $G_{\mu \nu}$ which has nonzero 
twist generating  a power suppression of the net  ${\rm IR}$ regime
contribution.

\section{Nonforward distributions in QCD}

\subsection{ Quark distributions.} 

Let us discuss now the nonforward parton distributions in 
the realistic QCD case.  
For quarks, we should take into account 
that the  field  $\psi_a(z)$ contains both the
$a$-quark annihilation operator and the $\bar a$-antiquark
creation operator, ${\rm i.e.},$ the matrix element of the same light-cone 
operator $\bar \psi_a(0)  \ldots \psi_a(z)$ 
determines  distribution functions both for the 
quark and antiquark. Another complication is related 
to spin.
There are two leading-twist operators
 $\bar \psi_a (0) \gamma_{\mu} E(0,z;A)  \psi_a (z)$ and 
 $\bar \psi_a (0) \gamma_{\mu} \gamma_5 E(0,z;A)  \psi_a (z)$,
where, as discussed above,  $E(0,z;A)$ is the path-ordered exponential 
(\ref{56}) 
which makes the operators gauge-invariant. 
 In the forward case, 
the first one gives the  spin-averaged distribution functions $f(x)$
while the second one is related to the spin-dependent structure
functions $g_1(x)$. In this paper, we
will concentrate on the 
$\bar \psi_a  \gamma_{\mu} E(0,z;A)  \psi_a$
operators and gluonic operators with
which it mixes under evolution.  
The relevant nonforward matrix element 
can be written as\footnote{Two other  definitions 
of the nonforward parton distributions in terms 
of  matrix elements of  composite operators
proposed by X.Ji \cite{ji} and Collins, Frankfurt and 
Strikman \cite{cfs} 
are discussed  in Sec.  IX.}
\begin{eqnarray} 
\hspace{-1cm} 
&&\lefteqn{
\langle \, p'  , s' \,  | \,   \bar \psi_a(0)  \hat z 
E(0,z;A)  \psi_a(z) 
\, | \,p ,  s \,  \rangle |_{z^2=0}
 } \label{57}  \\ \hspace{-1cm} &&  
= \bar u(p',  s')  \hat z 
 u(p,s) \, \int_0^1   
 \left ( e^{-iX(pz)}
 {\cal F}^a_{\zeta}(X;t)  -  e^{i(X-\zeta)(pz)}
{\cal F}^{\bar a}_{\zeta}(X;t)  \right ) 
 \,  dX \nonumber \\ \hspace{-1cm} &&  
+
\, \bar u(p',s')  \frac {\hat z  \hat r - \hat r \hat z}{2M}
 u(p,s) \, \int_0^1   
  \left ( e^{-iX(pz)}
 {\cal K}^a_{\zeta}(X;t)  -  e^{i(X-\zeta)(pz)}
{\cal K}^{\bar a}_{\zeta}(X;t)  \right ) 
 \,  dX
,
\nonumber
 \end{eqnarray} 
where $M$ is the nucleon mass and $s,s'$ specify the nucleon
 polarization. Throughout the paper, we use the ``hat''
(rather than ``slash'') convention $\hat z \equiv z^{\mu} 
\gamma_{\mu}$. 
In Eq.    (\ref{57}), the quark and  antiquark contributions
are  explicitly separated (cf. \cite{lnc}). The exponential 
$  e^{-iX(pz)}$ associated with the functions
  ${\cal F}^a_{\zeta}(X;t)$ and ${\cal K}^a_{\zeta}(X;t)$
indicates  that the field $\psi_a(z)$ corresponds to
the $a$-quark taking the momentum $Xp$ from the nucleon. 
When the momentum $Xp$ is taken   from the nucleon  by an  
$a$-antiquark, the corresponding annihilation operator
is in $\bar \psi_a(0)$, and the  functions ${\cal F}^{\bar a}(X;t)$ and 
${\cal K}^{\bar a}(X;t)$ are accompanied by the exponential 
$  e^{i(X-\zeta)(pz)}$ corresponding to  the momentum 
at the $\psi_a(z)$-vertex.
The antiquark  terms come with the  minus sign 
because the creation and annihilation operators 
for them appear in the reversed  order.

As emphasized by X. Ji \cite{ji}, 
the parametrization of this nonforward matrix element
must  include both the nonflip
term described by  the functions 
$ {\cal F}_{\zeta}(X;t)$
and the spin-flip term\footnote{
The possibility of a  spin-flip in 
nonforward matrix elements was discussed earlier
in refs.\cite{barahzh,ral}.}
 characterized by the functions which we denote
by  ${\cal K}_{\zeta}(X;t)$. 
Taking the $O(z)$ term of the Taylor expansion gives the 
sum rules (see \cite{ji})
\begin{eqnarray} 
\int_0^1   
 \left [
 {\cal F}^a_{\zeta}(X;t)  -  
{\cal F}^{\bar a}_{\zeta}(X;t)  \right ]
 \,  dX  =F^a_1(t) \, , \label{58}
\\
\int_0^1   
 \left [
 {\cal K}^a_{\zeta}(X;t)  -  
{\cal K}^{\bar a}_{\zeta}(X;t)  \right ] 
 \,  dX  =F^a_2(t)
\label{59} \end{eqnarray}
relating the nonforward distributions $ {\cal F}^a_{\zeta}(X;t)$,  
$ {\cal K}^a_{\zeta}(X;t)$ to 
the $a$-flavor components of the Dirac and Pauli
form factors:
\begin{equation}
\sum_a e_a F^a_1(t) = F_1(t)  \ \  , \  \  \sum_a e_a F^a_2(t) = F_2(t) \,  ,
\end{equation}
 respectively
(see also \cite{ral} and \cite{frankfurt73}).
The spin-flip terms disappear only if  $r=0$.  
 In the  weaker  $r^2 \equiv t =0$  limit,
they survive,
$e.g.,$ $F^a_2(0)= \kappa^a$ is the $a$-flavor 
contribution to the anomalous magnetic moment.
In the formal $t=0$ limit, the nonforward 
distributions  $ {\cal F}^a_{\zeta}(X;t)$,
$ {\cal K}^a_{\zeta}(X;t)$ 
reduce to the  asymmetric distribution functions
 $ {\cal F}^a_{\zeta}(X)$,  $ {\cal K}^a_{\zeta}(X)$. 
It is worth mentioning here that 
for a massive target  (nucleons in our case)
there is a kinematic restriction \cite{afs} 
\begin{equation}
-t \gtrsim \zeta^2 M^2/\bar \zeta. 
\end{equation}
Hence,  for fixed $\zeta$, the formal limit $t \to 0$
is not  physically reachable. 
However, many results (evolution equations being the
most important example)
obtained in the formal $t =0$, $M=0$  
limit are still applicable.

In the region $X \geq \zeta$,
the initial quark momentum $Xp$ 
is larger than the momentum transfer $r = \zeta p$, and 
we can treat ${\cal F}_{\zeta}^a  (X)$ 
as  a generalization of the 
usual  distribution function $f_a(x)$. 
When $\zeta  \to 0$, the  limiting curve 
for ${\cal F}_{ \zeta}(X)$ reproduces $f_a(X)$:
\begin{equation}
 {\cal F}^a_{\zeta=0} \, (X) =   f_a(X) \  \ ; \  \  
 {\cal F}^{\bar a}_{\zeta=0} \, (X) =   f_{\bar a}(X) 
 . 
\label{60} \end{equation}

The spin-flip asymmetric  distribution functions
${\cal K}_{ \zeta}(X)$  do not necessarily 
vanish in the  $\zeta  \to 0$ limit.
However, the relevant nucleon matrix  element
$\bar u(p')  (\hat z  \hat r - \hat r \hat z) u(p)$
is proportional to $\zeta$ and the spin-flip term
is invisible in the forward case.

In the region  $X < \zeta$, one can define $Y=X/\zeta$ and treat 
the function ${\cal F}_{\zeta}^a  (X)$ 
as a distribution amplitude $\Psi_{\zeta}^a(Y)$.
In particular, the nonflip part in this region can be written as  
\begin{equation}
\zeta \, \bar u(p')  \hat z 
 u(p) \, \int_0^1   
 \left [ e^{-iY(rz)}
 {\cal F}^a_{\zeta}(\zeta Y)  -  e^{-i(1-Y)(rz)}
{\cal F}^{\bar a}_{\zeta}(\zeta Y)  \right ] 
 \,  dY = \zeta \, \bar u(p')  \hat z 
 u(p) \, \int_0^1   
 e^{-iY(rz)}
 \Psi_{\zeta}^a(Y)
 \,  dY \, , 
\label{61} \end{equation}
where  the distribution amplitude  $\Psi_{\zeta}^a(Y)$ is defined by 
\begin{equation}
\Psi_{\zeta}^a(Y) = {\cal F}_{\zeta}^a (Y\zeta) 
- {\cal F}_{\zeta}^{\bar a} (\bar Y \zeta)\, .
\label{62} \end{equation}
The function  $\Psi_{\zeta}^a(Y)$ gives the probability 
amplitude that  the initial nucleon with momentum $p$ is composed 
of the final nucleon with momentum $(1-\zeta)p\equiv p-r$
and a $\bar qq$-pair in which the total pair momentum $r$ 
is shared in fractions
$Y$ and $1-Y \equiv \bar Y$.

\subsection{Gluon Distribution}

For gluons, the    nonforward  distribution 
can be defined  
through the matrix element
\begin{eqnarray} 
\hspace{-1cm} 
&&\lefteqn{
\langle p'  \,  | \,   
z_{\mu}  z_{\nu} G_{\mu \alpha}^a (0) E^{ab}(0,z;A) 
G_{ \alpha \nu }^b (z)\, | \,p  \rangle |_{z^2=0}
 } \label{63}  \\ \hspace{-1cm} &&  
= \bar u(p')  \hat z 
 u(p) \, (z \cdot p) \, \int_0^1   
\frac1{2}  \left [ e^{-iX(pz)}
 + e^{i(X-\zeta)(pz)} \right ] 
 {\cal F}^g_{\zeta}(X;t)  
 \,  dX 
\nonumber \\ \hspace{-1cm} &&  
+ \, 
\bar u(p')  \frac {\hat z  \hat r - \hat r \hat z}{2M}
 u(p) (z \cdot p) \, \int_0^1   
\frac1{2}  \left [ e^{-iX(pz)}
 + e^{i(X-\zeta)(pz)} \right ] 
 {\cal K}^g_{\zeta}(X;t) 
 \,  dX \,  . 
\nonumber
 \end{eqnarray} 
The exponentials $e^{-iX(pz)}$ and $ e^{i(X-\zeta)(pz)} $ are  
accompanied here by the same function  ${\cal F}^g_{\zeta}(X;t)$
reflecting the fact that gluon and ``antigluon'' is  the same thing.
 Again, the contribution from the region $0<X<\zeta$
can be written   as
\begin{equation}
\bar u(p')  \hat z 
 u(p) \, (z \cdot r) \, \int_0^1   
  e^{-iY(rz)} \, \Psi_{\zeta}^g(Y;t)
 \,  dY  + ``{\cal K}" \, {\rm term}, 
\label{64} \end{equation}
with the generalized  $Y \leftrightarrow \bar Y$ symmetric
distribution amplitude
$\Psi_{\zeta}^g(Y;t)$ given by
\begin{equation}
\Psi_{\zeta}^g(Y;t) = \frac12 \left ( {\cal   F}^g_{\zeta}(Y \zeta;t)
+ {\cal   F}^g_{\zeta}(\bar Y \zeta;t) \right ) \, .
\label{65} \end{equation}
In the formal $t=0$ limit, the nonforward distributions 
${\cal F}^g_{\zeta}(X;t)$, ${\cal K}^g_{\zeta}(X;t)$
convert into the asymmetric distribution functions
${\cal F}^g_{\zeta}(X)$, ${\cal K}^g_{\zeta}(X)$.
Finally, in the $\zeta =0$ limit, ${\cal F}^g_{\zeta}(X)$
reduces to the usual gluon density
\begin{equation}
{\cal F}^g_{\zeta=0}(X) = Xf_g(X).
\label{66} \end{equation}

\subsection{Flavor-singlet and valence quark distributions}

In our original definition (\ref{57})  of the quark distributions, 
the exponentials $\exp [-iX(pz)]$ and $\exp [i(X-\zeta)(pz)]$
are accompanied by different functions ${\cal F}_{\zeta}^a (X;t)$
and $ {\cal F}_{\zeta}^{\bar a} (X;t))$, respectively. 
In many cases, it is convenient 
to introduce  the flavor-singlet quark operator 
\begin{equation}
{\cal O}_Q(uz,vz) =   \sum_a {\cal O}_a^{(+)}(uz,vz) 
 \label{67} \end{equation}
where 
\begin{equation}
{\cal O}_a^{(+)}(uz,vz)  =  \frac{i}{2}
\biggl [ \bar \psi_a(uz) 
\hat z E(uz,vz;A)  \psi_a(vz)
- \bar \psi_a(vz) \hat z  E(vz,uz;A) \psi_a(uz) \biggr ] \, . 
 \label{68} \end{equation}
The nonforward  distribution function
${\cal F}_{\zeta}^Q (X;t)$ for the 
flavor-singlet quark combination (\ref{67})
\begin{equation} 
\langle \,  p',s' | \, {\cal O}_Q(uz,vz)\, | \, p,s \rangle |_{z^2=0}  = 
 \bar u(p',s')  \hat z 
 u(p,s)   \int_0^1    \, 
 \frac{i}{2}  \left [ e^{-ivX(pz)+iuX'(pz)} - e^{ivX'(pz) 
- iuX(pz)}\right ]
{\cal F}_{\zeta}^Q (X;t)
 \,  dX \, + ``{\cal K}"
\label{69} \end{equation}
(where $X' \equiv X - \zeta$) 
can be expressed as the  sum of ``$a+\bar a$'' 
distributions:
\begin{equation}
 {\cal F}_{\zeta}^Q (X;t) = \sum_a ({\cal F}_{\zeta}^a (X;t) +
{\cal F}_{\zeta}^{\bar a} (X;t)) \, . 
\label{70} \end{equation}
Writing the  contribution from the $0<X<\zeta$ region as
\begin{equation}
\zeta \bar u(p')  \hat z 
 u(p) \, (z \cdot r) \, \int_0^1   
  e^{-iY(rz)} \Psi_{\zeta}^Q(Y;t)
 \,  dY  + ``{\cal K}" \, {\rm term}, 
\label{64A} \end{equation}
we introduce the flavor-singlet quark
distribution amplitude $\Psi_{\zeta}^Q(Y;t)$ 
which has the  antisymmetry property 
$\Psi_{\zeta}^Q(Y;t) = - \Psi_{\zeta}^Q(\bar Y;t)$  with respect to
the $Y \leftrightarrow \bar Y$ transformation.

Another combination of  quark operators
\begin{equation}
{\cal O}_a^{(-)}(uz,vz)  = \frac12 
\biggl [ \bar \psi_a(uz) 
\hat z E(uz,vz;A)  \psi_a(vz)
+ \bar \psi_a(vz) \hat z  E(vz,uz;A) \psi_a(uz) \biggr ] 
 \label{71} \end{equation}
corresponds to the valence combinations
${\cal F}_{\zeta}^{V_a} (X;t) \equiv {\cal F}_{\zeta}^a (X;t) - 
{\cal F}_{\zeta}^{\bar a} (X;t)$:
\begin{equation} 
\langle \,  p',s' | \, {\cal O}_a^{(-)}(uz,vz)\, | \, p,s \rangle |_{z^2=0}  = 
 \bar u(p',s')  \hat z 
 u(p,s)   \int_0^1    \, 
 \frac{1}{2}  \left [ e^{-ivX(pz)+iuX'(pz)} + e^{ivX'(pz) 
- iuX(pz)}\right ]
{\cal F}_{\zeta}^{V_a} (X, t) 
 \,  dX \, + ``{\cal K}" .
\label{72} \end{equation}
In both cases (see Eqs.(\ref{69}),(\ref{72}) ), 
two possible  exponential factors are accompanied by the same 
distribution function, just like for the gluon distribution. 
In the region $0<X<\zeta$,  the function ${\cal F}_{\zeta}^{V_a} (X;t) $
can be written in terms of the flavor-nonsinglet distribution
amplitide $\Psi_{\zeta}^{V_a}(Y;t)$  which is symmetric
$\Psi_{\zeta}^{V_a} (Y;t) = \Psi_{\zeta}^{V_a} (\bar Y;t)$
with respect to the $Y \leftrightarrow \bar Y$ interchange.

\section{ Evolution equations for nonforward distributions}

\subsection{General formalism} 

Near  the light cone $z^2 \sim 0$, the bilocal operators $\phi(0)\phi(z)$ 
develop logarithmic singularities $\ln z^2$, so that the formal
limit $z^2 \to 0$ is singular.  Calculationally, these
singularities manifest themselves as  ultraviolet divergences
for the light-cone operators. 
The divergences are removed by  a 
subtraction  prescription characterized 
by some  scale $\mu$: 
${\cal F}_{\zeta} (X;t) \to {\cal F}_{\zeta} (X;t;\mu)$.
In QCD,  the gluonic 
operator
\begin{equation}
{\cal O}_g(uz,vz) =
z_{\mu}  z_{\nu} 
G_{\mu \alpha}^a (uz) E^{ab}(uz,vz;A) 
G_{\alpha \nu}^b (vz) 
 \label{73} \end{equation}
 mixes under renormalization with the flavor-singlet 
quark operator.
At one loop (${\rm i.e.},$ in the leading logarithm 
approximation), the easiest way to get the evolution equations 
for  nonforward distributions  is 
to use the 
evolution equation \cite{brschweig,bb}
for the light-ray 
operators\footnote{This procedure was also used 
in  a recent paper
\cite{bgr} . I was informed by  J. Blumlein that
its authors agree with my results for 
the $W_{\zeta} (X,Y)$ kernels given below.}.
For the flavor-singlet case, it 
reads\footnote{We prefer to use  the   kernels
$B_{ab}( u,v )$ 
which have the symmetry property 
$B_{ab}( u,v ) = B_{ab}( v,u )$  
and are related to the 
 $K^{ab}( u,v )$ kernels of Ref.  \cite{bb}  
by 
$B_{ab}( u,v ) = -K^{ab}(\bar u,v ) $.} 
\begin{equation}
 \mu \, \frac{d}{d \mu} \,  
{\cal O}_a(0,z)    =
\int_0^1  \int_0^{1}  
\sum_{b} B_{ab}(u,v ) {\cal O}_b( uz, \bar vz) \,
\theta (u+v \leq 1) \, du \, d v  \,  , 
\label{74} \end{equation}
where $\bar v \equiv 1-v$ and $a,b = g,Q$. For valence distributions,
there is no mixing, and their  
 evolution 
is generated  by the $QQ$-kernel  alone.
Inserting Eq.    (\ref{74})  between chosen  hadronic states
and parametrizing the matrix elements by appropriate
distributions, 
one can get the well-known evolution kernels 
such as DGLAP and BL-type   
kernels and also to calculate  the 
nonforward  kernels  $R^{ab}(x,y;\xi,\eta)$ and  $W_{\zeta}^{ab}(X,Z)$.
The kernels $ R^{ab}(x,y;\xi,\eta)$
  govern the evolution of  
the double  distributions:
\begin{equation} 
 \mu \frac{d}{d\mu}  \, F^a(x,y;t;\mu) =
\int_0^1 \int_0^1 \, \sum_b \,  R^{ab}(x,y;\xi,\eta) \, 
  F^b(\xi,\eta;t;\mu)\, \theta(\xi +\eta \leq 1) \,
d \xi \, d \eta \,  ,
\label{75} \end{equation}
where $a$ and $b$ denote $g$ or $Q$. 
Another set of kernels  $W_{\zeta}^{ab}(X,Z)$ 
dictates the evolution of the nonforward distributions 
and asymmetric distribution functions:
\begin{equation}
 \mu \frac{d}{d\mu}  {\cal F}_{\zeta}^a(X;t;\mu) =
\int_0^1  \, \sum_b \,  W_{\zeta}^{ab}(X,Z) \, 
{\cal F}_{\zeta}^b( Z;t;\mu) \, d Z \,  .
\label{76} 
 \end{equation}

The evolution of the double 
distributions will be briefly discussed later in Sec.  VI. 
Here we will discuss the structure of the
$ W_{\zeta}^{ab}(X,Z)$ kernels.   
Since the form of the equation is not affected by the
$t$-dependence,  ``$t$''  will not be explicitly
indicated in what follows.

Before  starting the actual  calculations,  
one should take into account  
that the gluon distribution  
${\cal F}_{\zeta}^g (X)$ is accompanied 
by the sum of two exponentials while the flavor singlet
quark distribution  ${\cal F}_{\zeta}^Q (X)$ 
with which it mixes is accompanied by the difference.
This  sign change is, in fact,  compensated by 
the  extra $(pz)$ factor in the right-hand side 
of  the  gluon distribution definition.
The set of   evolution equations for  ${\cal F}_{\zeta}^Q (X)$, 
${\cal F}_{\zeta}^g (X)$ can be obtained by 
substituting the definitions  of 
the gluon (\ref{63}) and quark (\ref{69})  distributions
 into Eq.    (\ref{74})  and 
performing  the  Fourier 
transformation  with respect to the $(pz)$-variable.
For this procedure,  the $(pz)$-factor 
is equivalent to differentiation $d/dX$ 
while $1/(pz)$  results in an integration over $X$.
Note, that both operations change 
the relative sign of the exponentials. 
Hence, it is convenient to introduce first the auxiliary kernels
$M^{ab}_{\zeta}(X,Z)$ which would appear in the absence of the $(pz)$ mismatch. 
They are  directly related  by 
\begin{equation}
M^{ab}_{\zeta}(X,Z) = \int_0^1  \int_0^1 B_{ab}(u,v) \, 
\delta(X- \bar u Z + v (Z- \zeta)) \,
\theta(u+v \leq 1)
\, du \, dv   
\label{77} \end{equation}
to  the    light-ray evolution 
 kernels  \cite{brschweig,bb},
which we write here in the form given in Ref.  \cite{gluon}:
\begin{eqnarray}
\begin{array}{rl} 
& \displaystyle  
B_{QQ}(u,v ) = \frac{\alpha_s}{\pi} C_F \left 
(1 + \delta( u) [\bar v/v]_+  + 
\delta(v) [\bar u/ u]_+ - \frac1{2} \delta( u)\delta(v) \right ) \, ,
\\
& \displaystyle  B_{gQ}(u,v ) = \frac{\alpha_s}{\pi} C_F \biggl 
(2 + \delta( u)\delta(v) \biggr  ) \, ,
\\ 
& \displaystyle  B_{Qg}(u,v ) = \frac{\alpha_s}{\pi} N_f \left 
(1 + 4uv -u -v \right ) \, , \\ 
& \displaystyle  B_{gg}(u,v ) = \frac{\alpha_s}{\pi} N_c \biggl (
4(1 + 3uv -u -v) + \frac{\beta_0}{2 N_c} \,
\delta(u)\delta(v) 
+ \left  \{ \delta(u) \biggl[ \frac{\bar v^2}{v} - \delta (v) \int_0^1
\frac{d z}{z} \biggr ] 
+ \{ u \leftrightarrow v \} \right \} 
 \biggr  )  \, .
\end{array}
\label{80} \end{eqnarray}

 The  $W$-kernels are related to the $M$-kernels  by 
\begin{eqnarray}
W^{gg}_{\zeta}(X,Z)=M^{gg}_{\zeta}(X,Z) \  \    ,   \  \ 
W^{QQ}_{\zeta}(X,Z)=M^{QQ}_{\zeta}(X,Z), \label{78} \\ 
\frac{\partial}{\partial X} W^{gQ}_{\zeta}(X,Z) =
-  M^{gQ}_{\zeta}( X,Z)
\,  \ \  , \  \
W^{Qg}_{\zeta}(X,Z)= - \frac{\partial}{\partial X} 
\, M^{Qg}_{\zeta}(X,Z) \,  . 
\label{79} \end{eqnarray}
Hence, to get $W^{gQ}_{\zeta}(X,Z)$ we should integrate 
$M^{gQ}_{\zeta}(X,Z)$ with respect to $X$.
The integration constant can be fixed from the requirement
that $W^{gQ}_{\zeta}(X,Z)$ vanishes  for $X>1$.
Then 
\begin{equation}
W^{gQ}_{\zeta}(X,Z)= \int_X^1 M^{gQ}_{\zeta}(\widetilde X,Z)
\, d \widetilde X  \,  . \label{79A}
\end{equation}

Integrating  the 
 delta-function in eq.(\ref{77}) 
produces   four different types of the $\theta$-functions,
each of which 
corresponds to a specific
evolution regime for  the nonforward 
distributions. 
In two extreme cases,  when
 $\zeta =0$ or $\zeta =1$, the evolution 
equation reduces to  known 
DGLAP and BL-type  equations, respectively.

\subsection{BL-type  evolution kernels} 

When  $\zeta =1$,  the initial momentum 
coincides with the momentum transfer and  
${\cal F}_{\zeta}(X)$ reduces to a distribution amplitude
whose  evolution is governed by the 
BL-type  kernels:
\begin{equation}
W_{\zeta =1}^{ab}(X,Z)= V^{ab}(X,Z). 
\label{81} \end{equation}

Taking $\zeta =1$ in Eq.    (\ref{77})  we obtain 
\begin{equation}
M^{ab}_{\zeta =1}(X,Z) \equiv U^{ab}(X,Z) 
= \int_0^1  \int_0^1 B_{ab}(u,v) \, 
\delta(X- \bar u Z - v (1-Z)) \,
\theta(u+v \leq 1)
\, du \, dv   \, . 
\label{82} \end{equation}
Eliminating the $\delta$-function, one would  observe that 
in the regions $X \leq Z$ and $X \geq Z$ the $U^{ab}(X,Z)$
 kernels are given 
by  different analytic expressions. 
However,  from the representation  (\ref{82})
and the symmetry property  $B_{ab}(u,v) =  B_{ab}(v,u)$
it follows that $U^{ab} (\bar X, \bar Z) = U^{ab}(X,Z)$. 
Hence,  it is sufficient to know 
the $U$-kernels in the $X \leq Z$ region only.
The basic function
$ U_0^{ab}(X,Z) \equiv \theta (X \leq Z) \, 
 U^{ab}(X,Z)$ can be calculated 
from 
\begin{equation}
U_0^{ab}(X,Z) 
= \frac1{Z} \int_0^{X}   \, 
B_{ab} \left  ( \bar v - (X-v)/Z  ,v \right ) dv \, . 
\label{83} \end{equation}
The total kernel $U^{ab}(X,Z)$ then can be written as 
$$
U^{ab}(X,Z) = \theta (X \leq Z) \,  U_0^{ab}(X,Z)
+ \theta (Z \leq X) U_0^{ab}(\bar X, \bar Z) \, .
$$
One can easily derive a table of $B \to U_0$ conversion formulas 
for all the structures present in the $B$-kernels:
\begin{eqnarray}
\delta(u) \, \delta(v) \to \delta(Z-X) \ \ , \  \  1 \to \frac{X}{Z} 
\ \ , \  \ 
\delta(u) \, \frac{\bar v}{v} \to 0 \ \ , \  \ 
\delta(u) \, \left ( \frac{\bar v}{v} \right )^2 \to 0 , \nonumber  \\
\delta(v) \,  \frac{\bar u}{u} \to \left (\frac{X}{Z}\right )  \frac1{Z-X}  \ \ , \  \ 
\delta(v) \,  \frac{\bar u^2}{u} \to \left (\frac{X}{Z} \right )^2  \frac1{Z-X} \, ,
\nonumber  \\
u+v \to \frac{X}{Z} \left ( 1 -  \frac{X}{2Z}  \right ) \ \ , \  \ uv \to 
\frac{X^2}{Z} \left ( \frac12 -  \frac{X}{6Z} - \frac{X}{3} \right ) \,  .
\label{84} \end{eqnarray} 
Using Eqs.(\ref{80})  and  this table, we can  get 
the BL-type  kernels $V^{ab}(X,Z)$.
Before doing this, we note that the BL-type kernels  
appear as a  part of the nonforward kernel $W_{\zeta }^{ab}(X,Z)$
even in the general $\zeta \neq 1,0$ case.
As explained earlier, if   $X$  is  in the
region $X \leq \zeta$, 
then the  function ${\cal F}_{\zeta}(X)$ 
can  be treated as a distribution amplitude
$\Psi_{\zeta}(Y)$ with $Y= X/  \zeta$. 
For this reason, when both $X$ and $Z$ are smaller than $\zeta$,
we would expect that the kernels 
$W_{\zeta}^{ab}(X,Z)$  must simply  reduce 
to the  BL-type  evolution kernels $V^{ab}(X/\zeta,Z/\zeta)$.
Indeed, the relation (\ref{77}) can be written as  
\begin{equation}
M^{ab}_{\zeta}(X,Z) = \frac1{\zeta} \int_0^1  \int_0^1 B_{ab}(u,v) \, 
\delta \left ({X}/{\zeta}- \bar u {Z}/
{\zeta}  - v ( 1-{Z}/{\zeta}) \right ) \,
\theta(u+v \leq 1)
\, du \, dv  \, . 
\label{85} \end{equation}
Comparing this expression with the representation 
for the $U_0^{ab}(X,Z)$ kernels, we conclude   that,  
in the region where $X/\zeta \leq 1$ and $Z/\zeta \leq 1$,
the kernels $M^{ab}_{\zeta}(X,Z)$ are given by 
\begin{equation}
M_{\zeta }^{ab}(X,Z) |_{0 \leq \{X,Z \} \leq \zeta}  =
\frac1{\zeta} \,  U^{ab}\left ({X}/{\zeta}, 
{Z}/{\zeta} \right )\, .
\label{86} \end{equation}
From the expressions  connecting the $W$-  and $M$-kernels, 
we obtain the following  relations between the nonforward evolution kernels
$W_{\zeta }^{ab}(X,Z)$ 
in the region $0 \leq \{X,Z\}  \leq \zeta$ 
(let us denote them by $L_{\zeta }^{ab}(X,Z) \equiv
 W_{\zeta }^{ab}(X,Z)|_{0 \leq \{X,Z\}  \leq \zeta}$)  
and the BL-type  kernels $V^{ab}(X,Z)$:
\begin{eqnarray}
 L_{\zeta }^{QQ}(X,Z) = \frac1{\zeta} \,  V^{QQ}\left ({X}/{\zeta}, 
{Z}/{\zeta} \right )  \ ;  \ 
L_{\zeta }^{gQ}(X,Z) =   V^{gQ}\left ({X}/{\zeta}, 
{Z}/{\zeta} \right )  \ ;  \nonumber \\ 
L_{\zeta }^{Qg}(X,Z) = \frac1{\zeta^2} \,  V^{Qg}\left ({X}/{\zeta}, 
{Z}/{\zeta} \right )  \ ;  \ 
L_{\zeta }^{gg}(X,Z) = \frac1{\zeta} \,  V^{gg} \left ({X}/{\zeta}, 
{Z}/{\zeta} \right ) \, . 
\label{87} \end{eqnarray}

Explicit calculations based on Eqs.(\ref{77})-(\ref{79A}), 
(\ref{81}), (\ref{87}) give 
\begin{eqnarray} 
&& \lefteqn{
V^{QQ}(X,Z) = \frac{\alpha_s}{\pi} \, 
C_F \, \left \{ \, \left [ \frac{X}{Z} 
\left ( 1 + \frac{1}{Z-X} \right )\, 
\theta\, (X <Z ) \right ]_+ \, + \, \{ X \to \bar X, Z \to \bar Z \} 
 \right \}  \, } \, , \label{88} \\ &&
V^{Qg}(X,Z) = - \frac{\alpha_s}{\pi} \, N_f \, 
 \left \{ \,  \frac{X}{Z} \left [ 4(1-X) +
 \frac{1-2 X}{Z} \right ] \, \theta\, (X <Z ) \, -
\, \{ X \to \bar X, Z \to \bar Z \}  \right \} \,  ,\label{89} \\ &&
V^{gQ}(X,Z) = \frac{\alpha_s}{\pi} \, C_F \, 
\left \{ \left (2 - \frac{X^2}{Z}\right ) \theta\, (X <Z ) \, + 
\frac{(1-X)^2}{1-Z} \  \theta\, (X >Z )
  \right \}   , \label{90} \\ &&
V^{gg}(X,Z) = \frac{\alpha_s}{\pi} \,  N_c \, 
\biggl \{ 2 \, {{X^2}\over{Z}} \left (3 -2X +\frac{1-X}{Z} \right ) \, 
+\frac1{Z-X}  \left (\frac{X}{Z} \right )^2   \nonumber \\ && \hspace{4cm} + 
\delta(X-Z) \left [ \frac{\beta_0}{2N_c} - 
\int_0^1 \, \frac{dz}{1-z}  \, \right ]
\biggr \} \, \theta\, (X <Z ) \,  + \, \{ X \to \bar X, Z \to \bar Z \}  \, .
\label{91} \end{eqnarray}

Note, that the $V^{gQ}(X,Z)$ kernel can be represented as 
the  sum 
\begin{eqnarray} 
V^{gQ}(X,Z) = \frac{\alpha_s}{\pi} \, C_F \, + \,  
\frac{\alpha_s}{\pi} \, C_F \, 
\left \{ \left (1 - \frac{X^2}{Z}\right ) \theta\, (X <Z ) 
\, - \, \{ X \to \bar X, Z \to \bar Z \} 
  \right \}   \label{90A}
\end{eqnarray}
of a constant term and a kernel which 
is explicitly antisymmetric with respect to the
$\{ X \to \bar X, Z \to \bar Z \}$ transformation.  
In fact, the  constant term  does not contribute 
to  evolution since the flavor-singlet distribution amplitude
$\Psi^Q(Z)$ with  which it is 
convoluted is  antisymmetric $\Psi^Q (Z) = - \Psi^Q (\bar Z)$.
For the same reason, the convolution of $V^{gQ}(X,Z)$ with 
$\Psi^Q(Z)$ determining the evolution correction to
${\cal F}_{\zeta}^g(X)$ behaves like $X^2$ for small $X$.

Furthermore, the BL-type kernels  also 
govern the evolution  in the region corresponding to transitions
from a fraction $Z$ which is larger
than $\zeta$ to a fraction $X$ which is smaller 
than $\zeta$. Indeed,  
using   the $\delta$-function to calculate the integral
over $u$, we get
\begin{equation}
M_{\zeta }^{ab}(X,Z) |_{X \leq \zeta \leq Z }
= \frac1{Z} \int_0^{X/\zeta}   \, 
B_{ab} \biggl  ( [1- X/Z -v(1-\zeta/Z)] \, ,v \biggr ) dv \, , 
\label{92} \end{equation}
which has the same analytic form (\ref{83}) 
as the expression for $M_{\zeta }^{ab}(X,Z) $
in the region $X \leq Z \leq \zeta$.
For $QQ, gg$ and $Qg$ kernels, this automatically means that
 $W_{\zeta }^{ab}(X,Z) |_{X \leq \zeta \leq Z }$ is given by the 
same  analytic expression as $L_{\zeta }^{ab}(X,Z) $ for $X<Z$.
Because of integration, to get  $W_{\zeta }^{gQ}(X,Z)$  one should also know 
$M_{\zeta }^{gQ}(X,Z) $ in the region $\zeta \leq X \leq Z$. 
However, our explicit calculation confirms that
 $W_{\zeta }^{gQ}(X,Z)$ in the transition region $X \leq \zeta \leq Z$ 
is given by the same expression as $L_{\zeta }^{gQ}(X,Z)$ for $X<Z$.

Note, that  the evolution jump through  the critical
fraction $\zeta$ is irreversible: 
the $\delta$-function in Eq.    (\ref{85})  requires that
$X/\zeta = v + (1-u-v) Z /\zeta $ or
$X \leq \zeta$ if  $Z \leq \zeta$. 
To put it in  words, when the parton momentum 
degrades in the evolution 
process to values smaller than the momentum transfer
$\zeta p \equiv r$,
further  evolution is like that for a distribution
amplitude: the momentum can decrease or increase
up to the $r$-value but cannot exceed this value.

\subsection{   Region  $Z  \geq \zeta$, $X  \geq \zeta$}

   Recall, that when  $X > \zeta$,  
the initial quark momentum $Xp$ 
is larger than the momentum transfer $r = \zeta p$,
and we can treat the asymmetric distribution
 function ${\cal F}_{\zeta}^a  (X)$ 
as  a generalization of the 
usual  distribution function $f_a(X)$ for a 
somewhat skewed kinematics. 
Hence, we  can expect that evolution in the region 
$\zeta < X \leq 1$   ,  
 $\zeta < Z \leq 1$ is similar
to that generated by the  DGLAP 
equation.
In particular,  it has the basic property that
  the evolved fraction
$X$ is always smaller than the original
fraction $Z$. The relevant kernels  are  
 given by 
\begin{equation} 
\displaystyle  
M_{\zeta }^{ab}(X,Z) |_{\zeta \leq X \leq Z \leq 1}
= \frac1{Z} \int_0^{\frac{1-X/Z}{1 -\zeta/Z}}    \, 
B_{ab} \biggl  ( [1- X/Z -v(1-\zeta/Z)] \, ,v \biggr ) dv \, . 
\label{93} \end{equation}
Changing  the integration variable to 
$w \equiv v(1- \zeta/Z)/(1 -X/Z) = v /(1-X'/Z')$,
we obtain  the  expression in which the arguments
of the $B$-kernels are treated in a more symmetric way
\begin{equation} 
\displaystyle  
M_{\zeta }^{ab}(X,Z) |_{\zeta \leq X \leq Z \leq 1}
= \frac{Z-X}{ZZ'} \int_0^{1}    \, 
B_{ab} \left   ( \bar w \, (1- X/Z)  \, , \, 
w\, (1- X'/Z')  \right ) dw \, , 
\label{94} \end{equation}
where $X' \equiv X - \zeta$ and $Z' \equiv Z - \zeta$
are the ``returning'' partners of the original
fractions $X,Z$. Moreover, since  $Z-X = Z' -X'$,  the kernels
$M_{\zeta }^{ab}(X,Z)$ are given by functions symmetric 
with respect to the interchange  of $X,Z$ with  $X',Z'$.  
This observation can be used to check the results of 
calculations.  However, since we are dealing with the 
asymmetric situation $X>X', Z>Z'$, 
  other practical applications
of this symmetry are not evident at the moment.
Again, we can easily obtain a table for transitions from the $B_{ab}$-kernels
to the $M^{ab}$-kernels for the region $\zeta \leq X \leq Z \leq 1$:
\begin{eqnarray} 
\delta(u) \, \delta(v)  \to \delta(Z-X)  \ \ ; \  \
1 \to \frac{Z-X}{ZZ'}   \ \ ; \  \
(u+v) \to \frac{Z-X}{2ZZ'} \left [2-  \frac{X}{Z} - \frac{X'}{Z'} \right ] \, ; 
\nonumber  \\ 
uv \to \frac{Z-X}{6ZZ'} \left (1-  \frac{X}{Z}\right ) 
\left (1 - \frac{X'}{Z'} \right )  \ \ ; \  \
\left (\delta(u) \,  \frac{\bar v}{v} + \delta(v) \, \frac{\bar u}{u}  \right ) 
\to \frac1{Z-X} \left [ \frac{X}{Z}+ \frac{X'}{Z'} \right ]  \, ;  
\nonumber  \\ 
\left (\delta(u) \,  \frac{\bar v^2}{v} + \delta(v) \, \frac{\bar u^2}{u}  \right ) 
\to \frac1{Z-X} \left [ \left (\frac{X}{Z} \right )^2
+ \left (\frac{X'}{Z'} \right )^2 \right ] \, . 
\label{95} \end{eqnarray}

Introducing the notation 
$P^{ab}_{\zeta}(X,Z)\equiv W_{\zeta }^{ab}(X,Z) 
|_{\zeta \leq X \leq Z \leq 1}$ and 
using the formulas given above, we calculate  the 
$P$-kernels\footnote{
Expressions for the 
 nonforward generalization
of the DGLAP evolution kernels (in  different notations) 
were  given  in  the  review \cite{glr} by L. Gribov, 
E. Levin and M. Ryskin. They discuss the generalized DGLAP  kernels
   in the context 
of the electroproduction amplitude with a 
timelike  photon (or $Z^0$) in the final state.
However,  as the  longitudinal momentum asymmetry  
parameter $\zeta$  for their kernels 
they took the ratio $\zeta_2 \equiv q_2^2/2(pq_1)$
involving only the invariant mass $q_2^2$ of the final photon.
As we have seen in Sec.  II F, the correct
value for $\zeta$ in this case is  $\zeta =\zeta_1+\zeta_2 $,
where $\zeta_1$ is the usual Bjorken parameter
$\zeta_1 \equiv -q_1^2/2(pq_1)$} :
\begin{eqnarray} 
&& \lefteqn{
P^{QQ}_{\zeta}(X,Z) = \frac{\alpha_s}{\pi} \, 
C_F \, \left \{ \,  \frac1{Z-X} \left [ 1 + \frac{XX'}{ZZ'} \right ]\,  
-\delta(X-Z) \int_0^1 \, \frac{1+z^2}{1-z} \ dz \right \}  \rightarrow 
\frac1{Z} P_{QQ}(X/Z) \, , }\label{96} \\ &&
P^{Qg}_{\zeta}(X,Z) = \frac{\alpha_s}{\pi} \, N_f \, \frac{1}{ZZ'}
 \left \{ \left (1 - \frac{X}{Z}\right ) \left (1 - \frac{X'}{Z'}\right )
 + \frac{XX'}{ZZ'} \right \} \, \rightarrow 
\frac1{Z^2} P_{Qg}(X/Z) \, , \label{97} \\ &&
P^{gQ}_{\zeta}(X,Z) = \frac{\alpha_s}{\pi} \, C_F \, 
\left \{ \left (1 - \frac{X}{Z}\right ) \left (1 - \frac{X'}{Z'}\right )
 + 1 \right \} \, \rightarrow 
\frac{X}{Z} P_{gQ}(X/Z) \,  , \label{98} \\ &&
P^{gg}_{\zeta}(X,Z) = \frac{\alpha_s}{\pi} \,  N_c \, 
\left \{ 2\left [ 1 + \frac{XX'}{ZZ'} \right ] \frac{Z-X}{ZZ'} 
+\frac1{Z-X} \left [ \left (\frac{X}{Z} \right )^2 +
 \left ( \frac{X'}{Z'} \right )^2 \right ] \right. \nonumber \\ && \left. 
\hspace{4cm} + \ 
\delta(X-Z) \left [ \frac{\beta_0}{2N_c} -
\int_0^1  \frac{du}{u} - 
\int_0^1  \frac{dv}{v}
 \, \right ]
\right \} \, \rightarrow 
\frac{X}{Z^2} P_{gg}(X/Z) \, .
\label{99} \end{eqnarray}
The formally divergent integrals over $u$ and $v$ 
provide here the usual ``plus''-type
regularization of the $1/(Z-X)$ singularities.
The prescription following 
from Eqs. (\ref{94}), (\ref{95}) is that  combining the $1/(Z-X)$
and $\delta(Z-X)$ terms into
$[{\cal F}_{\zeta}(Z)-{\cal F}_{\zeta}(X)]/(Z-X)$  in 
 the convolution of 
$ P_{\zeta}(X,Z)$ with ${\cal F}_{\zeta}(Z)$ 
one should change $u \to 1-X/Z$ and $v \to 1-X'/Z'$.

As expected, the  $ P^{ab}_{\zeta}(X,Z)$ kernels  
have a symmetric form. The arrows indicate
how the  nonforward   kernels $P^{ab}_{\zeta}(X,Z)$
 are related to the DGLAP kernels in the $\zeta=0$ limit
when $Z=Z'$ and $X=X'$.
Deriving these relations, one should take into account that
 the  asymmetric gluon distribution function 
${\cal F}_{\zeta}^g(X)$ reduces in the limit $\zeta=0$ 
to $Xf_g(X)$ rather than to $f_g(X)$. 

In the region $Z > \zeta$, the evolution is 
one-sided: the evolved fraction 
$X$ is smaller than $Z$. Furthermore, if $Z \leq \zeta$ then also $X\leq Z$,
${\rm i.e.},$ distributions in the $X > \zeta$ regions are not affected by 
the distributions in the $X < \zeta$ regions.
Hence, just like in the DGLAP case,  information  about 
the initial 
distribution in the $Z > \zeta$ region
is sufficient for  calculating  its evolution in this region.
This situation may  be contrasted with the evolution of distributions
in the $Z < \zeta$ regions: in that case one should know the 
asymmetric distribution functions in the whole domain $0 <Z <1$.

Qualitatively, the evolution in the $X, Z > \zeta$ region
proceeds just like in  the DGLAP evolution: the distributions shift to 
smaller and smaller values of $X$.
In the DGLAP case, the distributions approach the 
$\delta(x)$ form condensing at a single point $x=0$.
In the asymmetric case, the whole region $Z < \zeta$ works like a ``black hole''
for the partons:  after  they end up  there, they will never
come back to the $X > \zeta$ region.  Inside the $Z < \zeta$
region, the evolution is governed by the BL-equation
transforming the $\Psi_{\zeta}(Y)$ distribution amplitudes into 
their asymptotic forms like $Y \bar Y,  Y \bar Y (Y - \bar Y)$  
for the quarks
and $(Y \bar Y)^2, (Y \bar Y)^2(Y - \bar Y)$ for the gluons; 
a particular form is dictated by the symmetry properties
of the relevant operators.

\section{Asymptotic solutions of  evolution equations}

\subsection{Evolution of asymmetric distribution function }

To describe the qualitative  features
of the QCD evolution of the
nonforward distributions,
 we  will consider  the   simplest 
case, ${\rm i.e.},$ the evolution equation for the
 flavor-nonsinglet (valence) functions.  Then
 $Qg$, $gQ$ and $gg$ 
kernels do not contribute, 
and the evolution is completely determined by the
$QQ$-kernel.   
The multiplicatively renormalizable operators 
in this case were originally found in Ref.  \cite{tmf}
\begin{equation}
{\cal O}_n  = (z\partial_+)^n \, \bar \psi 
\lambda^a \hat z C_n^{3/2} 
( z\stackrel{\leftrightarrow}{\partial}/z\partial_+) \psi \, .
 \label{100} \end{equation}
 Here we  use 
the symbolic  notation 
$ (z\stackrel{\leftrightarrow}{\partial}/z\partial_+)$  of Ref.  \cite{tmf},
with 
$\stackrel{\leftrightarrow}{\partial}
=\stackrel{\rightarrow}{\partial} -
\stackrel{\leftarrow}{\partial}$ , $\partial_+
= \stackrel{\rightarrow}{\partial} +
\stackrel{\leftarrow}{\partial}$
and  $C_n^{3/2}(y)$ being   the Gegenbauer polynomials.
This  means that the Gegenbauer moments
\begin{equation}
{\cal C}_{\zeta}(n, \mu) = \int_0^1 C_n^{3/2} (2 Z/\zeta -1) 
{\cal F}_{\zeta} (Z; \mu) \, dZ
\label{101} \end{equation}
of  the asymmetric distribution function
${\cal F}_{\zeta} (X;\mu)$ have a simple evolution:
\begin{equation}
{\cal C}_{\zeta}(n, \mu)  = {\cal C}_{\zeta}(n, \mu_0)
\left [ \frac{ \ln \mu_0/\Lambda}{ \ln \mu/\Lambda} 
\right ]^{\gamma_n/ \beta_0} \,  ,
\label{102} \end{equation}
where  $\beta_0 = 11 -\frac23N_f$ is the 
lowest coefficient of the QCD $\beta$-function
and $\gamma_n$ are the nonsinglet anomalous dimensions \cite{gw,gp}
\begin{equation}
\gamma_n= C_F \left [ \frac1{2} - 
\frac1{(n+1)(n+2)} +2 \sum_{j=2}^{n+1} \frac1{j}
\right ].
\label{103} \end{equation}
For $n=0$,  the Gegenbauer  moment  coincides with the 
ordinary one and, 
since  $\gamma_0 =0$, the area under the curve remains constant.
Other Gegenbauer moments decrease as  $\mu$ increases.
For the ordinary moments of the nonforward
distribution 
\begin{equation}
{\cal M}_N({\zeta} ,\mu) \equiv 
\int_0^1 {\cal F}_{\zeta}(X;\mu) \, X^N \, dX 
\, , 
\label{104A} \end{equation}
using  explicit expression for the 
Gegenbauer polynomials
we can derive 
the following expansion  over the multiplicatively
renormalizable combinations ${\cal C}_{\zeta}(n,\mu)$:
\begin{equation}
{\cal M}_N({\zeta} ,\mu) =
\zeta^N \, N! \, (N+1)! \sum_{n=0}^N (-1)^n \frac{2(2n+3)}{(N+n+3)!(N-n)!}
\ {\cal C}_{\zeta}(n,\mu)  \, .
\label{104} \end{equation}
We can also write the expression 
which gives the evolved moments ${\cal M}_N({\zeta} ,\mu)$ 
in terms of the original ones:
\begin{equation}
{\cal M}_N({\zeta} ,\mu) =
\zeta^N \, N! \, (N+1)! \sum_{n=0}^N  \frac{(-1)^n 2(2n+3)}{(N+n+3)!(N-n)!}
\, \left [ \frac{ \ln \mu_0/\Lambda}{ \ln \mu/\Lambda} 
\right ]^{\gamma_n/ \beta_0} \,
\sum_{k=0}^n  \frac{(-1)^k (k+n+2)!}{2 \zeta^k \, k! \, (k+1)! \, (n-k)!}
\, {\cal M}_k({\zeta} ,\mu_0) \, .
\label{104B} \end{equation}

With increasing $N$, the number 
of contributing Gegenbauer  
moments ${\cal C}_{\zeta}(n,\mu)$  in Eq.    (\ref{104}) increases.
An important observation is 
 that the nonevolving (and $\zeta$-independent, but $t$-dependent)
term ${\cal C}(0)$ 
contributes to each  moment. As a result, in the $\mu \to \infty$ limit, 
all the moments tend to  constant values 
determined by the $n=0$ term in the sum (\ref{104}):
\begin{equation}
{\cal M}_N({\zeta},\mu\to \infty ) = 
\zeta^N \, \frac{6}{(N+2)(N+3)} \ 
{\cal C}(0)  \, = \,  
\int_0^{\zeta} \, \frac{{\cal C}(0)}{\zeta} \,  6 \, (X/\zeta) (1-X/\zeta ) \, 
  X^N \, dX  .
\label{105} \end{equation}
Note, that the last integral involves only 
the $X$-values smaller than $\zeta$.
This means that in the limit 
 $\mu \to \infty$, the function
 ${\cal F}_{\zeta}(X;\mu \to \infty)$
completely disappears from  the region $X \geq \zeta$, ${\rm i.e.},$ it 
reduces to the distribution amplitude $\Psi_{\zeta}(Y)$
which   ultimately tends 
to the usual asymptotic shape $6 \, Y(1-Y)$ in the $Y=X/\zeta $
variable:
 \begin{equation}
  {\cal F}_{\zeta}(X;\mu \to \infty) = 
6 \, {\cal C}(0) X  (1-X/\zeta )/ \zeta^2\,  . 
\label{106} \end{equation}

One may also be interested in finding expressions
showing how the function  ${\cal F}_{\zeta}(X;\mu)$ changes
its shape from an arbitrary original curve   ${\cal F}_{\zeta}(X;\mu_0)$
to the asymptotic one. 
Note, that    the Gegenbauer moments  for $\zeta < 1$
involve integration regions 
in which the argument  $C_n^{3/2} (2Z/\zeta-1)$
of the  polynomials  extends  beyond
the segment  $(-1 , 1)$ where they form an 
orthogonal set of functions. 
Hence, a formal inversion of  the Gegenbauer
 moments is only possible for $\zeta=1$. In this 
case, the inversion produces  the standard 
solution of the evolution equation for a distribution amplitude
\cite{tmf,bl}
\begin{equation}
{\cal F}_{\zeta=1} (X;\mu) = 
\sum \limits_{n=0}^{\infty} 
 \frac{4(2n+3)}{(n+1)(n+2)} \, X \bar X \,  C_n^{3/2} (2 X  -1)
\left [ \frac{ \ln \mu_0/\Lambda}{ \ln \mu/\Lambda} \right ]^
{\gamma_n/ \beta_0}\, 
\int_0^1 C_n^{3/2} (2 Z  -1) \, 
{\cal F}_{\zeta=1} (Z;\mu_0) \, dZ \,  . 
\label{107} \end{equation}
Thus,   if the initial distribution coincides with one of the
eigenfunctions  $X \bar X \,  C_n^{3/2} (2 X  -1)$, the 
evolution is very simple: the function  just decreases in magnitude
without changing its form.
 An  attractive feature of such a situation  is that 
approximating the initial distribution amplitude by a few 
lowest  Gegenbauer polynomials one obtains a simple  model
of its evolution.
 Inspired by this observation, one 
may be tempted  to construct  a  similar 
representation for the evolution of 
the asymmetric distribution function. 
Using   the expansion of the light-cone operator 
$\bar \psi(0) \lambda^a \hat z \psi (z)$ 
over the multiplicatively renormalizable  
operators ${\cal O}_n$ (see \cite{bb} ) 
\begin{equation}
\bar \psi(0) \lambda^a \hat z \psi (z) =
\sum \limits_{n=0}^{\infty} (-1)^{n} 
 \frac{2(2n+3)}{n!}
 \int_0^1 
(u\bar u)^{n+1} {\cal O}_n (uz)
\, du
\label{108} \end{equation}
and inserting it into the nonforward matrix element, we obtain   
\begin{equation}
{\cal F}_{\zeta} (X;\mu) = 
\sum \limits_{n=0}^{\infty} (-1)^{n} \, 
 \frac{2(2n+3)}{n!}
\left ( \frac{ \ln \mu_0/\Lambda}{ \ln \mu/\Lambda} 
\right )^{\gamma_n/ \beta_0}\, 
\, \zeta^n {\cal C}_{\zeta}(n, \mu_0) \, \int_0^1 
(u\bar u)^{n+1} 
\delta^{(n)}(X-u \zeta) \, du \,  . 
\label{109} \end{equation}
Integrating  
$(u\bar u)^{n+1} \delta^{(n)}(X-u \zeta)$  over $u$, we 
get the Gegenbauer polynomials 
$ C_n^{3/2} (2 X/\zeta  -1)$ accompanied by the spectral 
condition $X\leq \zeta$. 
This   means that the formal integration 
does not give a correct result for functions which do not 
vanish outside the region  $X \leq \zeta$. 
For such functions, one should first 
perform the summation over $n$ (which is, of course,  
practically impossible)  
and only then take the $u$-integral.

Another limit in which the integral over $u$
can be taken safely is $\zeta = 0$.
For small $\zeta$, the Gegenbauer polynomials
are dominated by the senior power $Z^n$ and in the $\zeta \to 0$ limit
we obtain 
\begin{equation}
{\cal F}_{\zeta=0} (X;\mu ) = 
\sum \limits_{n=0}^{\infty} 
 \frac{(-1)^{n}}{n!} \, \delta^{(n)}(X) \, 
\left ( \frac{ \ln \mu_0/\Lambda}{ \ln \mu/\Lambda} \right )^{\gamma_n/ \beta_0}\, 
\int_0^1 
{\cal F}_{\zeta=0} (Z;\mu_0) \, Z^n \,  dZ \,  ,
\label{110} \end{equation}
${\rm i.e.},$ the well-known 
result that the  moments of the usual parton densities 
have a simple DGLAP evolution. 
Note, that in this case, the functions which 
evolve without changing their shape are $\delta^{(n)}(x)$. 
From a pragmatic point of view, this observation is of little 
use. Modeling the solutions
of the DGLAP equations is known to be 
a rather complicated  excercise usually involving a numerical 
integration of the evolution equations.

Hence, the representation (\ref{109})  should   be understood only in the
sense of (mathematical) distributions in $X$ rather than
functions. To get meaningful results,
one should integrate them over $X$ with some
smooth function. 
In particular, integrating it with $X^N$,  one obtains 
the formula (\ref{104})  for the evolution of the $X^N$ moments of 
nonforward distributions.

\subsection{Evolution of double distribution}

Solving the evolution equation for the valence double distribution
$ F(x,y;\mu)$ defined by 
\begin{eqnarray} 
&& \lefteqn{\langle \,  p-r,s' | \, {\cal O}^{(-)}(0,z)|_{\mu}\, |
 \, p,s \rangle |_{z^2=0} }
\nonumber \\  && \hspace{0.5cm} = 
 \bar u(p',s')  \hat z 
 u(p,s)   \int_0^1    \, 
 \frac{1}{2}  \left [   e^{-ix(pz)-iy(r z)} + e^{ix(pz)-i\bar y(r z)}
 \right ]
F (x,y;\mu) 
 \,\theta(x+y \leq 1) \,  dx \, dy  \,  ,
\label{111} \end{eqnarray} 
we can give an alternative derivation of the asymptotic form 
of the valence nonforward distribution ${\cal F}_{\zeta} (X;\mu )$.
The $\mu$-dependence of  $ F(x,y;\mu)$  is governed by the
evolution equation
\begin{equation}
 \mu \, \frac{d}{d \mu} \,  F(x,y;\mu) =
\int_0^1 d \xi \int_0^1  R_{QQ}(x,y; \xi, \eta) 
F( \xi, \eta;\mu) d \eta.
\label{112} \end{equation}
Since the integration over $y$ converts $F(x,y)$ 
into the parton distribution function $f(x)$,
whose evolution is governed by the DGLAP equation
\begin{equation}
\mu \, \frac{d}{d \mu} \,  f(x;\mu) =
\int_x^1 \frac{d\xi}{\xi} P_{QQ}(x / \xi) 
f( \xi;\mu) d \xi, 
\label{113} \end{equation}
the kernel $R_{QQ}(x,y; \xi, \eta)$ must have  the property
\begin{equation}
\int_0^ {1-x}  R_{QQ}(x,y; \xi, \eta) d y =
\frac1{\xi} P_{QQ}(x/\xi).
\label{eq:rtop}
\label{114} \end{equation}
For a similar reason, integrating $R_{QQ}(x,y; \xi, \eta)$
over $x$ one should get  the BL-type   kernel:
\begin{equation}
\int_0^{1-y}   R_{QQ}(x,y; \xi, \eta) d x = V_{QQ}(y,\eta).
\label{115} \end{equation}

Explicit calculation  gives  for $R_{QQ}(x,y;\xi, \eta)$
 the following result 
\begin{eqnarray}
&& \hspace{-1.5cm} R_{QQ}(x,y;\xi, \eta) = 
\frac{\alpha_s}{\pi} C_F \frac1{\xi}
\left \{  \theta  (0 \leq x/\xi \leq 
{\rm min} \{ y/\eta, \bar y / \bar \eta \} ) -
\frac1{2} \delta(1-x/\xi) \delta(y-\eta) \right. \label{116}
 \\
 && \hspace{-1.5cm}  +\left.   
\frac{\theta (0 \leq x/\xi \leq 1) x/\xi}{ (1-x/\xi)} 
\left [ \frac1{\eta}\delta(x/\xi - y/\eta) + 
\frac1{\bar \eta} \delta(x/\xi - \bar y/ \bar \eta) \right]
-2\delta(1-x/\xi) \delta(y-\eta)
\int_0^1 \frac{z}{1-z} \, dz \right \}.
\nonumber
 \end{eqnarray}
It can also be obtained from the kernel $B_{QQ}(u,v)$
using the relation
\begin{equation}
 R_{QQ}(x,y; \xi, \eta) = \frac1{\xi} 
B_{QQ}(  y - \eta x/\xi, \bar y - \bar \eta x/\xi).
\label{117} \end{equation}

 It is easy to verify that the spectral constraint 
$x+y \leq 1$ is not violated by the evolution:
the kernel 
$R_{QQ}(x,y;\xi, \eta)$ has the property that 
$x+y \leq 1$ if $\xi+ \eta \leq 1$. 
Using our  expression for $R_{QQ}(x,y;\xi, \eta)$ and explicit
forms of the $P_{QQ}(x/\xi)$ and  $V_{QQ}(y, \eta)$ kernels 
(see Eqs. (\ref{96}), (\ref{88})  ) 
 one can  check that $R_{QQ}(x,y; \xi, \eta)$ 
satisfies the reduction formulas
(\ref{114}) and (\ref{115}).  
To solve  the evolution equation,  
we  combine
the standard methods used to find  solutions of  the underlying
DGLAP and BL  evolution equations.
To solve  the DGLAP equation, 
one should consider the moments with respect to $x$.
Multiplying Eq.    (\ref{112}) by  $x^n$,  integrating  
 over $x$ and utilizing 
the property 
 $R_{QQ}(x,y; \xi, \eta) = R_{QQ}(x/\xi,y; 1, \eta)/\xi$,
we get  
\begin{equation}
\mu \, \frac{d}{d \mu} \, F_n(y;\mu) =
\int_0^1  R_n (y,\eta) F_n(\eta;\mu) d \eta \,  , 
\label{118} \end{equation}
where  $F_n(y;\mu)$ is the $n$th $x$-moment of $F(x,y;\mu)$
\begin{equation}
F_n(y;\mu) = \int_0^{1-y} x^n  F(x,y;\mu) dx
\label{119} \end{equation}
and the kernel $R_n (y,\eta)$  is given by
\begin{equation}
R_n (y,\eta) = 
\frac{\alpha_s}{\pi} C_F 
\left \{ \left [  \left (\frac{y}{\eta} \right )^{n+1} 
\left [\frac1{n+1}+ \frac1{ \eta -y} \right ]
\theta(y \leq \eta) + 
\{y \to \bar y, \eta \to \bar \eta \} \right ] 
 + \delta(y-\eta) \left [ \frac1{2} -
\int_0^1 \frac{dz}{z} \,  \right ] \right \}.
\label{120} \end{equation}
It is straightforward to check 
that  $R_n (y,\eta)$ has the property
$$R_n (y,\eta) w_n(\eta) =
R_n (\eta,y) w_n(y),$$ where $w_n(y)= (y \bar y)^{n+1}$. 
Hence, the  eigenfunctions of $R_n (y,\eta)$ are orthogonal with
the weight $w_n(y)= (y \bar y)^{n+1}$, ${\rm i.e.},$ 
they are proportional to the Gegenbauer polynomials
$C^{n+3/2}_k(y-\bar y)$ (cf.\cite{bl,mikhrad}).
Now, we can write  the general solution of the evolution
equation
\begin{equation} 
F_n(y;\mu) =  2 \, (y \bar y)^{n+1} \frac{(2n+1)!\, (2n+2)!}{ n! \,  (n+1)!}
\sum_{k=0}^{\infty} 
\frac{(2k+2n+3) \, k! }{(2n+k+2)!}  \, 
C^{n+3/2}_k(y-\bar y) \,
\left [\frac{\log (\mu_0 /\Lambda)} {\log (\mu /\Lambda)} \right]^
{\gamma^{(n)}_k/\beta_0}\,  A_{nk}(\mu_0) \, ,
\label{121} \end{equation}
where 
\begin{equation} 
 A_{nk}(\mu_0) = \int_0^1 \, F_n(y;\mu_0) \, C^{n+3/2}_k(y-\bar y) \, dy
\label{A121} \end{equation}
and the anomalous dimensions $\gamma^{(n)}_k$ 
 are related to   the eigenvalues of the kernel $ R_n (y,\eta)$.
They coincide with the standard nonsinglet 
anomalous dimensions $\gamma_{N}$ (\ref{103}):  
$\gamma^{(n)}_k = \gamma_{n+k}$. 
Since $\gamma_0^{(0)}=0$, while  all other anomalous  dimensions 
are positive,   in the formal $\mu \to \infty$
limit  we have  $F_0(y, \mu \to \infty) \sim y \bar y$ and 
$F_n(y, \mu \to \infty) =0$  for all $n \geq 1$.
This means that 
\begin{equation}
F(x,y; \mu \to \infty)  \sim \delta(x) \, y \bar y,
\label{a121}  \end{equation}
${\rm i.e.},$ in each of its variables, the limiting function 
 $F(x,y; \mu \to \infty)$  
acquires the characteristic asymptotic form dictated by
the nature of the variable:
$\delta(x)$ is specific for the distribution functions \cite{gw,gp},
while  the $y \bar y$-form  is  
the asymptotic shape   for the lowest-twist two-body 
distribution amplitudes \cite{tmf,bl}.
For the asymmetric   distribution function this gives
${\cal F}_{\zeta}(X, \mu \to \infty) \sim (X/\zeta^2)(1-X/\zeta)$.
This result was already obtained in the previous subsection.

\section{Basic uses of nonforward distributions}

\subsection{Deeply virtual Compton scatterring}

Using the parametrization for the matrix elements
of the quark operator, we can easily write  a parton-type representation
for the handbag contribution to the DVCS amplitude:
\begin{eqnarray} 
T^{\mu \nu} (p,q,q') =   \frac{ 1}{2\, (p q')} \,
\sum_a 
e_a^2\, & &  \left [
\left (-g^{\mu \nu} + \frac1{p \cdot q' } 
(p^{\mu}q'^{\nu} +p^{\nu}q'^{\mu}) \right ) \,
 \left \{\  \bar u(p') \hat q'  u(p) T_F^a(\zeta )  +  \frac1{2M} \bar u(p') 
(\hat q' \hat r - \hat r \hat q' )u(p)  T_K^a(\zeta ) \right \} \right.
\nonumber \\ & & \left.
+ i \epsilon^{\mu \nu \alpha \beta} \frac{p_{\alpha} q'_{\beta}}{(pq') }
\,  
\left \{ \bar u(p') 
\hat q' \gamma_5  u(p) 
\,  T_G^a(\zeta ) 
+  \frac{(q'r)}{2M} \bar u(p') \gamma_5 
  u(p)  T_P^a(\zeta )  \right \}  \right ]
\label{122} \end{eqnarray}
where $\hat q' \equiv  \gamma_{\mu}q'^{\mu}$,  
and 
$T^a(\zeta )$ are the 
invariant amplitudes 
depending on  the scaling variable $\zeta $.
In particular, 
\begin{eqnarray} 
T_F^a(\zeta ) =
- \int_0^{1} 
\left  [ \frac1{X-\zeta +i\epsilon}
+ \frac1{X- i \epsilon} \right ]  \left ( {\cal F}^a_{\zeta}(X;t)
+ {\cal F}^{\bar a}_{\zeta}(X;t) \right ) \, dX \,  . 
\label{123}  \end{eqnarray} 
 
Since the nucleon is the lowest bound state in  the 3-quark 
system,  the nonforward distribution function 
for $t <0$ is real. Hence, the imaginary part of  $T_F^a(\zeta )$
 can be produced only by singularities of the terms
in the square brackets.
Taking into account  that  the nonforward distributions
vanish for $X=0$, we conclude that
only the term  containing $1/(X-\zeta +i\epsilon)$ 
generates the imaginary part:
\begin{eqnarray} 
\frac1{\pi}\, {\rm Im} \, T_F^a(\zeta ) = {\cal F}^a_{\zeta}(\zeta;t) \, + \, 
{\cal F}^{\bar a}_{\zeta}(\zeta;t) 
\label{125} \end{eqnarray} 
with a similar expressions for $ {\rm Im} \, T_{K,G,P}^a(\zeta )$.
As discussed  in Sec.  I,  the function 
 ${\cal F}^a_{\zeta}(\zeta;t)$  does 
not coincide with the usual parton distribution $f_a(\zeta)$, 
even in the formal $t \to 0$
limit.
To get the real part of the $1/(X-\zeta +i\epsilon)$ terms,
one should use the principal value prescription
\begin{eqnarray} 
{\rm Re} \, T_F^a(\zeta ) = - {\rm P} \int_0^1
({\cal F}^a_{\zeta}(X;t) \, + \, 
{\cal F}^{\bar a}_{\zeta}(X;t) ) \frac{dX}{X-\zeta} \, .
\label{126} \end{eqnarray} 
Since the principal value prescription is 
based on  cancellation of $X< \zeta$ and $X> \zeta$
parts of the integral, it makes  sense to 
preserve  ${\cal F}^a_{\zeta}(X;t)$  as a single function.
Splitting it into 
 $X< \zeta$ and  $X> \zeta$ components, one would simply get
two divergent expressions for the real part of the amplitude.

Let us  study  how these formulas are modified 
by the evolution. At one loop,  the $\ln Q^2$ term can 
be easily calculated using the coordinate representation:
\begin{eqnarray} 
T_1(p,q,q') = \frac{\alpha_s}{2\pi}
\int d^4 z  e^{-i(qz)} \int_0^1 \int_0^1
\langle \, p' \, | \, \bar \psi(uz) \frac{\hat z}{2i\pi(z^2)^2}
 \psi (\bar v z) \,  | \, p
\rangle \ln z^2 \, B_{QQ}(u,v) \, du \, dv  \, . 
\label{127} \end{eqnarray}
Parametrizing the matrix element by
the nonforward distribution (\ref{57}), we obtain for the 
 $s$-channel short-distance amplitude
\begin{eqnarray} 
t^{(s)}_1(p,q,q') =  \ln Q^2 \, \int_0^1 \int_0^1
(qp) \, \frac{B_{QQ}(u,v) \, du \, dv}{(q + (\bar v X -u X')p)^2+ i \epsilon} =
\frac{1}{2} \ln Q^2 \int_0^1 \int_0^1 \frac{B_{QQ}(u,v) 
\, du \, dv}{(1-u) X' -vX + i \epsilon} \, , 
\label{128} \end{eqnarray}
where $X' = X - \zeta$.
Using  explicit expression for the $B_{QQ}(u,v)$ kernel,
we obtain 
\begin{eqnarray} 
-  \frac1{X' +i\epsilon} \to  t_1^s(X)  = - \frac1{X' +i\epsilon}
\left \{ 1 + \frac{\alpha_s}{2 \pi} \, C_F \,  
\left [ \frac32 + 
\ln \left ( \frac{X' + i \epsilon}{ - \zeta + i \epsilon} \right ) 
 \right ]\, \ln Q^2  \right \} \, .
\label{129} \end{eqnarray}
A similar expression can be derived for the
 evolution of the $u$-channel-type term: 
\begin{eqnarray} 
- \frac1{X- i \epsilon} \to t_1^u (X) 
 = - \frac1{X- i \epsilon} \left \{ 1+ \frac{\alpha_s}{2\pi}  \, C_F \,  
\left [ \frac32 + 
\ln \left ( X / \zeta  \right ) 
 \right ]\, \ln Q^2 \right \}  \, .
\label{130} \end{eqnarray}
Clearly, the $u$-channel term  can be obtained from the $s$-channel one 
by the change $X' \to X$, $\zeta \to - \, \zeta$.
In the region $X < \zeta$, both $t_1^u$ and  $t_1^s$   are real.
Furthermore, it is easy to establish that the correction terms 
in both cases vanish  when integrated
with the asymptotic distribution $6X(1-X/\zeta)/\zeta$, 
explicitly showing that the latter does   not evolve with $Q^2$. 
Note that  $t_1^u(X)$ is purely real in the whole range $0 \leq X \leq 1$, 
while $t_1^s(X)$  is  purely real only in the region $X < \zeta$. 
 For   $X \geq  \zeta$, the latter  has an imaginary part:
\begin{eqnarray} 
\, t_1^s(X)   = - {\rm P} \, \frac1{X-\zeta}  \, + i \pi 
\delta (X-\zeta) + \frac{\alpha_s}{2\pi}\, C_F \, 
 \left \{ \,\frac32   \left ( - {\rm P} \, \frac1{X-\zeta }
+  i \pi  \delta (X-\zeta) \right )
\right.
\nonumber \\ \left.
 + i \pi \left [ \frac{ \theta(X \geq \zeta)}{(X-\zeta)} \right ]_+
 - \left [\frac{\ln  |X/\zeta-1|}{X-\zeta} \right ]_+ 
 \right \}\, \ln Q^2 \, .
\label{131} \end{eqnarray}

This information can be used  to write down the expression showing  the 
leading logarithm evolution of the
function  ${\cal F}_{\zeta}(\zeta;Q^2)$ determining the imaginary part 
of the amplitude: 
\begin{eqnarray} 
{\cal F}_{\zeta}(\zeta;Q^2) = {\cal F}_{\zeta}(\zeta;Q_0^2)
+ \frac{\alpha_s}{2\pi} \, C_F \, \ln Q^2/Q_0^2  \, \int_{\zeta}^1  
 \left \{ \,\delta (X-\zeta) \left ( \,\frac32 \, -
 \int_0^1   \frac{ dz}{1-z} \right ) 
+ \frac1{X- \zeta}  
\right \} {\cal F}_{\zeta}(X;Q_0^2) \, dX \, .
\label{132} \end{eqnarray} 
Evidently, the expression in the braces is 
given by the nonforward  evolution kernel
$P_{\zeta}^{QQ}(\zeta , X)$ (\ref{96}).   For the usual 
distribution function the analogous equation
contains the DGLAP kernel $P(\zeta/X)$:
\begin{eqnarray} 
f(\zeta;Q^2) = f(\zeta;Q_0^2)
+ \frac{\alpha_s}{2\pi} \, C_F \, \ln Q^2/Q_0^2 \, \int_{\zeta}^1  
 \left \{ \, \delta (X-\zeta) \left ( \,\frac32 \, - 2\int_0^1   \frac{ dz}{1-z} \right ) 
+ \frac{ 1+(\zeta/X)^2}{X-\zeta} 
\right \} f(X;Q_0^2) \, dX \, .
\label{133} \end{eqnarray} 
The  comparison of the two expressions  shows that
evolution of the function ${\cal F}_{\zeta}(\zeta;Q^2)$
is not identical to that of $f(\zeta;Q^2)$.
Recall also that in the forward case the lowest-order amplitude is proportional 
to $1/(X- \zeta + i  \epsilon ) + 1/(X + \zeta - i  \epsilon ) $.

\subsection{Gluonic contribution to 
hard exclusive meson electroproduction}

The kinematics of hard exclusive 
meson  electroproduction  processes
$\gamma^* p \to M p'$  is very close to that
of the virtual Compton scattering,
especially in a situation when one can neglect the 
mass of the  final meson.
Again, one can use the $\alpha$-representation
rules to determine possible  regimes capable
of producing a powerlike contribution for large
$Q^2$. The basic difference is the absence of the
regime analogous to short-circuiting 
a subgraph containing the photon vertices,
since instead of the final photon described 
by an  elementary field we have now a bound state.
Hence, the leading  short-distance  regime
corresponds to contraction into point
of a subgraph which  contains the virtual
photon vertex   and  located in the 
middle between the  two long-distance-sensitive 
$pp'$- and $q'$-components  of the diagram.
The $pp'$-component is described by 
the nonforward distribution function
while the $q'$-part is parametrized by the
meson distibution amplitude. 

Depending on the type of lines 
connecting the short-distance subgraph
with the $\langle p' | \ldots | p \rangle$
matrix element, one deals either with 
 quark  (Fig.  9$a$) or  gluonic  (Fig.  10) contributions
to the lowest-order amplitude.
The structure of the quark contribution is
similar to that of the hard-gluon-exchange 
contribution to a  meson electromagnetic form factor,
with the  distribution amplitude of the initial state 
substituted by 
the quark nonforward distribution. 
There is also an  analogue of the soft contribution 
to the meson form factor (see Fig.  9$b$).
It corresponds to the infrared regime $\alpha_{\sigma_3} \to \infty$.

Let us concentrate here on the gluonic contribution 
which requires a proper handling of 
restrictions imposed by 
gauge invariance. 
Using the coordinate representation for 
the hard propagators,
we can write the contribution of Fig.  10$a$ as 
\begin{equation} 
T^{\alpha}_g (p,p',q') = \int  \, 
\langle q', M| \bar \psi(0) \ \gamma^{\alpha} S^c(-z_1)  \tau^a \gamma^{\mu}
 S^c(z_1-z_2) \tau ^b \gamma^{\nu}   \psi (z_2) |0  \rangle  \, 
\, \langle p' |  A_{\mu}^a (z_1)  A_{\nu}^b (z_2) | p \rangle \, 
d^4 z_1 d^4 z_2 \,,
\label{133A}
\end{equation}
where $\tau^a, \tau^b$ 
are the $SU(3)$ color matrices. 
The first matrix element here can be expressed through 
the meson distribution amplitude $\varphi(\tau)$ 
while the second one is related to 
the asymmetric gluon distribution.
Other 3 lowest-order diagrams 
can be written in a similar way. 
Applying  formally   the power counting  
 (see Eq.    (\ref{50}) and discussion
preceding it),   we may  conclude
that   each   gluonic  contribution has 
an extra $Q^2$
factor  compared to the quark term, 
since the quarks have twist 1 while the twist of 
 the gluon vector potential $A_{\mu}$ is zero.
Technically, the enhancement appears when
the $p_{\mu} p_{\nu}$ factor from the matrix element
$\langle p' |  A_{\mu}^a (z_1)  A_{\nu}^b (z_2) | p \rangle $
combines with the $q'_{\mu}$,  $q'_{\nu}$ 
factors from hard propagators and polarization vectors, thus producing 
the estimate  $\langle p' |  A 
A | p \rangle  \sim Q^2$.  
However, the  power counting  formulas like  (\ref{50}) 
only give an upper estimate for the relevant 
contribution. The actual behavior is determined 
by the twist $t_{\cal O}$ of the composite operator
${\cal O}$ constructed 
from the elementary fields corresponding
to the external lines of the ${\rm SD}$-subgraph. 
It is well-known that the simplest {\it gauge-invariant}
composite operator containing two gluonic fields
is  $G_{\mu \rho } G_{\nu}^{ \rho }$, and its 
twist  equals 2 rather than 0,
just like for the lowest-twist $\bar \psi \ldots \psi$
operator.   Diagrammatically, this means that, in the Feynman gauge, 
the leading power terms of four lowest-order diagrams
completely cancel each other and the total result
is suppressed by $1/Q^2$ compared to leading
 contributions of separate  diagrams. 
In general, picking out  nonleading 
power terms (higher twist contributions)
is a notoriously difficult
problem of perturbative QCD.
However, in  our  case, the cancellation
of  leading terms is guaranteed by  gauge invariance
of the total result.  
Hence, choosing a gauge in which 
the combination  
$ q'_{\mu}  q'_{\nu} \langle p' |  A_{\mu}^a (z_1) 
 A_{\nu}^b (z_2) | p \rangle $ is prevented from 
producing the $(q'p)^2$ factor, we would 
eliminate the artificially enhanced
terms on diagram by diagram basis.
 This goal is achieved if one uses   the  gauge
$q'^{\mu} A_{\mu} (z;q') = 0$.
Then    $A_{\mu}$
can be expressed in terms of  the field-strength tensor 
$G_{\mu \rho}$ (see, $e.g.,$ \cite{ikr}) 
\begin{equation} 
A_{\mu}(z;q') = q'^{\rho} \int_0^{\infty} 
G_{\mu \rho}(z+\sigma q') \, e^{-\epsilon \sigma}
\, d \sigma . 
\label{134} \end{equation}
This representation also 
makes it easy to parametrize the matrix element
$\langle p' |  A_{\mu}^a (z_1)  A_{\nu}^b (z_2) | p \rangle $
in terms of the gauge-invariant gluon distribution:
\begin{eqnarray} 
\hspace{-1cm} &&\lefteqn{\langle p' \,  | \,   
A_{\mu}^a (z_1;q')
A_{\nu}^b (z_2;q')\, | \,p  \rangle |_{(z_1 -z_2)^2=0}
= \frac{\delta^{ab}}{N_c^2-1}\,  \frac {\bar u(p')  \hat q' 
 u(p)}{2 (q' \cdot p)}
 \, \biggl (- g_{\mu \nu} + 
\frac{p_{\mu}q'_{\nu}+p_{\nu} q'_{\mu}}{(p \cdot q')} \biggr ) 
} \label{135}  \\ &&  
\times \int_0^1  \frac12  \left [ e^{-iX(pz_1)+iX'(pz_2)}
 + e^{iX'(pz_1)-iX(pz_2)} \right ] 
 \frac{{\cal F}^g_{\zeta}(X)}{(X-i\epsilon)(X' +i \epsilon)}  
 \,  dX  \ + \  ``{\cal K} ".
\nonumber 
\end{eqnarray} 

In Ref.  \cite{bfgms},   the amplitude 
of hard diffractive electroproduction was calculated 
for the longitudinal polarization of both the 
virtual photon ($\epsilon_{\gamma^*}^{\mu} = 
(q'^{\mu}+ \zeta p^{\mu})/Q$)
and produced vector meson ($\epsilon_{V}^{\mu} =q'^{\mu}/m_V$).
In this case,  the contribution of Fig.  10$a$ in the $(q'A)=0$
gauge can be written as
\begin{eqnarray} 
T_{LL}^a (p,q',r)  \sim \bar u(p') \hat q' u(p) 
\int_0^1  d \tau \, \varphi_V(\tau) \int_0^{1}  \, 
{\rm Sp} \left \{ \hat \epsilon_{V}  \hat \epsilon_{\gamma^*}
\frac{\zeta \hat p - \tau  \hat q'}{(\zeta p - \tau q')^2} \gamma_{\mu} 
\frac{(X-\zeta) \hat p + \tau \hat q'}{((X-\zeta)  p + \tau q')^2}
\gamma_{\nu}  \right \}  \nonumber  \\ 
\times \biggl (- g_{\mu \nu} + 
\frac{p_{\mu}q'_{\nu}+p_{\nu} q'_{\mu}}{(p \cdot q')} \biggr ) 
 \frac{{\cal F}^g_{\zeta}(X) }{(X-i\epsilon)(X-\zeta+i\epsilon)} \, dX \, , 
\label{136A} \end{eqnarray}
where $\varphi_V(\tau)$ 
is the distribution amplitude of the longitudinal 
 vector meson.  This gives 
\begin{equation} 
T_{LL}^a (p,q,r)  \sim   \frac{\bar u(p') \hat q' u(p)}{Q m_V} \int_0^1 
\varphi_{V}(\tau) 
 \,\frac{ d \tau} {\tau }
\int_0^{1} 
  \frac{{\cal F}^g_{\zeta}(X) }{X(X-\zeta+i\epsilon)} \, dX \, .
\label{136} \end{equation}
Other diagrams give similar contributions, 
differing only  in the  $\tau$-dependent factor.
For Fig.  10$b$, one should  substitute $1/\tau$ by $1/\bar \tau$,
while Figs.10$c,d$ both have  $1/\tau \bar \tau $ factor. 
Since $1/\tau + 1/\bar \tau = 1/\tau \bar \tau $,
the total contribution also has the $ 1/\tau \bar \tau $ structure
\begin{equation} 
T_{LL} (p,q,r)  \sim   \frac{\sqrt{1-\zeta}}{Q m_V} \int_0^1 
\varphi_{V}(\tau) 
 \,\frac{ d \tau} {\tau \bar \tau }
\int_0^{1} 
  \frac{{\cal F}^g_{\zeta}(X) }{X(X-\zeta+i\epsilon)} \, dX \, ,
\label{136B} \end{equation}
where $\sqrt{1-\zeta}$  comes from 
$\bar u(p')  = \sqrt{1-\zeta} \, \bar u(p)$. 
The  amplitude $T_{LL} (p,q,r) $
has  imaginary part due to the factor  $1/(X-\zeta +i\epsilon)$:
\begin{eqnarray} 
 \frac1{\pi}\, {\rm Im} \, T_{LL}(\zeta ) 
\sim    \frac{\sqrt{1-\zeta} }{\zeta Q m_V}
\, {\cal F}^g_{\zeta} \, (\zeta)\, 
\int_0^1 \frac{\varphi_V(\tau)}
{\tau \bar \tau} \, d \tau  \, .
\label{137} \end{eqnarray} 
In Ref.  \cite{bfgms},
the gluonic matrix element was approximated by the 
gluon distribution function $f_g(\zeta)$. 
To get our result from that of Ref.  \cite{bfgms},
one should substitute there $f_g(\zeta)$ by  $\sqrt{1-\zeta} \,
{\cal F}^g_{\zeta}(\zeta) /\zeta$.

Though 
the asymmetric  distribution function 
${\cal F}^g_{\zeta}(X)$ coincides with 
$X f_g(X)$ in the limit $\zeta =0 $, in general 
these two functions  differ
 when $\zeta \neq 0$.
As discussed earlier,   the imaginary part 
appears for $X= \zeta$, ${\rm i.e.},$ in an
asymmetric  configuration in which the second gluon 
carries a vanishing fraction
of the original hadron momentum,
while 
$\zeta f_g(\zeta)$  corresponds to a symmetric 
 configuration  in which the final gluon
has  the momentum equal to that of  the initial one.

\section{Factorization and end-point effects}

\subsection{General remarks}

The standard question about  PQCD applications 
for hard  processes is whether   factorization
of short- and long-distance contributions
is maintained in higher orders. 
Since the Feynman integrals
can be written in different 
representations,  one can approach the factorization problem
in various  ways. 
In particular,  the classic studies 
of  deep inelastic scattering in QCD \cite{opemu,gw,gp}
relied on the operator product expansion 
in which  the coordinate representation
plays a crucial role. 
The claims  that factorization 
also holds for a more complicated Drell-Yan process
\cite{ardy,poldy} were supported by  studies 
\cite{apv,libbyst,egmpr} based  on the 
analysis in  the momentum representation 
(see, however, \cite{erad378}). 
The early studies  of exclusive processes in QCD 
which started with the analysis of the large-$Q^2$
behavior of the pion EM form factor 
also incorporated both the OPE-like
coordinate representation methods 
\cite{cz77,ar77}  and momentum-representation
oriented approaches \cite{farjack,bl}. 
 Factorization was intensively studied in the  following   years
(see \cite{css,blpqcd} and references therein).
Referring an interested reader to Ref.  \cite{cfs} 
for a recent momentum-representation analysis 
of factorization for hard exclusive  electroproduction 
 processes,  here we   briefly discuss 
 possible sources 
of factorization breaking analysing them 
within our approach \cite{tmf} based on the combined use of 
the $\alpha$-representation and the OPE-type methods.

\subsection{Structure of the lowest-order term}

 Exclusive processes   are rather vulnerable to factorization 
breaking. In contrast to   inclusive  cross sections, 
 factorization for exclusive ampitudes may fail even 
at the tree 
level. Hence it is a good idea just to write down the lowest-order contribution 
and carefully look at it. Take the DVCS amplitude (\ref{123}).
It has terms $1/(X-i \epsilon)$ and 
$1/(X-\zeta +i \epsilon)$ which 
are singular for $X=0$ and $X=\zeta$, respectively.
An immediate question is whether these singularities appear 
within the region of integration and if yes, whether they are inside
that region or at its end-points. 
To be prepared to address this question,
we performed a  detailed study of  spectral 
properties of  nonforward distributions.
Our $\alpha$-representation analysis shows that
$0 \leq X \leq 1$. Since the singularity $1/(X-\zeta +i \epsilon)$
is inside the integration region, we can write it as 
${\rm P} \{ 1/(X-\zeta ) \}  - i \pi \delta(X-\zeta)$:
it generates both real and imaginary parts of the amplitude.
On the other hand, the $1/(X - i \epsilon)$ singularity is at the end-point,
and the relevant real part is given by a divergent integral unless the 
nonforward distribution ${\cal F}_{\zeta}(X)$ vanishes at $X=0$.
Hence, to claim factorization for the real part, it is 
absolutely necessary  to 
give the arguments that ${\cal F}_{\zeta}(0) =0 $.
In our analysis, we derived ${\cal F}_{\zeta}(X)$ from the
double distribution $F(x,y)$.  The basic expression 
for ${\cal F}_{\zeta}(X)$ shows that ${\cal F}_{\zeta}(X) \sim X$ 
for any $F(x,y)$ which is finite as $x,y \to 0$.
One can  get ${\cal F}_{\zeta}(0) \neq 0 $ only if $F(x,y)$
is singular for $x=0$, $e.g.,$ if it behaves like 
$\delta(x)$ and does not vanish when $y =0$. 
If $F(x,y)$ has such a behavior, there should be a 
very special reason for it. 

Similarly, for the meson electroproduction, 
the integral over $\tau$ contains the  factor
$1/\tau (1- \tau)$ singular
at the endpoints $\tau =0 $ , $\tau =1$.
Again, the factorization formula makes sense only
if the distribution amplitude $\varphi(\tau)$ 
vanishes for $\tau =0 , \, 1$.
Since $\varphi(\tau)$ is analogous to 
the $\zeta=1$ limit of a nonforward distribution,
we may expect that it also vanishes 
 at $\tau =0$  because of  small phase space
for the $\tau \to 0$ configuration. 
Furthermore, since for massless quarks $\varphi(1- \tau) = \pm \varphi( \tau) $,
if $\varphi ( \tau)$ vanishes at $\tau =0$, it also vanishes
for $\tau =1$. 

Of course,  even if  the vanishing at end-points  holds for any diagram 
of perturbation theory, this still does not mean that
the nonperturbative functions have the same property.
So, a cautious statement  might be that if in perturbation
theory some  function
does  not vanish at a particular  end-point,
it is unlikely that it will vanish nonperturbatively.
If it vanishes perturbatively, there is some hope that
this  property is preserved for the nonperturbative function.

The standard procedure to get an educated guess
concerning the end-point behavior of 
hadron distribution amplitude $\varphi(\tau, \mu)$ 
is to study the asymptotic $\mu \to \infty$ limit 
of their evolution. This 
idea is equivalent to saying that 
$\varphi(\tau, \mu)$ has the same behavior 
at the end-points as the relevant BL evolution kernel $V(\tau,\tau')$.
In particular, in refs.\cite{cz} it was shown that
$\varphi^{\rm as} (\tau) \sim  \tau (1-\tau)$
both for the longitudinally and transversely polarized $\rho$-mesons.

Similar estimates of the end-point 
behavior of the  distribution amplitudes follow from 
QCD sum rule considerations. 
In particular, if   perturbative term 
$\Pi^{pert}(\tau, M^2) $ 
of the QCD sum rule ($M^2$  is the SVZ-Borel parameter) 
\begin{equation}
f_{\rho} \varphi (\tau)e^{-m_{\rho}^2/M^2} + \text{higher states} = 
\Pi^{pert}(\tau, M^2) +\text{condensates}
\end{equation}
vanishes for $\tau =0$ and $ \tau =1$, one can argue that
because of quark-hadron duality,
$ \varphi (\tau)$ should also vanish at the end-points.
For the correlators corresponding to the 
leading-twist $\rho$-meson distribution amplitudes,
we have indeed $\Pi^{pert} (\tau, M^2) \sim \tau , (1- \tau)$
at the end-points.

Note, furthermore,  that both quark and gluon propagators 
of  the simplest  hard subraph 
have  denominators  proportional to $\tau$.
However, for a longitudinally polarized 
virtual photon, only the $O(\tau)$  term 
in the numerator of the quark
propagator survives which converts  the 
$1/ \tau^2$ singularity  of the hard amplitude
into $1/\tau$. This will not happen
if the $\rho$-meson  is transversely polarized.
Hence, for transverse polarization 
one would face the integral with $\varphi_T(\tau)/\tau^2$
which logarithmically diverges if 
  $\varphi_T(\tau) \sim \tau$
for small $\tau$.
This result excludes the transverse case from 
straightforward PQCD applications.
This fact was repeatedly emphasized in refs.
\cite{bfgms,afs,cfs,fks}.

\subsection{Double-flow regime}

One of the lessons from the  discussion  above
is that  taking into account only  the denominators of the
``hard'' quark and gluon propagators 
one is guaranteed to get a  $1/\tau^2$ factor
capable of destroying factorization 
from the very start. 
It is the cancellation of one power of $\tau$ by 
a numerator factor  in case of a
longitudinally  polarized virtual photon
 which makes the factorization possible.
In the absence of this cancellation,
$e.g.,$ for transversely polarized $\rho$-meson, 
even if we take $\varphi_T (\tau) \sim \tau (1-\tau)$,
the integral would logarithmically diverge.
One may object that in such a situation factorization
still works if  $\varphi_T(\tau)$ vanishes faster
than $\tau$ as $\tau \to 0$. 
Note, however, that  evolution 
generates  terms proportional to 
$\tau$: 
$$\varphi_T(\tau,\mu^2) 
= \varphi_T(\tau,Q^2) + \ln Q^2/\mu^2 
\int_0^1  V(\tau, \tau') \varphi_T (\tau',\mu^2) \, 
d \tau'  \, , 
$$ 
since $V(\tau, \tau') \sim \tau$ for small $\tau$. 
In the presence of nonzero masses $m$ or other
infrared cutoffs,  one should change $1/\tau$
by $1/(\tau +m^2/Q^2)$.
As a result, the  logarithmic divergence converts  the $\tau$ integral
into an extra  $\ln Q^2/m^2$.
Together with the evolution logarithm  $\ln Q^2/\mu^2$,
they would amount to a double logarithm 
in a one loop diagram of Fig.  11 type. 
It should be emphasized  that this is not 
a Sudakov double logarithm. 
In particular, in two loops one would only get
$\ln^3 Q^2$ ($\ln^2 Q^2$ from evolution and $\ln Q^2$
from the $\tau$-integral) rather than $\ln^4 Q^2$.
The possibility to get an extra logarithm in 
the form-factor-type amplitudes was discovered 
 a long time ago in a scalar model
(see $e.g.,$ Ref.  \cite{poggio}).
In a scalar model,  there are no numerator 
factors to moderate the $1/\tau^2$ singularity,
 hence  such a possibility is always realized.
In Ref.  \cite{tmf}, the  diagram of Fig.   11 type 
for a scalar analogue of the pion form factor was studied 
with the help of the 
$\alpha$-representation and the Mellin transformation. 
It was shown that,  in the superrenormalizable 
$\phi^3_{(4)}$ model,  this diagram has the  
$\ln [Q^2/m^2]/Q^4$ behavior
despite the fact that there is no logarithmic evolution
in this model. The logarithm appears because   
the leading  ${\rm SD}$-pole $1/(J+2)$  for 
the Mellin transform of  this diagram
can be obtained in two ways:  from the 
 small-$\rho_L$ integration ($\rho_L = \alpha_1 + \alpha_2$)
and from the  small-$\rho_R$ integration ($\rho_R = \alpha_4 + \alpha_5$).
There are no other possibilities.
In particular, small-$\lambda$ integration 
($\lambda = \alpha_1 + \alpha_2 +\alpha_3 +\alpha_4 +\alpha_5)$
gives a nonleading pole $1/(J+3)$. 
Hence, the leading term comes from a configuration
in which the large momentum $Q$ flows simultaneously
through two subgraphs $V_L=\{\sigma_1,\sigma_2\}$
and $V_R=\{\sigma_4,\sigma_5\}$ while the momentum through
the  intermediate  line $\sigma_3$  is small.
Such a configuration
 was called in Ref.  \cite{tmf} the double-flow regime.

In a renormalizable $\phi^3_{(6)}$ model
the  diagram shown in Fig.  11 has the  $\ln^2 [Q^2/m^2]/Q^4$ behavior 
because the leading  ${\rm SD}$-pole $1/(J+2)$ 
can be obtained in three ways:  from small-$\lambda$ integration, 
from small-$\rho_L$ integration 
and from small-$\rho_R$ integration. 
The factorization for a  scalar analog
of the pion form factor in the  $\phi^3_{(6)}$ model 
was studied in more detail in Ref.  \cite{fi3}.
It was shown there, in particular, that 
the $\ln^2 Q^2/m^2$ behavior 
of the one-loop diagram results from the overlap of the evolution 
and the double-flow regime.
In Ref.  \cite{tmf},  it was emphasized that the presence
of the double-flow regime is a natural feature
of exclusive ampitudes. Hence, to establish factorization,
one should first check whether it is present or not.
For the pion form factor in QCD (and other renormalizable
models with spin-$\frac12$ quarks) its absence 
to all orders was demonstrated in
Ref.  \cite{tmf}.

A rather peculiar  double-flow contribution
appears in a two-loop  PQCD  diagram for the
nucleon form factors \cite{dumul}. Its specifics is that
it  works for a term in which 
one takes only quark masses in the numerators
of the propagators of the intermediate lines.
Proceeding by a routine calculation, it is rather 
difficult to detect such a contribution
among a wide variety of two loop terms. 
However, it is rather easy to find it 
if one has a guiding principle,
such as  the requirement that 
both the left and right components 
of a double-flow configuration 
should simultaneously give the leading  power
behavior.

\section{Comparison with other approaches and notations}

In  our definitions of various distribution
functions,  we took  the   relevant matrix element
and  expressed it through  an integral 
representation over the momentum fractions,
incorporating the spectral condition  $0 \leq X \leq 1$.
Another approach is to introduce 
distribution functions  by 
making  a Fourier transform of the matrix element 
with respect to $(pz)$ (cf. \cite{colsop,efp,jaffe}).
One can easily derive the result of such a procedure 
by rewriting our representations in a form 
with a universal exponential in the r.h.s.
Consider, $e.g.$, the matrix element for the
quark operator:
\begin{eqnarray} 
\hspace{-1cm} 
&&\lefteqn{
\langle \, p'  , s' \,  | \,  \bar \psi_a(0)  \hat z 
E(0,z;A)  \psi_a(z) 
\, | \,p ,  s \,  \rangle |_{z^2=0}
 } \label{qsim}  \\ \hspace{-1cm} &&  
= \bar u(p',  s')  \hat z 
 u(p,s) \, \int_{-1 + \zeta}^1   
 e^{-i \widetilde X (pz)} \left [
 {\cal F}^a_{\zeta}(\widetilde X;t) \, \theta(0 \leq \widetilde X  \leq 1) 
-   {\cal F}^{\bar a}_{\zeta}(\zeta -\widetilde X ;t) 
\, \theta(-1 + \zeta  \leq X  \leq \zeta)   \right ] 
 \,  d\widetilde X 
+ 
`` {\cal K}"  \, .
\nonumber
 \end{eqnarray} 
The Fourier transformation
would project out the function
\begin{equation} 
\widetilde {\cal F}^a_{\zeta}(\widetilde X;t) =
{\cal F}^a_{\zeta}(\widetilde X;t)\, \theta(0 \leq\widetilde  X  \leq 1) 
-   {\cal F}^{\bar a}_{\zeta}(\zeta -\widetilde X ;t) 
\, \theta(-1 + \zeta  \leq \widetilde X  \leq \zeta)  \label{092}
\end{equation}
which {\it a)} coincides with the quark 
distribution for $\zeta \leq \widetilde X \leq 1$,
{\it b)} reduces to  the (minus) antiquark distribution for 
 $-1+ \zeta \leq \widetilde X \leq 0$
and {\it c)} 
is given by their difference for $0 \leq \widetilde X \leq \zeta$.
 The   $\widetilde X$-variable 
changes within the segment   $(-1 + \zeta,1)$  
centered at $\widetilde X = \zeta/2$,
with the total range length equal to $2-\zeta$. 
To avoid the 
nonsymmetric and $\zeta$-dependent limits,
one can introduce the variable (cf.\cite{ji}) 
\begin{equation} 
\tilde x \equiv  \frac{ \widetilde X - \zeta/2}
{1- \zeta/2} \label{093}
\end{equation}
which changes from $-1$ to 1.
The ratio 
\begin{equation} 
\xi \equiv \frac{\zeta}{1- \zeta /2}
\label{094}
\end{equation}
is an alternative parameter 
characterizing  the longitudinal momentum 
asymmetry of the nonforward matrix element. 
For $t=0$ and a massless hadron, it varies between 0 and  2.
The reversed relations are
\begin{equation} 
\zeta = \frac{ \xi} {1+ \xi /2} \  \ ,  \  \  
\widetilde X =  \frac{\tilde x +  \xi /2}{1+ \xi /2}  \  \ ,  \  \  
\widetilde X  -\zeta =  \frac{\tilde x -  \xi /2}{1+ \xi /2} \, . 
\label{095}
\end{equation}

Using translation invariance (cf. Eq.    (\ref{69})),
one can easily derive that 
the   operator 
with the quark fields taken at symmetric points $-z/2, z/2$
has a rather compact 
representation in terms of the $\tilde x$-variable:
\begin{eqnarray} 
\langle \, p'  , s' \,  | \,  \bar \psi_a(-z/2)  \hat z 
E(-z/2,z/2;A)  \psi_a(z/2) 
\, | \,p ,  s \,  \rangle |_{z^2=0}
= \bar u(p',  s')  \hat z 
 u(p,s) \, \int_{-1}^1   
 e^{-i\tilde x  (Pz)} 
H_a (\tilde x, \xi ;t)  
 \,  d\tilde x 
+ 
``E_a"  \, ,
 \label{096}  
 \end{eqnarray} 
where $P=(p+p')/2$ is the average 
momentum of the initial and final hadron
(note, that $(Pz) = (1-\zeta/2)(pz) = (pz)/(1+ \xi/2) \, $).
This representation is equivalent 
to the definition of the 
{\it off-forward parton distributions} 
$H_a(\tilde x, \xi ; t)$,  $E_a(\tilde x, \xi ; t) $
introduced  by X. Ji \cite{ji} (see also \cite{drm}).
Basically, the latter are related 
to our  nonforward distributions 
by 
\begin{equation} 
\widetilde  {\cal F}_{\zeta}^a(\widetilde X;t) = (1+ \xi/2) 
H_a(\tilde x, \xi ; t)  \, , \label{097}  
\end{equation}
and similarly for other functions. 
The off-forward distributions $H_a(\tilde x, \xi ; t)$, ${\rm etc.}$ 
are defined both for positive and negative $\tilde x$.
Depending on the value of $\tilde x$,
one can  distinguish three different components:
  quark  ($\xi / 2 \leq \tilde x \leq 1$),
antiquark  ($-1 \leq \tilde x \leq - \xi / 2 $)
and   mixed ``quark minus antiquark'' 
($- \xi / 2 \leq \tilde x \leq  \xi / 2 $) components of $H$. 
  The mixed component corresponds  evidently to  the region
$0 \leq \widetilde X \leq \zeta$ of the $\widetilde X$-variable
in which the  nonforward distributions
can be treated as distribution amplitudes.
Since $\widetilde X (pz) = (\tilde x+\xi/2)(Pz)$
and $(\widetilde X - \zeta )  (pz) = (\tilde x- \xi/2)(Pz)$,
the partons in this picture carry  momenta $(\tilde x+\xi/2)P$
and $(\tilde x-\xi/2)P$. Using Eqs.(\ref{095}),(\ref{097}), etc.
one can convert  our  evolution kernels  $W_{\zeta}(X,Z)$ 
into those obtained  earlier by X. Ji in Ref.  \cite{ji2}.
The gluonic matrix element can be also represented 
in the form of Eq.    (\ref{096}):
\begin{eqnarray} 
\hspace{-1cm} 
&&\lefteqn{
\langle p'  \,  | \,   
z_{\mu}  z_{\nu} G_{\mu \alpha}^a (-z/2) E^{ab}(-z/2,z/2;A) 
G_{ \alpha \nu }^b (z/2)\, | \,p  \rangle |_{z^2=0}
 }  \nonumber \\ \hspace{-1cm} &&  
= \frac12  \, \bar u(p')  \hat z 
 u(p) \, (  P z) \,  \int_{-1}^1   
  e^{-i\tilde x(Pz)}
H_g(\tilde x,\xi;t)  
 \, d \tilde x + 
``E_g"  \, .
 \label{099}  
 \end{eqnarray} 
Due to the symmetry  property $H_g(\tilde x,\xi;t)  = H_g(-\tilde x,\xi;t)$,
integration over $\tilde x$ in this case can be restricted to 
the $0 \leq \tilde x \leq 1$ region.  
Note, that in the forward limit  $\xi =0$, $t=0$, the function $H_g(\tilde x,\xi;t)$ 
reduces to $\tilde xf_g (\tilde x)$ (cf. Eq.    (\ref{66})).
To get an off-forward distribution 
reducing to   $f_g(\tilde x)$, X. Ji \cite{ji2} uses 
the definition equivalent to adding a factor of $\tilde x$
in the integrand on  the right-hand side of Eq.    (\ref{099}):
$H_g(\tilde x,\xi;t)  \to \tilde x H_g^{\rm Ji}(\tilde x,\xi;t)$.
However,   $\tilde x=0$ corresponds 
to $X= \zeta/2$ or  to the middle-point $Y=1/2$  
of the distribution amplitude  $\Psi^g_{\zeta} (Y)$ 
(see Eq.    (\ref{095})), ${\rm i.e.},$ to a situation  when 
the gluons   carry   equal  fractions $\zeta p/2$
of the original momentum $p$.  
Since $H_g(\tilde x,\xi;t)$ is an even function of $\tilde x$,
there are no evident  reasons that it vanishes
for $\tilde x=0$. Hence, dividing $H_g(\tilde x,\xi;t)$ by $\tilde x$
produces  an artificial   singularity  of 
$H_g^{Ji}(\tilde x,\xi;t)$ for $\tilde x=0$.

Another parametrization for  the nonforward 
matrix element of the gluon operator  was proposed   by Collins, 
Frankfurt and Strikman \cite{cfs}. 
Their definition of the {\it nondiagonal}
gluon distribution $f_g(x_1,x_2 ; t)$  is 
also based on the Fourier transformation.
For positive values, their variables  $x_1, x_2$ 
correspond to our fractions  $X$ and $X - \zeta \equiv X'$,
respectively. 
In  our notations, the function 
$f_g(x_1= X,x_2= X-\zeta; t)$ can be written as
\begin{equation}
f_g( X, X-\zeta;t) = \frac1{ X( X-\zeta)} \,  
\widetilde {\cal  F}_{\zeta}^g ( X;t) \, . \label{910}
\end{equation}
The  factor $1/X(X-\zeta)$  was  motivated  
by the necessity to cancel the  inverse factor 
which may   emerge 
from the derivatives present in the field-strength tensor
$G_{\mu \nu}$.  
Actually,  this  expectation is  not supported  
by perturbative  calculations.
Take, $e.g.,$ the evolution kernel $P_{\zeta}^{gQ}(X,Z)$.
It can be treated as a  perturbative,  leading logarithm
approximation for the   gluon distribution inside a quark
(cf. \cite{hbhoy}).
According to Eq.    (\ref{98}),  $P_{\zeta}^{gQ}(X,Z)$ does not 
vanish   for $X = \zeta$.
If ${\cal  F}_{\zeta}^g (X)$ 
does not vanish for  $X= \zeta$,   the 
 function $\widetilde {\cal  F}_{\zeta}^g ( X;t)$
does not vanish both for $X=0$ and $X=\zeta$ 
and  $f_g(x_1,x_2)$  is singular 
both for $x_1=0$ and $x_2=0$. 

In fact,  the combination ${\cal  F}_{\zeta}^g (X)/(X- i \epsilon)
(X-\zeta + i \epsilon )$
appears  in our parametrization (\ref{135}) 
for the matrix element of the  operator constructed 
from two vector potentials  $A_{\mu}$,  $A_{\nu}$
taken in the light-cone gauge.
In this sense,  $f_g(x_1,x_2)$ or,
what is the same,  ${\cal  F}_{\zeta}^g (X)/X(X-\zeta)$
can be treated as a   basic  gluon distribution 
given by   the matrix element 
of the product of fundamental  gluonic fields $A_{\mu}A_{\nu}$
rather than by that of  the secondary fields $G_{\mu \rho}G^{\rho}_{\nu}$.
Note, however, that if $f_g (x_1,x_2)$, 
${\rm i.e.}$ ${\cal  F}_{\zeta}^g (X)/X(X-\zeta)$,  
has no singularities, then the meson electroproduction 
amplitude has  no imaginary part at leading twist.
Since this is  
impossible, $f_g (x_1,x_2)$ {\it must} have singularities,
and one may  wish to explicitly display them specifying 
 their nature, $e.g.,$ $1/(x_2- i \epsilon )$, 
$1/(x_1 +i \epsilon )$. This goal is achieved automatically 
if   ${\cal  F}_{\zeta}^g (X)$ is used as the basic distribution.

In our approach,  the starting point is the double distribution
$F_g(x,y;t)$  defined through the nonforward
matrix element of the  gauge-invariant gluonic 
operator
\begin{eqnarray} 
\hspace{-1cm} 
&&\lefteqn{
\langle p'  \,  | \,   
z_{\mu}  z_{\nu} G_{\mu \alpha}^a (0) E^{ab}(0,z;A) 
G_{ \alpha \nu }^b (z)\, | \,p  \rangle |_{z^2=0}
 }   \nonumber \\ \hspace{-1cm} &&  
=  \, \bar u(p')  \hat z 
 u(p) \, (  p z) \,  \int_0^1  dx  \int_0^1  \, \frac1{2} 
  \left (  e^{-ix(pz)-iy(r z)} + e^{ix(pz)-i\bar y(r z)}
\right ) F_g(x,y;t) \, \theta(x+y \leq 1) \,   dy \, . 
 \label{911}  
 \end{eqnarray} 
As explained earlier, in perturbation 
theory
the spectral properties
$0 \leq \{ x,y ,  x+y \}  \leq 1$ 
can be proved  to any order with the help of the $\alpha$-representation. 
Furthermore, 
the function $F_g(x,y;t)$
does not depend on the $\zeta$-parameter.
The family of 
$\zeta$-dependent nonforward gluon distributions
  ${\cal  F}_{\zeta}^g (X;t)$  is obtained from 
$F_g(x,y;t)$  by  integration over $y$ (see (\ref{40})):
\begin{equation}
{\cal F}_{\zeta}^g (X;t) = 
 \int_0^{ {\rm min} \{ X/\zeta, \bar X / \bar \zeta  \} }
 F_g(X-y \zeta,y;t) \, dy \,  .
\label{912} \end{equation}
Recall that the double distribution
$F_g(x,y;t)$ can be treated as a distribution
function with respect to $x$ and as a distribution
amplitude with respect to $y$. 
This physical interpretation suggests that
$F_g(x,y;t)$ is a regular function for all values 
of $y$ and for at least nonzero values of $x$. 
We  made this reservation
because  the evolution asymptotically  makes $F_g(x,y;t;\mu)$ 
(we added the dependence on the
factorization scale $\mu$)
proportional to $\delta(x)$ as $\mu \to \infty$.  
In this  situation,  $F_g(x,y;t;\mu)$  is 
singular at  $x=0$. However,  the 
$\delta(x)$-term still  produces a regular  nonforward 
distribution  ${\cal  F}_{\zeta}^g (X;t)$, 
though     confined to the restricted region $0 \leq X \leq \zeta$. 

Assuming that  the double distribution
$F_g(x,y;t;\mu)$ is finite  everywhere, we 
conclude that the nonforward 
distribution  ${\cal  F}_{\zeta}^g (X;t;\mu)$
in this case 
is  also finite for all $0 \leq X \leq 1$ and,
moreover, that  it vanishes for $X=0$.
As discussed earlier, the latter property 
is vital for factorization.
If it is not fulfilled,  the $X$-integral
in the  lowest-order expression diverges at the
end-point $X=0$, where the $1/(X- i \epsilon)$
prescription is of no help. One may think that
this problem can be avoided if one uses 
the function $\widetilde {\cal  F}_{\zeta}^g ( \widetilde  X ;t)$
defined through the Fourier transformation
with the variable  $ \widetilde  X $ changing from
$-1 + \zeta$ to 1. Since  the point $\widetilde  X =0$ 
is inside the integration region,  the 
$1/(\widetilde  X - i \epsilon)$ prescription apparently may help.
Note, however, that if our function ${\cal  F}_{\zeta}^g ( X ;t)$
does not vanish for $ X =0$, the  Fourier transform 
$\widetilde {\cal  F}_{\zeta}^g ( \widetilde  X ;t)$ 
is not continuous both for $\widetilde  X =0$ and $\widetilde  X =\zeta$.
As a  result, the singularities  of
$\widetilde {\cal  F}_{\zeta}^g ( \widetilde  X ;t)/(\widetilde  X - i \epsilon)
(\widetilde  X -\zeta + i \epsilon)$  are not integrable.

\section{Conclusions} 

In this paper, we discussed basic properties
of nonforward parton distributions, a new type
of functions accumulating nonperturbative information 
about hadron dynamics.
We  demonstrated that
there are two basic ways to describe  asymmetric matrix
elements $\langle   p' | {\cal O}(0,z) |p \rangle$
of quark and gluon light-cone operators ${\cal O}(0,z)$.
One possibility is to  introduce 
    double distributions
$F(x,y;t)$ which are independent 
of the longitudinal momentum asymmetry  (or skewedness) 
parameter $\zeta = 1-(p'z)/(pz)$
of the matrix element and refer to the light-cone  
fractions $xp$, $yr$ of the original hadron momentum 
$p$ and momentum transfer $r =p' - p$ carried 
by the active parton. Another approach is to use 
 nonforward  distribution
functions ${\cal F}_{\zeta}(X;t)$  which 
specify the light-cone projection of the
total momentum   $Xp=xp+yr$ carried  
by the parton.  These functions ${\cal F}_{\zeta}(X;t)$
explicitly 
depend on  $\zeta$. 
Both types of distributions have hybrid properties, 
in some aspects resembling usual parton distribution
functions and in  other ones the distribution
amplitudes. Their $t$-dependence
is analogous to that of  hadronic form factors.  
The use of  ${\cal F}_{\zeta}(X;t)$ is more convenient for 
ultimate applications to hard PQCD processes, resulting in  
a formalism that is  very similar to the standard 
PQCD parton picture. On the other hand, 
the  double distributions $F(x,y;t)$ 
have more transparent   spectral properties 
which has serious advantages  
  at the foundation  stages  of the PQCD analysis.
In this paper, we concentrated on general aspects
of the theory of nonforward distributions and their
uses.  There are many  interesting applications
to deeply virtual Compton scattering and  hard exclusive
electroproduction  processes which require 
further,  more specific  studies of the 
nonforward distribution functions including modeling 
their nonperturbative low-energy shape, logarithmic 
PQCD evolution, calculation of nonlogarithmic 
higher-order corrections, {\rm etc.}
Work in this direction has already 
been started \cite{ji}-\cite{ffgs}.

\section{Acknowledgments}

I thank   A. Afanasev, I. Balitsky, J. Blumlein, A. Freund, L. Frankfurt,
I. Halperin, 
P. Hoodbhoy,  C. Hyde-Wright, X. Ji, G. Korchemsky, 
E.Levin, L. Mankiewicz, I. Musatov, D.Robaschik, 
A. Szczepaniak, G. Sterman, M.  Strikman, W. Lu 
and  A. Zhitnitsky 
for  useful discussions
and  correspondence.  
I am grateful to N.  Isgur for continued 
encouragement and advice.
This work was supported
by the US Department of Energy 
under contract DE-AC05-84ER40150.


\newpage

\begin{figure}[htb]
\mbox{
   \epsfxsize=15cm
 \epsfysize=5cm
 \hspace{2cm}  
  \epsffile{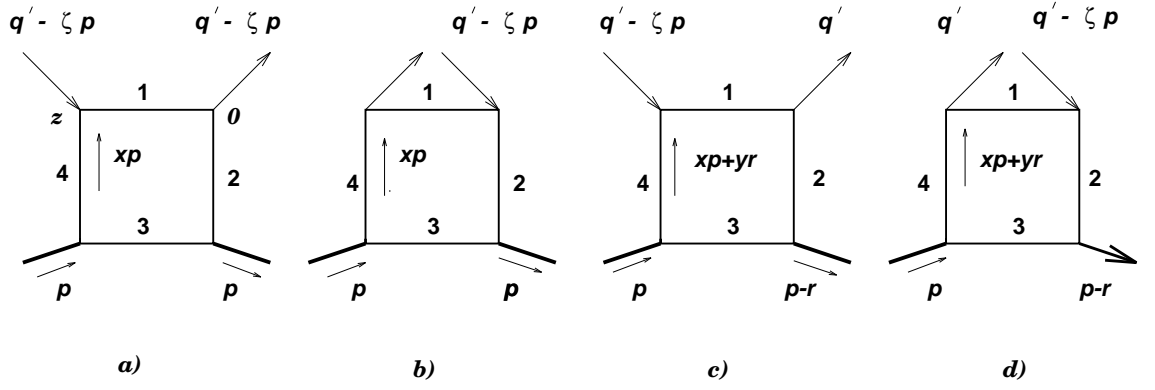}  }
  \vspace{0.5cm}
{\caption{\label{fig:1} Scalar model analogs of
$a),b)$ virtual forward Compton amplitude 
and $c),d)$ deeply virtual Compton scattering.
   }}
\end{figure}

\begin{figure}[htb]
\mbox{
   \epsfxsize=12cm
 \epsfysize=4.5cm
 \hspace{2cm}  
  \epsffile{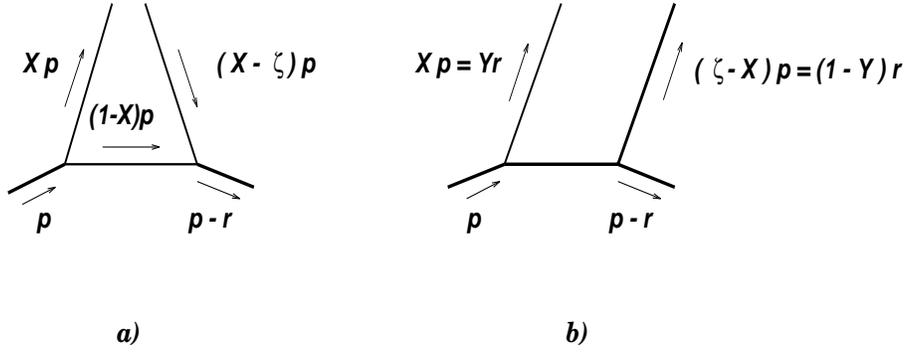}  }
  \vspace{0.5cm}
{\caption{\label{fig:2}  Longitudinal momentum flow
for two components of the
asymmetric distribution function ${\cal F}_{\zeta}(X)$:
{\it a)} $X > \zeta$ and {\it b)} $X < \zeta$.
   }}
\end{figure}

\begin{figure}[htb]
\mbox{
   \epsfxsize=12cm
 \epsfysize=5cm
 \hspace{2cm}  
  \epsffile{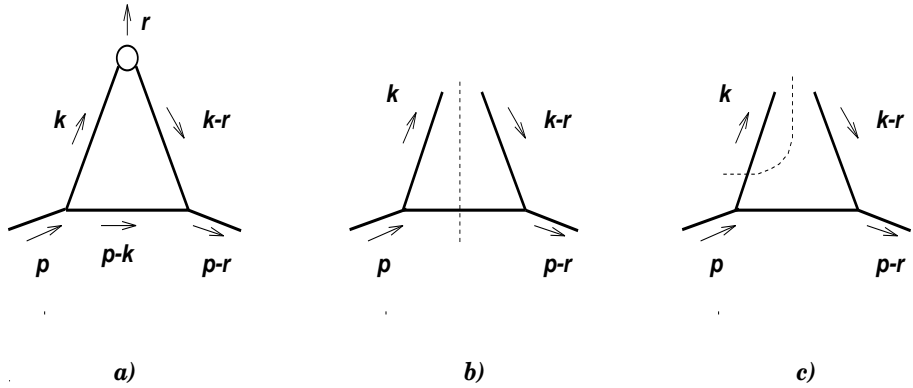}  }
  \vspace{0.5cm}
{\caption{\label{fig:3} $a)$ Structure of momentum integral defining 
the asymmetric distribution function ${\cal F}_{\zeta}(X)$.
$b)$ Cut of parton-hadron amplitude  corresponding to
the residue for the region $X > \zeta$.
$c)$ Cut of parton-hadron amplitude  corresponding to
the residue for the region $X < \zeta$.
}}
\end{figure}

\begin{figure}[htb]
\mbox{
   \epsfxsize=9cm
 \epsfysize=5cm
 \hspace{2cm}  
  \epsffile{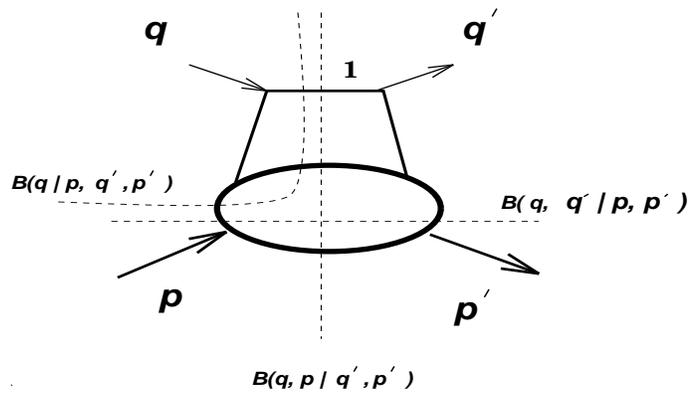}  }
  \vspace{0.5cm}
{\caption{\label{fig:4} Handbag diagram 
for deeply virtual Compton scattering.
   }}
\end{figure}

\begin{figure}[htb]
\mbox{
   \epsfxsize=12cm
 \epsfysize=5cm
 \hspace{2cm}  
  \epsffile{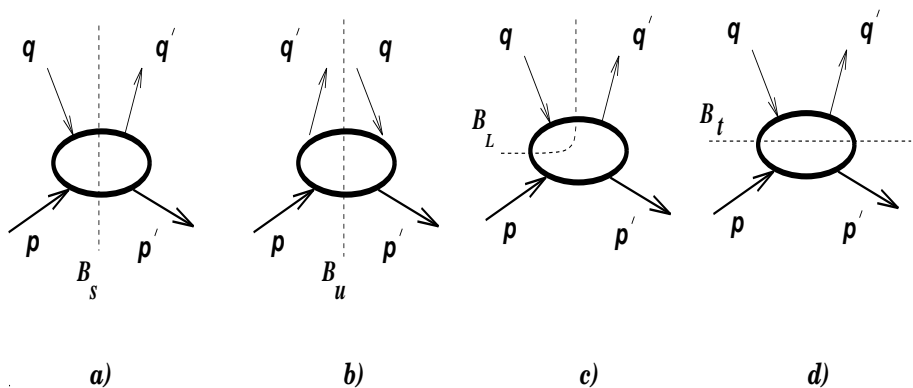}  }
  \vspace{0.5cm}
{\caption{\label{fig:5} Four-point amplitude corresponding to 
the  deeply virtual Compton scattering.
   }}
\end{figure}

\begin{figure}[htb]
\mbox{
   \epsfxsize=15cm
 \epsfysize=5cm
 \hspace{2cm}  
  \epsffile{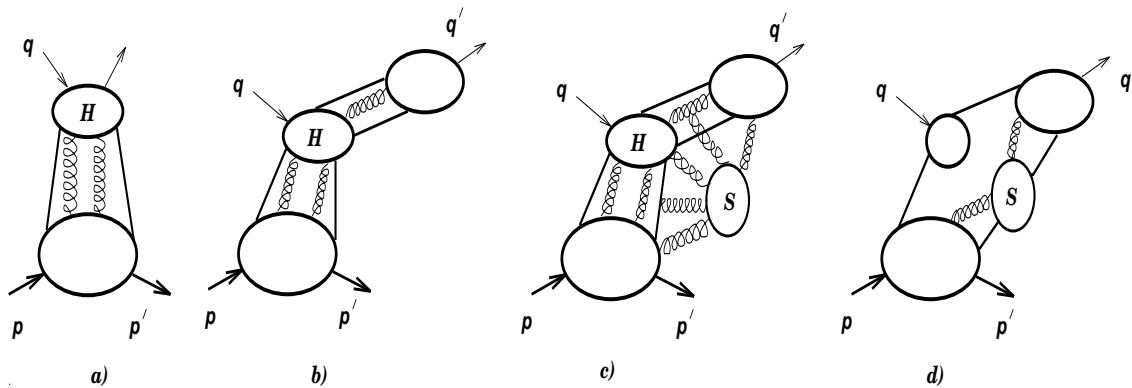}  }
  \vspace{0.5cm}
{\caption{\label{fig:6} Some regimes
responsible for powerlike contributions to
the DVCS amplitude.
   }}
\end{figure}

\begin{figure}[htb]
\mbox{
   \epsfxsize=12cm
 \epsfysize=3cm
 \hspace{2cm}  
  \epsffile{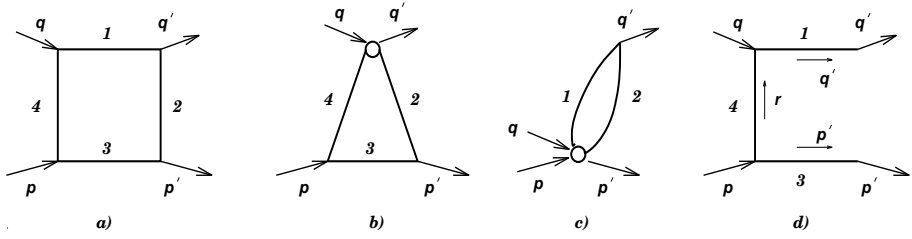}  }
  \vspace{0.5cm}
{\caption{\label{fig:7} $a)$ Scalar one-loop 
analog of the DVCS amplitude.
   Reduced graphs corresponding to $SD$-regimes
$b)$ $\alpha_1 \sim 0$ , $c)$ $\alpha_3 + \alpha_4 \sim 0$ 
and $d)$  $IR$-regime $\alpha_2  \sim  \infty$. 
   }}
\end{figure}

\begin{figure}[htb]
\mbox{
   \epsfxsize=12cm
 \epsfysize=5cm
 \hspace{2cm}  
  \epsffile{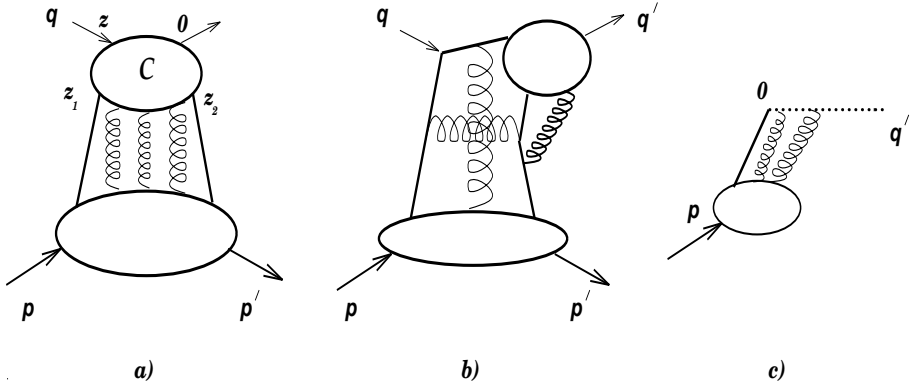}  }
  \vspace{0.5cm}
{\caption{\label{fig:8} {\it a)} General structure of the
leading $SD$ contribution to the DVCS amplitude in QCD.
{\it b)} $SD$ configuration with two
long-distance parts.{\it  c)} Matrix element 
with double-logarithmic 
dependence on the $UV$ cut-off parameter $\mu$. 
   }}
\end{figure}

\begin{figure}[htb]
\mbox{
   \epsfxsize=12cm
 \epsfysize=5cm
 \hspace{2cm}  
  \epsffile{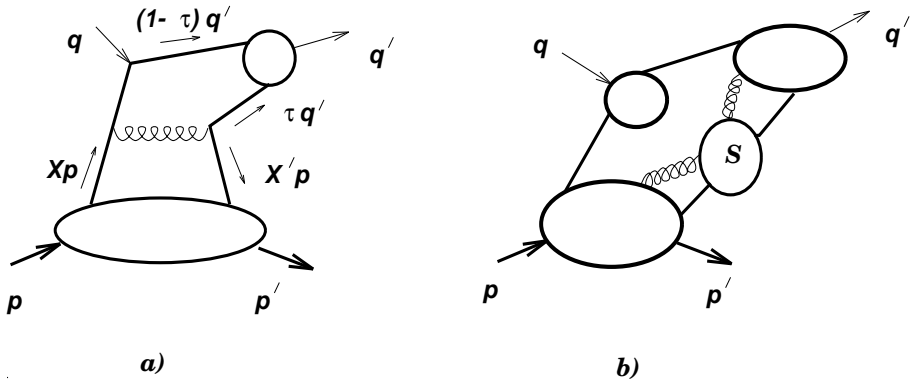}  }
  \vspace{0.5cm}
{\caption{\label{fig:9} Hard exclusive meson electoproduction 
process: {\it a)}  Leading $SD$-contribution 
with quark nonforward distribution;  {\it b)}
Soft contribution. 
   }}
\end{figure}

\begin{figure}[htb]
\mbox{
   \epsfxsize=15cm
 \epsfysize=5cm
 \hspace{2cm}  
  \epsffile{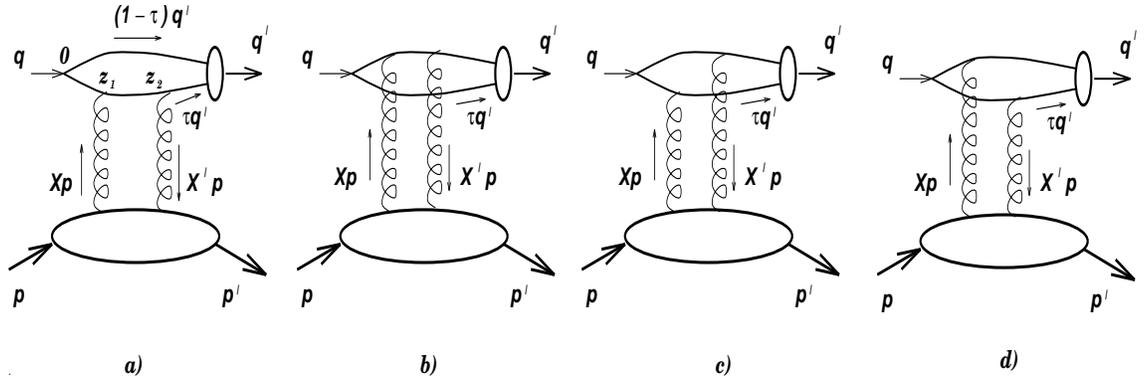}  }
  \vspace{0.5cm}
{\caption{\label{fig:10} Gluon contribution 
to hard exclusive meson electoproduction 
amplitude.
   }}
\end{figure}

\begin{figure}[htb]
\mbox{
   \epsfxsize=6cm
 \epsfysize=3cm
 \hspace{2cm}  
  \epsffile{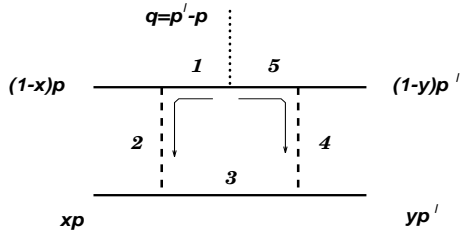}  }
  \vspace{0.5cm}
{\caption{\label{fig:11} Double-flow
regime  for the  scalar analog of a
meson form factor.
   }} \end{figure}

\end{document}